\documentclass[apj]{emulateapj}

\newcommand\rotate{\@pt@rottrue}%

\usepackage{graphicx}
\usepackage{amssymb}
\usepackage{natbib}
\usepackage{afterpage}
\usepackage{upgreek}

\newcommand{\spitzer}{\emph{Spitzer}}
\newcommand{\lumunit}{erg~s$^{-1}$}

\newcommand{\msun}{$M_\odot$}

\newcommand{\mbh}{$M_{\rm BH}$}
\newcommand{\nh}{$N_{\rm H}$}
\newcommand{\neii}{[Ne\,{\sc ii}]}
\def\micron{\hbox{$\upmu$m}} 

\shorttitle{Nuclear SF activity and BH accretion in nearby Seyfert galaxies}
\shortauthors{P. Esquej et al.}

\begin{document}

\title{Nuclear star formation activity and black hole accretion \\ in nearby Seyfert galaxies}

\author{
Pilar Esquej\altaffilmark{1,2,3},
Almudena Alonso-Herrero\altaffilmark{2,4},
Omaira Gonz\'{a}lez-Mart\'{\i}n\altaffilmark{5,6},
Sebastian F. H\"{o}nig\altaffilmark{7,8},
Antonio Hern\'{a}n-Caballero\altaffilmark{2,3},
Patrick F. Roche\altaffilmark{9},
Cristina Ramos Almeida\altaffilmark{5,6},
Rachel E. Mason\altaffilmark{10},
Tanio D\'{\i}az-Santos\altaffilmark{11},
Nancy A. Levenson\altaffilmark{12},
Itziar Aretxaga\altaffilmark{13},
Jos\'{e} Miguel Rodr\'{\i}guez Espinosa\altaffilmark{5,6},
Christopher Packham\altaffilmark{14}
}

\altaffiltext{1}{Centro de Astrobiolog\'ia, INTA-CSIC, Villafranca del Castillo, 28850, Madrid, Spain}
\altaffiltext{2}{Instituto de F\'{\i}sica de Cantabria, CSIC-Universidad de Cantabria, 39005 Santander, Spain}
\altaffiltext{3}{Departamento de F\'{\i}sica Moderna, Universidad de Cantabria, Avda. de Los Castros s/n, 39005 Santander, Spain}
\altaffiltext{4}{Augusto G. Linares Senior Research Fellow}
\altaffiltext{5}{Instituto de Astrof\'{\i}sica de Canarias (IAC), C/V\'{\i}a L\'{a}ctea, 38205, La Laguna, Spain}
\altaffiltext{6}{Departamento de Astrof\'isica, Universidad de La Laguna (ULL), 38205, La Laguna, Spain}
\altaffiltext{7}{UCSB Department of Physics, Broida Hall 2015H, Santa Barbara, CA, USA}
\altaffiltext{8}{Institut f{\"u}r Theoretische Physik und Astrophysik, Christian-Albrechts-Universit{\"a}t zu Kiel, Leibnizstr. 15, 24098 Kiel, Germany}
\altaffiltext{9}{Department of Physics, University of Oxford, Oxford OX1 3RH, UK}
\altaffiltext{10}{Gemini Observatory, Northern Operations Center, 670 N. A'ohoku Place, HI 96720, USA}
\altaffiltext{11}{Spitzer Science Center, 1200 East California Boulevard, Pasadena, CA 91125, USA}
\altaffiltext{12}{Gemini Observatory, Casilla 603, La Serena, Chile}
\altaffiltext{13}{Instituto Nacional de Astrof\'{\i}sica, \'{O}ptica y Electr\'{o}nica (INAOE), Aptdo. Postal 51 y 216, 72000 Puebla,  Mexico}
\altaffiltext{14}{Department of Physics and Astronomy, University of Texas at San Antonio, One UTSA Circle, San Antonio, TX 78249, USA}


\begin{abstract}
Recent theoretical and observational works indicate the presence of a
correlation between the star formation rate (SFR) and the active galactic
nuclei (AGN) luminosity (and, therefore, the black hole accretion rate, $\dot M_{\rm BH}$)
of Seyfert galaxies.  This suggests a physical connection between the gas
forming stars on kpc scales and the gas on sub-pc scales that is feeding
the black hole. We compiled the largest sample of Seyfert galaxies to
date with high angular resolution ($\sim\,0.4-0.8$\,\arcsec) mid-infrared (8--13\,\micron)
spectroscopy. The sample includes 29 Seyfert galaxies drawn from the
AGN Revised Shapley-Ames catalogue. At a median distance of 33 Mpc, our data allow us to
probe nuclear regions on scales of $\sim$\,65\,pc (median value). 
We found no general evidence of suppression of the 11.3\,\micron~polycyclic 
aromatic hydrocarbon (PAH) emission in the vicinity of these AGN, 
and used this feature as a proxy for the SFR.
We detected the 11.3\,\micron~PAH feature in the nuclear spectra of 45\% of our sample.
The derived nuclear SFRs are, on average, five times lower
than those measured in circumnuclear regions of $600\,$pc in size (median value). 
However, the projected nuclear SFR
densities (median value of 22\,\msun\,yr$^{-1}\,{\rm kpc}^{-2}$)  are a factor of 20 higher 
than those measured on circumnuclear scales.
This indicates that the SF activity per unit area in the central $\sim$\,65\,pc of Seyfert galaxies 
is much higher than at larger distances from their nuclei.
We studied the connection between the nuclear SFR
and $\dot M_{\rm BH}$ and showed that numerical simulations reproduce fairly well
our observed relation.

\end{abstract}


\section{Introduction}\label{sec:sec1}
One of the most important challenges in modern cosmology is to disentangle the physics behind the processes underlying galaxy formation
and evolution. Observations over the past decades have revealed that supermassive black holes (SMBHs) likely
reside at the centers of all galaxies with a bulge and that the properties of these black holes and their host galaxies are 
tightly correlated \citep[e.g.][]{Magorrian1998,Ferrarese2000,Gebhardt,Kormendy2013}. The co-evolution of galaxies and their corresponding SMBHs depends on some physical mechanism, referred to as feedback, that links accretion and ejection of gas residing on a sub-pc scale in galactic nuclei
to the rest of the galaxy \citep{Silk1998,King2010,Nayakshin2012}. The connection between star formation (SF) activity on different physical scales in
a galaxy and the presence of an active galactic nucleus (AGN) has been a long discussed topic. However, there still are many uncertainties under
consideration to disentangle the processes behind such a relation \citep[see e.g.][and references therein]{Hopkins2010}.

In the standard unification model,  the powering mechanism of AGN is
gas accretion onto a central SMBH. However, the physics of angular
momentum transfer to the vicinity of the black hole is still unclear
\citep[see][for a recent review]{Alexander2012}. Given that the
angular momentum of inflowing gas produced by galaxy mergers or other
large scale 
structures (e.g., bars) cannot be removed instantaneously, many
studies  proposed that the inflowing gas could form a 
circumnuclear disk where SF can take place. \citet[][and references
therein]{Kawakatu2008} put forward a model for  
such a circumnuclear disk, which might be coincident with the putative torus
of the unification model of AGN \citep{Antonucci1993}.  This
model predicts that SF would mostly take place in the outer parts of a
100\,pc-size torus \citep{Wada2002}. 
\citet{CidFernandes1995} proposed the presence of a starburst in the obscuring 
torus as a solution for the absence of conspicuous broad lines in Seyfert 2s.
The starburst disk 
model of \citet{Thompson2005} estimates that most of the gas is
supplied from outside the inner 200\,pc, but this is better suited for 
ultra-luminous infrared galaxies due to the high star formation rates
(SFRs) considered. \citet{Ballantyne2008} presented an update of the
\citet{Thompson2005} model with typical maximum SFRs of $\sim\,1\,M_\odot\,{\rm yr}^{-1}$,  that  
could also potentially obscure the AGN. These nuclear pc-sized
starbursts will mostly be associated with low luminosity AGN
(i.e. Seyferts and  
low-ionization nuclear emission-line regions -- LINERs).
From an observational point of view, nuclear starbursts have been detected in Seyfert 2 galaxies and LINERs using UV images obtained with the Hubble Space Telescope \citep{Heckman1995,GonzalezDelgado1998,Colina2002}.

The numerical simulations of \citet{Hopkins2010} predict a
relation with some scatter between the SFR of the galaxy on different
scales 
-- going from 10\,kpc-scale to the central parsec -- and the black hole accretion rate ($\dot M_{\rm BH}$). This correlation appeared to be more prominent on smaller physical scales. However, simulations also indicate dynamical delays between the peaks of the SF and the BH growth \citep{Hopkins2012},
in agreement with results from observational works \citep[e.g.,][]{Davies2007,Wild2010,Ramos2013}.

Mid-infrared (mid-IR) spectroscopy is a powerful tool to explore the
nature of AGN and SF activity in galaxies.  
Among the most remarkable characteristics of the mid-IR spectra of
galaxies is the presence of polycyclic aromatic hydrocarbons (PAH)
emission, with the most prominent features being at  6.2,
7.7, 8.6, 11.3 and 17\,\micron. 
They are due to the stretching and bending
vibrations of aromatic hydrocarbon materials, where the shortest
wavelength features are dominated by the smallest PAHs
\citep[e.g.][]{Tielens2010}. 
This type of emission mostly originates in photo-dissociation regions
where aromatic molecules are heated by the radiation field 
produced by young massive stars \citep{Roche1985, Roche1991}. Therefore,  PAHs are often used as indicators of the \emph{current}
SFR of galaxies. Note that they can also be excited by UV emission from B stars and thus PAH emission probes SF over a few
tens of million years \citep[e.g.,][]{Peeters2004,DiazSantos2010}.

PAH features are detected in AGN, although
they generally appear weak when compared with those of star forming 
galaxies \citep{Roche1991}. It has been    
proposed that the PAH molecules might be destroyed in the vicinity of
an active nucleus due to the presence of a hard radiation 
field \citep{Voit1992}.  There is also evidence that different PAHs
might behave differently. \citet{DiamondStanic2010} showed that the
11.3\,\micron\, PAH feature emission is a reliable indicator of the SFR in AGN,
at least for Seyfert-like AGN luminosities and kpc scales, while the 6.2, 7.7, and 8.6\,\micron\,features appear suppressed.
Sings of variations between the different features have been
reported by many authors \citep[e.g.][]{Peeters2004,Galliano2008}. 
For instance, \citet{Smith2007} found that the ratio of the PAH emission
at 7.7 and 11.3\,\micron~is relatively constant among pure starbursts, while it 
  decreases by up to 
factor of 5 for galaxies hosting a weak AGN.  
They interpreted this as a selective destruction of the smallest
PAH carriers by the hard radiation arising from the accretion disk, ruling out
the explanation in terms of ionization of the molecules \citep[see
also][]{Siebenmorgen2004}.

A number of works have studied the  SF activity using PAH emission and its relation to the AGN activity.
\citet{Shi2007} demonstrated that the SF contribution increases from Palomar-Green QSO, to 2MASS
QSO, and radio galaxies. Using measurements of the 7.7 and 11.3\,\micron~PAH using {\it Spitzer}/IRS data,
they found higher SFRs for more intense nuclear activity, which
indicates that the AGN selection technique influences the level of SF activity detected in the corresponding host galaxies.
\citet{Watabe2008} investigated the nuclear vs. circumnuclear SF for a sample of Seyfert galaxies using ground-based  observations
of the 3.3\,\micron~PAH feature. 
Assuming that the this PAH traces the SF activity, 
they found that both SF and AGN activity are correlated \citep[see also,][]{Imanishi2003, Imanishi2004}.
Such a relation implied that SF in the inner region of the AGN (within a few
hundred parsecs from the center) might have a greater influence on $\dot M_{\rm BH}$.
On the other hand, \citet{Mason2007} found weak or absent PAH emission in the central 20\,pc of the Seyfert~1 galaxy NGC\,1097, whilst in the circumnuclear region, strong 3.3 and 11.3\,\micron~PAH bands were detected. 
In the case of NGC\,1097, the absence of PAH emission may be related to destruction/ionization of 
PAH molecules by hard photons from the nuclear star cluster.

\citet{DiamondStanic2012} recently found a strong correlation between the kpc-scale SF derived using the 11.3\,\micron\,PAH feature
and 24\,\micron\,observations for Seyfert galaxies. However, the limited angular resolution of their 
\spitzer\,data  ($\sim\,4-5\,$\arcsec) did not allow them to resolve nuclear ($\sim\,100\,$pc) scales, 
and it is unclear if the measured PAH feature is associated with the galaxy or to the nuclear environment.

In this work we compile a sample of 29 Seyfert galaxies from the revised Shapley-Ames (RSA) galaxy catalogue
\citep{Sandage1987} with published ground-based mid-IR high angular resolution spectroscopy obtained on 8m-class telescopes. 
At a median distance of 33\,Mpc, this sample allows us to study the nuclear SF activity around AGN on scales of  
$\sim$65\,pc. We also use mid-IR spectra taken with the Infrared Spectrograph  \citep[IRS, ][]{Houck2004} on board {\it Spitzer}
for all objects in our sample to investigate the extended ($\sim\,600\,$pc) SF in the host galaxy. This enables
us to study the relation SFR-$\dot M_{\rm BH}$ at different scales in the local Universe.

This paper is structured as follows: Section~2 describes the sample
selection and data analysis. In Section~3 we study the nuclear 
11.3\,\micron\, PAH feature emission. Section~4 compares the circumnuclear and nuclear SF activity and its
relation with $\dot M_{\rm BH}$. Finally, our conclusions are summarized in Section~5. Throughout this work we assumed a $\Lambda$CDM cosmology with ($\Omega_{\rm M}$,~$\Omega_{\Lambda}$)~=~(0.3,~0.7) and ${H}_{0}$~=~70~ ${\rm km}~{\rm s}^{-1}~{\rm Mpc}^{-1}$.

\begin{table*}

\caption{Sample properties} 

\begin{center}
\small
\begin{tabular}{lrccrrrrrl}
\hline\hline
  \multicolumn{1}{c}{Object} &
  \multicolumn{1}{c}{$D_L$$^{\rm (a)}$} &
  \multicolumn{1}{c}{b/a} &
  \multicolumn{1}{c}{Type} &
  \multicolumn{1}{c}{log\,$L_{\rm 2-10 keV}$$^{\rm (b)}$} &
  \multicolumn{1}{c}{log\,$L_{\rm agn}$$^{\rm (c)}$} &
  \multicolumn{1}{c}{log(\mbh)} &
  \multicolumn{1}{c}{Refs.} \\ 
  & (Mpc) & & & (\lumunit) & (\lumunit) & (\msun) \\[0.5ex]
\hline

  Circinus   	& 4.2 	& 0.4 & Sy2 & 42.6$^{\rm (*)}$ 	& 43.8 & 6.42 & (1)\\ 
  ESO 323--G077 & 65.0 	& 0.7 & Sy1 & 42.7 				& 43.9 & 7.40 & (2)\\ 
  IC 5063  		& 49.0 	& 0.7 & Sy2 & 42.8 				& 44.0 & 7.74 & (1)\\		
  Mrk 509 		& 151.2 & 0.8 & Sy1 & 43.9 				& 45.4 & 7.86 & (3,4,5)\\  
  NGC 1068 		& 16.3 	& 0.9 & Sy2 & 43.0$^{\rm (*)}$ 	& 44.3 & 7.59 & (1)\\
  NGC 1365 		& 23.5 	& 0.5 & Sy1 & 42.1$^{\rm (d)}$ 	& 43.1 & 8.20 & (6)\\    
  NGC 1386 		& 12.4 	& 0.4 & Sy2 & 41.6$^{\rm (*)}$ 	& 42.6 & 7.42 & (1)\\
  NGC 1808		& 14.3 	& 0.6 & Sy2 & 40.4 				& 41.2 & \ldots & (7)\\               
  NGC 2110		& 33.6 	& 0.7 & Sy1 & 42.6 				& 43.7 & 8.30 & (3,4,8)\\            
  NGC 2992 		& 33.2 	& 0.3 & Sy1 & 43.1 				& 44.4 & 7.72 & (8) \\ 
  NGC 3081 		& 34.4 	& 0.9 & Sy2 & 42.5 				& 43.6 & 7.13 & (1)\\  
  NGC 3227 		& 16.6 	& 0.7 & Sy1 & 42.4 				& 43.5 & 7.62 & (9,4,10)\\
  NGC 3281		& 46.1 	& 0.5 & Sy2 & 42.6 				& 43.8 & 7.91 & (1)\\
  NGC 3783 		& 42.0 	& 0.9 & Sy1 & 43.2 				& 44.5 & 7.48 & (3,4,10)\\ 
  NGC 4151 		& 14.3 	& 0.7 & Sy1 & 42.1 				& 43.2 & 7.66 & (11,12)\\  
  NGC 4388		& 36.3 	& 0.2 & Sy2 & 42.9 				& 44.1 & 7.23 & (1)\\
  NGC 4507		& 51.0 	& 0.8 & Sy2 & 43.1 				& 44.4 & 7.65 & (1)\\
  NGC 4945 		& 3.6 	& 0.2 & Sy2 & 42.3$^{\rm (*)}$ 	& 43.4 & 6.15 & (13,14)\\ 
  NGC 5128 		& 3.7 	& 0.8 & Sy2 & 41.9 				& 42.9 & 7.84 & (4)\\   
  NGC 5135 		& 59.3 	& 0.7 & Sy2 & 43.1$^{\rm (*)}$ 	& 44.4 & 7.29 & (1)\\
  NGC 5347 		& 33.6 	& 0.8 & Sy2 & 42.4$^{\rm (*)}$ 	& 43.5 & 6.97 & (1,15)\\ 
  NGC 5506 		& 26.6 	& 0.3 & Sy1 & 43.0 				& 44.3 & 7.95 & (1)\\
  NGC 5643 		& 17.2 	& 0.9 & Sy2 & 41.4 				& 42.3 & 7.40 & (16,15)\\ 
  NGC 7130 		& 70.0 	& 0.9 & Sy2 & 43.1$^{\rm (*)}$ 	& 44.4 & 7.59 & (1)\\
  NGC 7172 		& 37.4 	& 0.6 & Sy2 & 42.2 				& 43.3 & 7.67 & (17)\\ 
  NGC 7213 		& 25.1 	& 0.9 & Sy1 & 42.1 				& 43.1 & 7.74 & (5)\\ 
  NGC 7469 		& 70.8 	& 0.7 & Sy1 & 43.3 				& 44.7 & 7.08 & (3,4,6)\\ 
  NGC 7479 		& 34.2 	& 0.7 & Sy1 & 42.0 				& 43.0 & 7.68 & (7,10)\\ 
  NGC 7582 		& 22.6 	& 0.4 & Sy1 & 41.9$^{\rm (d)}$ 	& 42.9 & 7.13 & (1,18)\\  [1ex]
\hline\hline\end{tabular}\\
\end{center}
{\bf Notes.}-- ${^{\rm (a)}}$Distances from NED.
{${^{\rm (b)}}$Compton-thick sources according to \citet{Marinucci2012}. Hard X-ray luminosities are corrected by a factor 70.}
{${^{\rm (c)}}$AGN bolometric luminosities calculated from X-ray luminosities after applying the bolometric corrections of \citet{Marconi2004}.}
{${^{\rm (d)}}$Changing-look AGN \citep[e.g.][]{Bianchi2005}. Data from an intermediate state.}
{${^{\rm (*)}}$Compton-thick sources.}\\

{\bf References.} (1) Marinucci et al. (2012), (2) Malizia et al. (2007), (3) Dadina et al. (2007), (4)  Tueller et al. (2008), (5) Asmus et al. (2011), (6) Risaliti et al. (2005), (7) Brightman et al. (2011), (8) Woo \& Urry (2002), (9) H\"{o}nig et al. (2010), (10) Diamond-Stanic \& Rieke (2012), (11) Wang et al. (2011), (12) Beckmann et al. (2006), (13) Guainazzi et al. (2000), (14) M\"{u}ller et al. (2003), (15) Beifiori et al. (2009), (16) Guainazzi et al. (2004), (17) Akylas et al. (2001), (18) Piconcelli et al. (2007).

\nocite{Marinucci2012}
\nocite{Risaliti2005}
\nocite{Brightman2011}
\nocite{Dadina2007}
\nocite{Tueller2008}
\nocite{Woo2002}	
\nocite{Hoenig2010}
\nocite{DiamondStanic2012}
\nocite{Wang2011}
\nocite{Beckmann2006}
\nocite{Guainazzi2000}	
\nocite{Mueller2003}
\nocite{Guainazzi2004}
\nocite{Akylas2001}
\nocite{Asmus2011}
\nocite{Malizia2007}
\nocite{Beifiori2009}
\nocite{Piconcelli2007}
\normalsize
\label{table:tab1} 
\end{table*}

\section{Sample selection and data analysis}\label{sec:sec2}

\subsection{Sample}\label{sec:sample}
Our sample (see Table~\ref{table:tab1}) is drawn from the galaxy-magnitude-limited RSA Seyfert sample, 
which includes the 89 Seyfert galaxies brighter than $B_{T}$=13\,mag from \citet{Maiolino1995} and \citet{Ho1997}. We selected galaxies
with existing high angular resolution ($\sim\,0.4-0.8$\arcsec)  mid-IR spectra observed on
8m-class telescopes. The sample contains a total of 29 Seyfert galaxies, of which 16  (55\%)
are Type~2 and 13  (45\%) are Type~1 AGN. We included in the Seyfert 1 category those
galaxies classified as Seyfert 1.5, 1.8, and 1.9, as well as those with broad near-IR lines.

We used the hard 2$-$10\,keV X-ray luminosity (see Table~\ref{table:tab1} for references) as a proxy for the AGN bolometric luminosities
after correcting for absorption and applying the bolometric corrections of \citet{Marconi2004}. 
The high column density in Compton-thick objects (defined as those having \nh $>10^{24}\,{\rm cm}^{-2}$, see Table~\ref{table:tab1}) prevents us from measuring the intrinsic nuclear luminosity below 10 keV. Instead, one can only derive the reflection component from model fitting. 
Assuming that the [O\,{\sc iii}] forbidden line is a tracer of the AGN intrinsic luminosity and comparing it with the observed hard X-ray emission of Compton-thick AGN, \citet{Marinucci2012} derived a correction factor of 70, that we used to correct the observed $2-10\,$keV luminosities of these objects\protect\footnote[1]{The X-ray luminosity of NGC\,7479 is not corrected by such a factor because, despite its high \nh, its Compton-thick nature is not confirmed. Therefore, its luminosity could be up to a factor of 60--70 times higher \citep{Panessa2006}.}
This large correction factor is also theoretically justified by the torus model proposed by \citet{Ghisellini1994}.
This is the method commonly used for Compton-thick sources and applied in other several works, as in e.g. \citet{Bassani1999,Panessa2006}. 
We expect the nuclear mid-IR and X-ray luminosities to be well correlated \citep[e.g.][]{Levenson2009,Asmus2011},
which is fulfilled for sources in our sample \citep{Hoenig2010,Gonzalez2013}.
The only significant outlier, NGC\,1808, indicates that for this source the AGN does not dominate the continuum mid-IR emission \citep[see figure~5 in][]{Gonzalez2013}. The uncertainties in
$L_{\rm agn}$  are driven by the scatter on the
relationship for the bolometric correction which, in general, is
significantly larger than the error on the X-ray luminosities. Based
on the $L_{\rm agn}$ determination, \citet{Young2010} 
derived typical uncertainties of 0.4\,dex \citep[see also][]{Marinucci2012}.

\begin{figure}
\center

\hspace{-0.5cm}
\includegraphics[scale=0.37,angle=90]{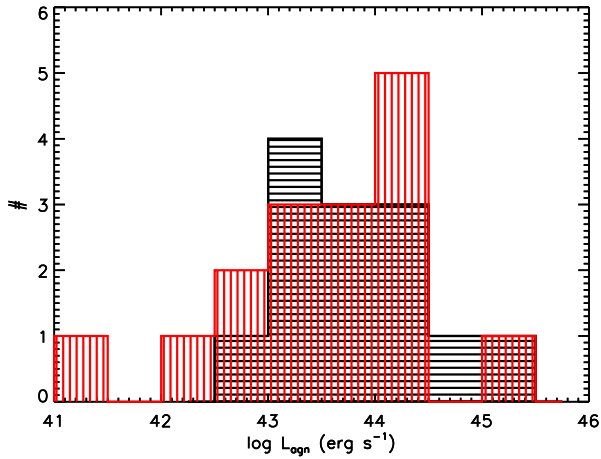} 

\caption{\label{fig:fig_Lagn_hist} Distribution of the AGN bolometric
  luminosities for Sy1 (black histogram, horizontal filling lines) and
  Sy2 (red histogram, vertical lines) galaxies in our sample.} 
\end{figure}

Our sample spans AGN bolometric luminosities in the range
$\log {L_{\rm agn}}=41.2-45.5\,{\rm erg \,s}^{-1}$, with a median value of 43.7\,\lumunit. 
This is a fair representation of the full RSA Seyfert sample \citep[see figure~1 of][]{DiamondStanic2012}$^{2}$.
As can be seen from Figure~\ref{fig:fig_Lagn_hist}, 
Type 1 and Type 2 sources have similar distributions of $L_{\rm agn}$, with median values (in logarithm scale) of
43.7 and 43.8 erg~s$^{-1}$ for Sy1 and Sy2s, respectively.
We also list in Table~\ref{table:tab1} the BH masses of the galaxies
in our sample and corresponding references. There is no available
black hole mass measurement for NGC~1808. The  median value for our
sample is 3.9$\times 10^7$\,\msun, which is similar to the 3.2$\times
10^7$\,\msun\, median value for the complete RSA Seyfert sample \citep{DiamondStanic2012}. 
In terms of the Eddington ratio, we sample values of  ${L_{\rm agn}/L_{\rm edd}=10^{-4}-0.3}$. 

\protect\footnotetext[2]{Note that in this figure they used the luminosity of the mid-IR line [O\,{\sc iv}] as the proxy for the AGN bolometric luminosities.}

\begin{table}

\caption{Ground-based high angular resolution mid-IR spectroscopy} 

\begin{center}
\small
\begin{tabular}{lccc}
\hline\hline
{Galaxy} &
{Instrument} &
{slit} &
 {Refs}\\
  && ($\arcsec$) \\
\hline
ESO~323$-$G077 & VISIR    & 0.75 & (4)\\
IC~5063       & T-ReCS   & 0.67 & (5,2) \\
              & VISIR    & 1.00 & (4)\\
Mrk~509       & VISIR    & 0.75 & (4)\\
NGC~1068      & Michelle & 0.36 & (6) \\
              & VISIR    & 0.40    & (4)\\
NGC~1365      & T-ReCS   & 0.35 & (7,2) \\  
NGC~1386      & T-ReCS   & 0.31 & (2) \\
NGC~1808      & T-ReCS   & 0.35 & (8,2)\\
NGC~2110      & Michelle & 0.36 & (9)\\
              & VISIR    & 0.75 & (4) \\
NGC~2992      & Michelle & 0.40 & (10)\\
NGC~3081      & T-ReCS   & 0.65 & (2)\\
NGC~3281      & T-ReCS   & 0.35 & (2,11)\\
NGC~3227      & VISIR    & 0.75 & (4)\\
NGC~3783      & VISIR    & 0.75 & (4)\\
NGC~4151      & Michelle & 0.36 & (12)\\
NGC~4388      & Michelle & 0.40 & (10)\\
NGC~4507      & VISIR    & 1.00 & (4)\\
NGC~5135      & T-ReCS   & 0.70 & (13,2)\\  
NGC~5347      & Michelle & 0.40 & (10)\\
NGC~5506      & T-ReCS   & 0.36 & (14,2)\\
NGC~5643      & T-ReCS   & 0.36 & (2)\\
              & VISIR    & 0.75 & (4)\\
NGC~7130      & T-ReCS   & 0.70 & (13,2)\\  
NGC~7172      & T-ReCS   & 0.36 & (14,2)\\
NGC~7213      & VISIR    & 0.75 & (4)\\
NGC~7469      & VISIR    & 0.75 & (4)\\
NGC~7479      & T-ReCS   & 0.35 & (2)\\
NGC~7582      & T-ReCS   & 0.70 & (2)\\
              & VISIR    & 0.75 & (3)\\
\hline\hline\end{tabular}\\
\end{center}
{\bf References.} (1) Roche et al. (2006), (2) Gonz\'{a}lez-Mart\'{\i}n et al. (2013), (3) H\"{o}nig et al. (2008), (4) H\"{o}nig et al. (2010),
(5) Young et al. (2007), (6) Mason et al. (2006), (7) Alonso Herrero et al. (2012),  (8) Sales et al. (2013),
(9) Mason et al. (2009), (10) Colling (2011), (11) Sales et al. (2011),
(12) Alonso Herrero et al. (2011), (13) D\'{\i}az Santos et al. (2010), (14) Roche et al. (2007).
\nocite{Hoenig2008}
\nocite{Roche2006}
\nocite{Gonzalez2013}
\nocite{Hoenig2010}
\nocite{Young2007}
\nocite{Mason2006} 
\nocite{AlonsoHerrero2012}
\nocite{Sales2013}
\nocite{Mason2009}
\nocite{Colling2011}
\nocite{Sales2011}
\nocite{Alonso2011}
\nocite{DiazSantos2010}
\nocite{Roche2007}
\normalsize
\label{table:tab2} 
\end{table}

\subsection{Observations} 

Ground-based mid-IR spectroscopic observations of the 29 Seyfert galaxies were taken with three different instruments. They operate on 8m-class telescopes and cover the $N$-band, $\sim 8-13\,\mu$m. Table~\ref{table:tab2}
summarizes details of the mid-IR spectroscopic observations, along with references where the data were
originally published.
Observations taken with the Thermal-Region Camera Spectrograph
\citep[T-ReCS][]{Telesco1998} on the 8.1\,m Gemini-South Telescope used the low resolution mode, which provides a spectral
resolution of $R=\Delta \lambda/\lambda \sim 100$, and slit widths between 0.31 and 0.70\arcsec. Observations performed by
Michelle \citep{Glasse1997} on the 8.1\,m Gemini-North telescope, which has a higher spectral resolution by a factor of two ($R\sim 200$), were obtained with slit widths of $\sim 0.4\arcsec$. Finally, observations  with the
VLT spectrometer and imager for the mid--infrared \citep[VISIR,][]{Lagage2004} instrument mounted on the 8.2\,m VLT UT3 telescope at the ESO/Paranal observatory were obtained
with the low spectral resolution mode ($R\sim 300$) and a slit width
of 0\arcsec .75 or 1\arcsec \, (and 0.4\arcsec\,for NGC\,1068).
For the typical distances of our sample the ground-based slit widths probe typical physical scales of $\sim$\,65\,pc.
These range from  $\sim$\,7--255\,pc for all objects except for Mrk\,509 (545 pc), which is by far the most distant galaxy in the sample.

Sixteen sources were observed with {\it Gemini}/T-ReCS \citep[][and references therein]{Gonzalez2013}. Thirteen sources have VLT/VISIR observations
\citep[see][for details]{Hoenig2010}, with 4 overlapping with  the T-ReCS sample. Finally, 6 Seyfert 
galaxies were observed with Gemini/Michelle \citep{Mason2006,Alonso2011,
Colling2011}, of which two sources also have VISIR observations. 
We refer the reader to the original papers for details on the observations and the data reduction. 

\begin{figure}
{\includegraphics[scale=0.5]{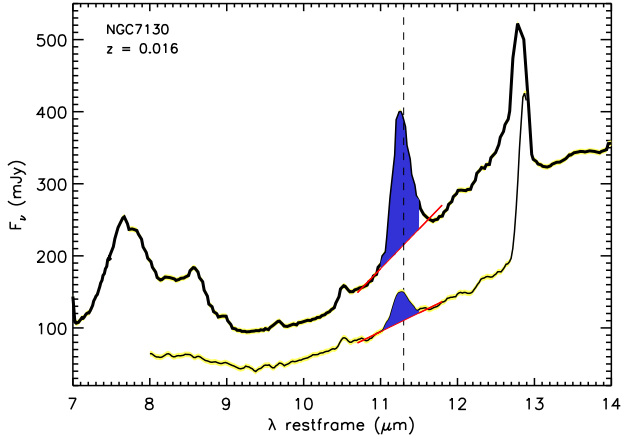}}\\
\caption{{\it Spitzer}/IRS SL spectrum (thick line) from CASSIS
  compared with the ground-based nuclear T-ReCS spectrum (thin line)
  from \citet{Gonzalez2013} 
of NGC~7130, one of the galaxies in our
  sample. We show the location of the 11.3\,\micron\,PAH feature, with
the blue shaded area indicating the spectral range used for obtaining the
integrated flux. The red lines are the fitted local continua. The rms
of the spectrum is shown in yellow. 
We note that the [Ne\,{\sc ii}]\,12.81\,\micron~emission line is contaminated by the 12.7\,\micron\,PAH feature, which cannot be resolved. The complete 
sample is shown in Figure~\ref{fig:all_spectra} of the Appendix.}
\label{fig:fig_example_spectrum}
\end{figure}

We retrieved mid-IR {\it Spitzer}/IRS spectra (for all sources except for NGC\,1068) from the Cornell
Atlas of \spitzer/IRS Source \citep[CASSIS v4,][]{Lebouteiller2011}. We used staring mode observations  
  taken with the short-low (SL) module covering the spectral range
  $\sim 5-15$\,\micron. The spectral resolution is   $R \sim 60-120$. The CASSIS database provides spectra
with  optimal extraction regions to ensure the best signal-to-noise ratio and are fully reduced. We only needed
to apply a small offset to stitch together the two short-wavelength modules SL1 and SL2 (but note that this does not affect the PAH measurement, see Section~\ref{sec:sec_measurePAH}). NGC\,1068 is part of the GOALS 
programme \citep{Armus2009} and IRS SH data have been obtained from the NASA/IPAC Infrared Science Archive (IRSA).
Assuming a typical spatial resolution of 3.7\,\arcsec\,for the SL module of IRS given by the slit
width, this corresponds to a physical scale of about $\sim$600\,pc for our sample, 
i.e. a factor of 10 less resolved than for the ground-based data.
Figure~\ref{fig:fig_example_spectrum} shows a comparison of \spitzer\,against ground-based data of NGC~7130 for illustration.
We present the spectra of the full sample in the Appendix (see Figure~\ref{fig:all_spectra}).

\subsection{Measuring the $11.3\,\mu$m PAH feature}\label{sec:sec_measurePAH}
A number of methods have been developed to provide accurate measurements of the PAH feature fluxes, specially for the relatively large
spectral range covered by IRS. These include, among others, {\tt PAHFIT} \citep{Smith2007}, {\tt DecompIR}
\citep{Mullaney2011}, and spline fit \citep[e.g.][]{Uchida2000,Peeters2002}. They are useful for decomposing IR spectra, especially when the AGN emission is contaminated by extra-nuclear emission. These techniques, however, might not be appropriate for the limited wavelength coverage 
of ground-based data and/or weak PAH features \citep[see][]{Smith2007}.

We  measured the flux of the 11.3\,\micron\, PAH feature following the method described
in \citet{Hernan2011}. We fitted a local continuum by linear interpolation of  the average flux in two narrow bands on both
sides of the PAH. We then subtracted this fitted local continuum and integrated the residual data in a spectral range centered on
11.3\,\micron\ ($\lambda_{\rm rest}$\,=\,11.05$-$11.55\,\micron), to obtain the PAH flux.
Figure~\ref{fig:fig_example_spectrum} illustrates the method. As can be seen from this figure, this procedure
slightly underestimates the PAH feature flux due to losses at the wings of the line profiles and overlaps between adjacent PAH
bands. We corrected for these effects by assuming that the line has a Lorentzian profile of known width and applying a multiplicative
factor \citep[see][for details]{Hernan2011}. We measured the equivalent width (EW) of the 11.3\,\micron~PAH feature 
by dividing the integrated PAH flux by the interpolated continuum at the center  of the
feature. We derived the uncertainties in the measurements by performing Monte Carlo
simulations. This was done by calculating the dispersion around the measured
fluxes and EWs in a hundred simulations of the original spectrum with
random noise distributed as the rms.

Additionally, PAH fluxes measured using a local continuum tend to be
smaller than those using a continuum fitted over a large spectral
range.  
To scale up our flux values to the total emission in the PAH features,
we used the multiplicative factor of 2. 
This value was derived by \citet{Smith2007} for the
11.3\,\micron\,PAH, after comparing results obtained by the  
spline fitting and the {\tt PAHFIT} full decomposition.
For consistency in all measurements, we used the same 
method described above for both the ground-based and IRS spectra. 
We detected the 11.3\,\micron~PAH feature at the $2\sigma$ level
or higher significance in all objects of our sample observed with IRS, 
except for NGC~3783 (see Figs.~\ref{fig:fig_example_spectrum} and \ref{fig:all_spectra}).
Our measurements of the $11.3\,\mu$m PAH fluxes agree well with those of \citet{DiamondStanic2012}
using {\tt PAHFIT}, even though they used their own spectral extraction from IRS data.

\section{Nuclear 11.3\,\micron~PAH feature emission}\label{sec:sec3}

\begin{table*}[ht]

\caption{Nuclear measurements of the sample.} 
\begin{center}
\hspace{-0.5cm}
\small
\begin{tabular}{lcrrrrrr}
\hline\hline
  \multicolumn{1}{c}{Name} &
  \multicolumn{1}{c}{Ins} &
  \multicolumn{1}{c}{Slit/size} &
  \multicolumn{1}{c}{$L_{\rm 11.3\,\mu m \,PAH}$} &
  \multicolumn{1}{c}{$EW_{\rm 11.3\,\mu m\, PAH}$} &
  \multicolumn{1}{c}{$SFR_{\rm nuclear}$} &
    \multicolumn{1}{c}{$\frac{ {\rm SFR}_{\rm nuclear} }{ {\rm SFR}_{\rm circ}}$} &
    \multicolumn{1}{c}{$\frac {f_{11.3\mu{\rm m \,
        PAH}}}{f_{\rm [NeII]}}$}\\
  & & (\arcsec/pc) & (10$^{40}$ \lumunit) & ($10^{-3}$\,\micron) & (\msun\,yr$^{-1}$)\\
\hline
Circinus${^{\rm (a)}}$	&	IRS	&	3.70/75	&	     5.1$\pm$  0.3	&	    61$\pm$  1	&	 0.13	&	 0.23	&     0.83\\
ESO323-G077	&	VISIR	&	0.75/235	&	$<$  9.3	&	$<$  9	&	$<$ 0.23	&	$<$ 0.07	&	$<$0.56\\
IC5063	&	T-ReCS	&	0.65/153	&	$<$  8.4	&	$<$ 12	&	$<$ 0.21	&	$<$ 0.53	&	\ldots\\
Mrk509	&	VISIR	&	0.75/545	&	    44.5$\pm$ 17.6	&	    10$\pm$  4	&	 1.11	&	 0.18	&  \ldots\\
NGC1068	&	Michelle	&	0.36/28	&	    14.4$\pm$  2.7	&	     9$\pm$  1	&	 0.36	&	 0.36	&  \ldots\\
NGC1365	&	T-ReCS	&	0.35/40	&	$<$  1.9	&	$<$ 18	&	$<$ 0.05	&	$<$ 0.06	&	$<$0.97\\
NGC1386	&	T-ReCS	&	0.31/19	&	$<$  0.5	&	$<$ 31	&	$<$ 0.01	&	$<$ 0.26	&	$<$0.91\\
NGC1808	&	T-ReCS	&	0.35/24	&	     8.2$\pm$  0.5	&	   365$\pm$ 22	&	 0.21	&	 0.17	&     0.74\\
NGC2110	&	VISIR	&	0.75/121	&	$<$  1.3	&	$<$  7	&	$<$ 0.03	&	$<$ 0.12	&	\ldots\\
NGC2992	&	Michelle	&	0.40/64	&	$<$  3.8	&	$<$ 30	&	$<$ 0.09	&	$<$ 0.14	&	$<$0.33\\
NGC3081	&	T-ReCS	&	0.65/107	&	$<$  1.9	&	$<$ 18	&	$<$ 0.05	&	$<$ 0.25	&	$<$0.34\\
NGC3227	&	VISIR	&	0.75/60	&	     2.9$\pm$  0.5	&	    63$\pm$ 11	&	 0.07	&	 0.18	&     0.58\\
NGC3281	&	T-ReCS	&	0.35/78	&	$<$  6.2	&	$<$  9	&	$<$ 0.16	&	$<$ 0.98	&	$<$0.40\\
NGC3783	&	VISIR	&	0.75/151	&	$<$  3.3	&	$<$  6	&	$<$ 0.08	&	$<$ 0.79	&	$<$0.22\\
NGC4151	&	Michelle	&	0.36/25	&	$<$  2.4	&	$<$ 17	&	$<$ 0.06	&	$<$ 1.03	&	$<$0.36\\
NGC4388	&	Michelle	&	0.40/70	&	$<$  7.7	&	$<$ 68	&	$<$ 0.19	&	$<$ 0.38	&	$<$0.18\\
NGC4507	&	VISIR	&	1.00/245	&	$<$  4.5	&	$<$  5	&	$<$ 0.11	&	$<$ 0.12	&	$<$0.10\\
NGC4945${^{\rm (b)}}$	&	IRS	&	3.70/64	&	     0.4$\pm$  0.1	&	   358$\pm$ 17	&	 0.01	&	 0.13	&     0.13\\
NGC5128${^{\rm (a)}}$	&	IRS	&	3.70/66	&	     0.6$\pm$  0.1	&	    65$\pm$  2	&	 0.01	&	 0.18	&     0.43\\
NGC5135${^{\rm (c)}}$	&	T-ReCS	&	0.70/200	&	     5.9$\pm$  2.2	&	    34$\pm$ 12	&	 0.15	&	 0.04	&     0.63\\
NGC5347	&	Michelle	&	0.40/65	&	$<$  6.5	&	$<$ 56	&	$<$ 0.16	&	$<$ 0.76	&	$<$0.64\\
NGC5506	&	T-ReCS	&	0.35/45	&	     6.4$\pm$  2.8	&	    15$\pm$  6	&	 0.16	&	 0.32	&     0.40\\
NGC5643	&	VISIR	&	0.75/62	&	     1.6$\pm$  0.2	&	    49$\pm$  4	&	 0.04	&	 0.24	&     0.29\\
NGC7130${^{\rm (c)}}$	&	T-ReCS	&	0.70/236	&	    47.1$\pm$  3.3	&	   166$\pm$ 11	&	 1.18	&	 0.22	&     1.11\\
NGC7172	&	T-ReCS	&	0.35/63	&	$<$  2.8	&	$<$ 43	&	$<$ 0.07	&	$<$ 0.16	&	$<$0.27\\
NGC7213	&	VISIR	&	0.75/91	&	$<$  0.7	&	$<$  8	&	$<$ 0.02	&	$<$ 0.14	&	$<$0.06\\
NGC7469	&	VISIR	&	0.75/255	&	    47.7$\pm$  4.7	&	    31$\pm$  3	&	 1.19	&	 0.10	&     0.82\\
NGC7479	&	T-ReCS	&	0.35/58	&	$<$  5.6	&	$<$ 37	&	$<$ 0.14	&	$<$ 0.82	&	\ldots\\
NGC7582	&	T-ReCS	&	0.70/76	&	     3.9$\pm$  0.7	&	    50$\pm$  9	&	 0.10	&	 0.30	&     0.41\\
\hline\hline\end{tabular}\\
\end{center}
{\bf Notes.}--- \\
\small
The 11.3\,\micron~PAH luminosities include a multiplicative
factor of 2 for comparison with {\tt PAHFIT} measurements (see Section~\ref{sec:sec_measurePAH}). \\
The values of the EW and $f_{11.3\mu{\rm m \,PAH}}/f_{\rm [NeII]}$ are derived from measurements 
using the fitted local continuum. Upper limits at a 2$\sigma$ significance are included for non detections with the $<$ symbol.\\
${^{\rm (a,b)}}$Circumnuclear SFRs from {\it ISO} 11.3\,\micron~PAH measurements from
${^{\rm (a)}}$\citet{Siebenmorgen2004} and ${^{\rm (b)}}$\citet{Galliano2008}, for 24\arcsec~and 20\arcsec~apertures, respectively.\\
${^{\rm (c)}}$Values for the \neii\,line from \citet{DiazSantos2010}.\\
\label{table:tab3}
\normalsize
\end{table*}

In this section we investigate the nuclear 11.3\,\micron~PAH feature emission 
in our sample of galaxies. Hereinafter, nuclear scales generally refer to the 
physical regions observed with the T-ReCS/Michelle/VISIR instruments, 
whereas circumnuclear scales are those probed with the IRS spectroscopy.
The only exceptions are the most nearby (distances of $\sim 4\,$Mpc) galaxies
Circinus, NGC~4945, and NGC~5128. 
To explore similar physical regions in comparison to the rest of the sample
we used the IRS observations as our nuclear spectra for the three galaxies. 
For these sources the circumnuclear data are from \citet{Siebenmorgen2004} 
and \citet{Galliano2008} (see Table~\ref{table:tab3} for more information).

The median value of the nuclear physical sizes probed with our data is
65\,pc (see Table~\ref{table:tab3}). 
If two nuclear spectra exist for the same galaxy, we used the one sampling a physical 
scale closest to the median value, for consistency with the rest of the galaxies.
This information along with the physical scales probed with the nuclear spectra are given 
in Table~\ref{table:tab3}. Note that this is approximately a factor of
10 improvement in physical resolution with respect to the circumnuclear median value of 600\,pc. 
Hereinafter, we will use the term \emph{size} as referring to the physical scale probed, which is determined by the slit widths of the observations.

\subsection{Detection of the $11.3\,\mu$m PAH feature}\label{sec:detection_PAH}
The 11.3\,\micron\,PAH feature is weak or it might not even be present in a large fraction of galaxies in our sample, 
as can be seen from the nuclear spectra in Figs.~\ref{fig:fig_example_spectrum} and
\ref{fig:all_spectra}.
We deemed the feature as detected if the 
integrated flux is, at least, two times above the corresponding measured error. This is
equivalent to having the PAH feature detected with a significance of $2{\rm \sigma}$ or
higher. The non-detections are given as upper limits at a $2\sigma$ level, that is,
with a 95\% probability that the real flux is below the quoted value. 

Table~\ref{table:tab3} gives the nuclear luminosities and EWs of the
11.3\,\micron~PAH detections, as well as $2\sigma$ upper limits 
for the remaining objects. Note that the flux of the 11.3\,\micron\,PAH feature is not 
corrected for extinction. Thus, its proper characterisation might be hampered in cases 
of high extinction, i.e. when the PAH molecules are embedded behind the silicate grains
and the feature is buried within the silicate absorption at $\sim$9.7\,\micron.
This also depends on the location of the material causing the extinction relative 
to the PAH emitting region. Another additional complication is the presence of crystalline silicate absorption at
around 11\,\micron, which has been detected in local ultraluminous infrared galaxies 
\citep{Spoon2006} and in local Seyferts \citep{Roche2007}. In particular, 
Colling (2011) detected crystalline silicate absorption that could be 
blended with the 11.3\,\micron\,PAH feature in some of the galaxies in our sample, 
namely, NGC~4388, NGC~5506, NGC~7172, and NGC~7479 (see also Section~\ref{sec:sec_stacking}).

Using T-ReCS/VISIR/Michelle data we detected nuclear 11.3\,\micron~PAH emission in 
10\,galaxies. For the three most nearby Seyferts, the 11.3\,\micron~PAH feature is detected in 
the IRS observations, while in the corresponding T-ReCS/VISIR spectra the feature is below 
the detection limits. Taking this into account, we detected nuclear 11.3\,\micron~PAH feature  
emission in 13 out of 29 galaxies ($45\%$ of the sample). The
detection rate is similar for Seyfert 1s and Seyfert 2s (40\% and 50\%, respectively). The
observed EWs of the feature (using the fitted local continuum) are
between $\sim 0.01-0.4$\,\micron. These values are much
lower than those typical of high metallicity star forming
galaxies \citep[see][]{Hernan2011}. This is expected given that we are probing 
smaller regions around the nucleus, and probably the continuum
  emission is  mostly arising from dust heated by  the AGN.

To study a possible extra-nuclear origin of the PAH feature we
  investigated the morphology of the galaxies in our sample. We
 compiled the {\emph{b/a}} axial ratio (measurements from NED, RC3 $D_{25}/R_{25}$ isophotal $B$-band diameters) 
 to determine the
  inclination of the host galaxy, where $b$ and $a$ are the minor
  and  mayor axis, respectively. Axial ratios  $b/a\,<$\,0.5 are
  considered as edge-on galaxies, whereas face-on galaxies have 
  $b/a\,>$\,0.5 (see Table~\ref{table:tab1}).
With this definition we find 11 edge-on and 18 face-on galaxies.
Out of the 13 sources
  with detection of the nuclear 11.3\,\micron\,PAH feature, we find 5 (44\%) edge-on
  and 8 (45\%) face-on galaxies. We do not find that a positive detection
  predominates in edge-on galaxies, where material of the host galaxy
  along our line of sight could be misinterpreted as nuclear SF. 
However, we cannot rule out a dominant contribution from extra-nuclear
SF in the most edge-on galaxies in our sample. 
\citet{Gonzalez2013} found that the host galaxies could significantly contribute to the 
nuclear component for sources with the deepest silicate absorption features.

The majority of the nuclear 11.3\,\micron~PAH detections in our sample are  galaxies with 
well-documented nuclear starbursts and/or {\it recent} SF activity based on UV and optical 
observations \citep[NGC~5135, NGC~7130]{GonzalezDelgado1998}, modelling of the optical 
spectra \citep[NGC~5135, NGC~5643, NGC~7130, NGC~7582]{StorchiBergmann2000}, and near-IR integral 
field spectroscopy \citep[Circinus, NGC~1068, NGC~3227, NGC~3783, NGC~5128]{Davies2007,Tacconi2013}.

\begin{figure}
\center

\includegraphics[scale=0.5]{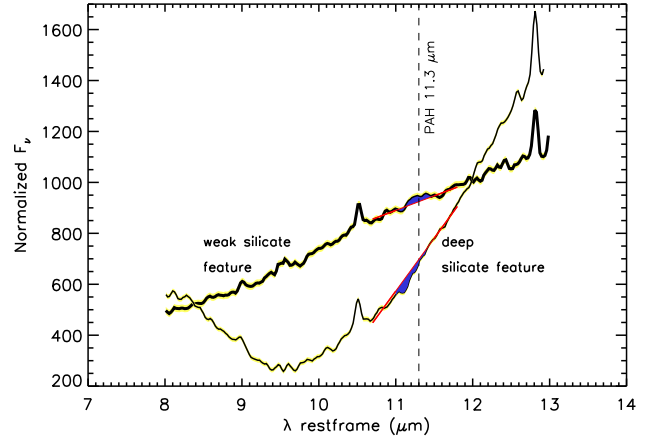} 

\caption{\label{fig:fig_stacked} Result from the stacking of nuclear
  spectra (from T-ReCS, Michelle, and VISIR) without 11.3\,\micron~PAH
  detections. The stacked spectrum for sources with weak silicate
  features (six galaxies, thick line) shows a $2\sigma$
    detection of the 11.3\,\micron~PAH feature. In the stacked spectrum (thin line)
  of the seven galaxies with deep silicate features the PAH feature
  remains undetected.
The flux density units are arbitrary. The individual spectra were normalized
  at 12\,\micron.  See Section~\ref{sec:sec_stacking} for details.}  
\end{figure}

\subsection{Stacking nuclear spectra with undetected $11.3\,\mu$m PAH emission}\label{sec:sec_stacking}
In this section we further investigate those galaxies with weak
or no detected  11.3\,\micron~PAH feature emission. We stacked the individual
spectra deemed to have undetected 11.3\,\micron~PAH features according to our
$2\sigma$ criterion (see Table~\ref{table:tab3} and Section~\ref{sec:detection_PAH}). 
We divided them in two groups. The first includes galaxies with a weak silicate feature: 
ESO~323-G077, NGC~1365, NGC~3081, NGC~3783, NGC~4151, and NGC~4507. 
The second group contains galaxies with relatively deep silicate features: 
NGC~1386, NGC~2992, NGC~3281, NGC~4388, IC~5063, NGC~7172 and NGC~7479.
 We excluded from the stacking NGC~2110 and NGC~7213 because the silicate feature 
is strongly in emission, and NGC~5347 because the spectrum is very noisy. 
We normalized the spectra at 12\,\micron\, and then used the  
{\tt IRAF} task {\tt scombine} with the \emph{average} option to
combine the different observations.

Figure~\ref{fig:fig_stacked} shows the stacked spectra for the two groups.
We applied the same method as in Section~\ref{sec:detection_PAH} to
determine if the PAH feature is detected. We  found that the 11.3\,\micron~PAH appears detected in the
stacked nuclear spectrum of the galaxies with weak silicate
  features at a $2\sigma$ level. The derived EW of the 11.3\,\micron~PAH is $8\times 10^{-3}$\,\micron. The feature remains undetected in the stacked nuclear spectrum of sources with deep silicate features. This could be 
explained in part as due to extinction effects, given that the
silicate absorption in these galaxies likely comes from cold
foreground material \citep{Goulding2012}. We also note that the
  minimum around 11\,\micron~in the stacked spectrum of galaxies with
  deep silicate features
could be from crystalline silicates. Indeed, \citet{Colling2011}  found
that inclusion of crystalline silicates
improved the fit of the silicate features in NGC~7172, NGC~7479, and
NGC~4388. Therefore, it would be expected
to also appear in the stacked spectrum.

\subsection{Is the $11.3\,\mu$m PAH feature suppressed in the vicinity of AGN?}
It has been known for more than 20 years now that PAH emission is weaker in
local AGN than in high metallicity star forming galaxies, although 
some AGN do also show strong PAH features on circumnuclear scales \citep{Roche1991}. 
It is not clear, however, if the decreased detection of PAH emission and
the smaller EWs of the PAH features in AGN are due to 1) an increased mid-IR 
continuum arising from the AGN, 2) destruction of the PAH carriers in 
the harsh environment near the AGN \citep{Roche1991, Voit1992} or 3)
decreased SF in the nuclear region \citep{Hoenig2010}. Additionally, there is a
prediction that smaller PAH molecules would be destroyed more easily in strong 
radiation fields \citep[see e.g.][]{Siebenmorgen2004}, also indicating that 
different PAHs may behave differently.

The  effects of an increasing continuum produced by the AGN is clearly 
seen in the 3.3\,\micron~PAH map of the central region of NGC~5128. The ratio  
of the feature-to-continuum  (i.e., the EW of the feature) decreases towards the AGN,
whereas the feature peaks in the center 
\citep[see][for more details]{Tacconi2013}. This 
implies that the PAH molecules are not destroyed in  the harsh
environment around the AGN of this galaxy (see also 
Section~\ref{sec:sec_PAHshielded}). Similarly,
  \citet{DiazSantos2010}  showed that at least the molecules
  responsible for the 11.3\,\micron~PAH feature can survive 
within $<$100\,pc from the AGN.

Some recent observational works reached apparently opposing
conclusions on the PAH emission of AGN on physical scales within a
few kpc from the nucleus, but note that these are for much larger physical scales than those probed here.
\citet{DiamondStanic2010} demonstrated for the RSA Seyferts that the
11.3\,\micron\,PAH feature is not suppressed, whereas other mid-IR PAH
features are. \citet{LaMassa2012}, on the other hand, combined optical and mid-IR 
spectroscopy of a large sample of AGN and star forming galaxies and concluded that in 
AGN-dominated systems (higher luminosity AGN) the 11.3\,\micron\,PAH feature does 
get suppressed.

\begin{figure}
\center

\hspace{-0.5cm}
\includegraphics[scale=0.37,angle=90]{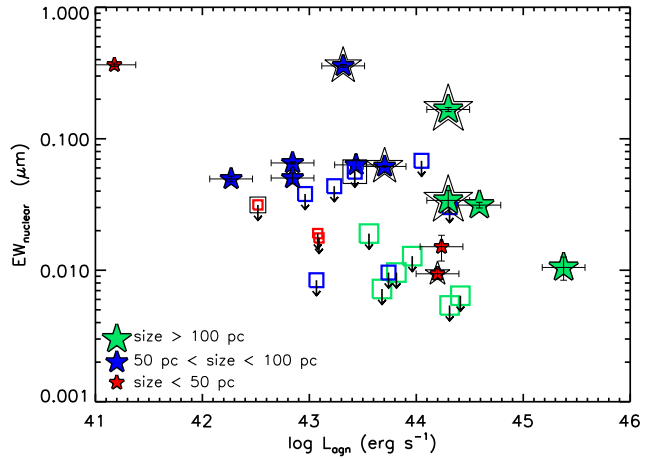}

\caption{\label{fig:fig_Lagn} Nuclear EW of the 11.3\,\micron\, PAH feature
  versus the AGN bolometric luminosity. Filled star symbols are
  detections whereas open squares are upper limits. The sizes/colors of the
  symbols (see figure legend) indicate the different physical sizes
  probed, which are determined by the slit widths of the
  observations. Hereinafter, we have marked Compton-thick objects in all plots using a double-star or a double-square for detections or upper limits, respectively.} 
\end{figure}

To investigate this issue, we plotted in Figure~\ref{fig:fig_Lagn}
the observed nuclear EW of the 11.3\,\micron~PAH feature against the AGN bolometric 
luminosity for our sample, with smaller symbols indicating 
regions closer to the AGN. This figure does not
show any clear trend. If the decreased nuclear EW of the 11.3\,\micron\,PAH feature were 
due to the AGN mid-IR continuum in more luminous AGN, we would expect a trend of 
decreasing EW for increasing AGN bolometric luminosities.  Alternatively, 
we would expect the same trend if the PAH molecules responsible for the 11.3\,\micron~feature 
were suppressed/destroyed more easily in luminous AGN. From this figure
we can see that at a given AGN luminosity we sometimes detect nuclear 11.3\,\micron~PAH 
emission, whereas in other cases we do not. In other words, we do not
see clear evidence in our sample for the 11.3\,\micron~PAH feature to be suppressed in more luminous AGN, at least for the AGN bolometric 
luminosities covered in our sample of Seyfert galaxies.

As can be seen from Figure~\ref{fig:fig_Lagn}, there is no clear influence 
of the probed physical region sizes on the observed EWs. Hence, we do not see 
a tendency for the EW of the PAH feature to decrease for smaller physical regions.
This would be the case if we were to expect a higher AGN continuum contribution and/or PAH destruction as 
we get closer to the AGN. No trend is either
  present when plotting the observed EWs  
with respect to luminosity densities. 
We note, however, that for the three closest Seyferts (Circinus, 
NGC~4945, NGC~5128) the 11.3\,\micron~PAH feature is not detected in ground-based high-resolution T-ReCS spectra, which probe 
scales of $\sim7-15\,$pc for these sources \citep{Roche2006,Gonzalez2013}.

\begin{figure*}
\center
\includegraphics[scale=0.35,angle=90]{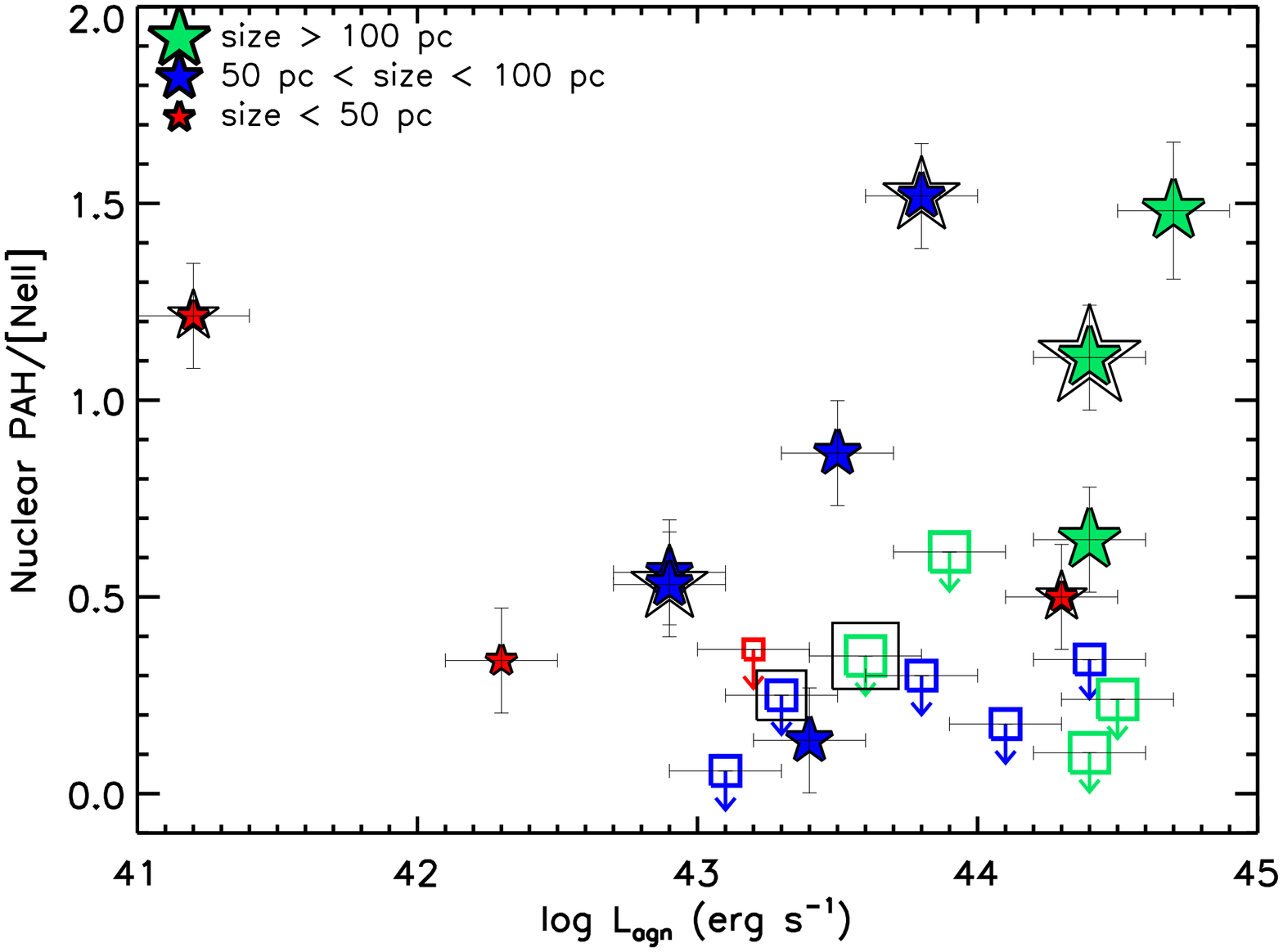} ~
\includegraphics[scale=0.35,angle=90]{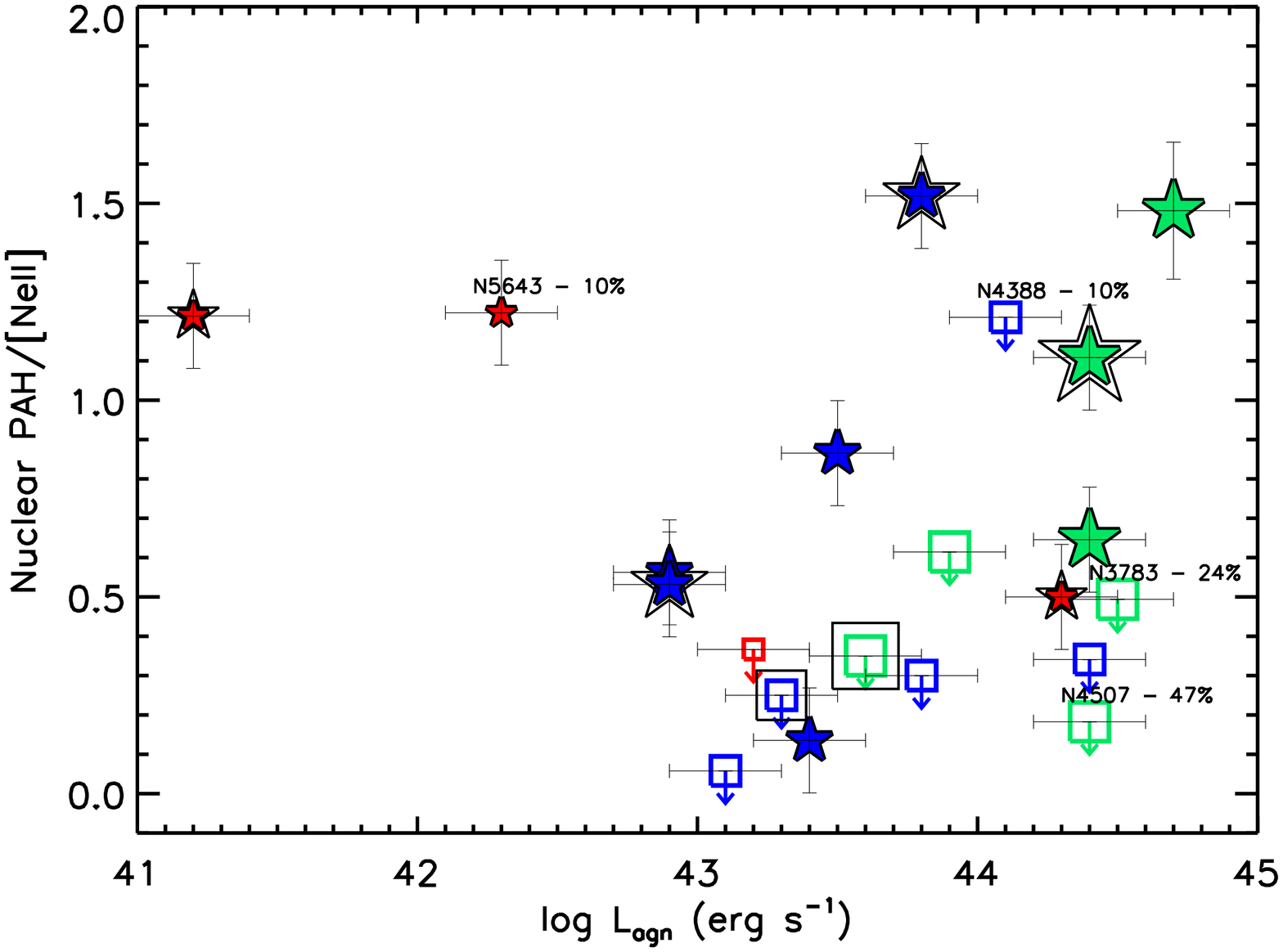} ~
\caption{Flux ratio of the 11.3\micron~PAH and the \neii~lines plotted against the AGN luminosity for the total (AGN+SF) [Ne\,{\sc ii}] emission (left) and corrected when possible for the AGN contribution (right panel), using values from \citet{Melendez2008} (these are shown with name tags on the figure, with the corresponding percentage of star formation). See text for details.}
\label{fig:fig_lineratio_Lagn} 
\end{figure*}

\subsection{Nuclear PAH molecules shielded by the dusty torus?}\label{sec:sec_PAHshielded}
As we have shown in the previous sections and as presented by others
\citep[e.g.,][]{Miles1994,Marco2003,DiazSantos2010,Hoenig2010,Gonzalez2013,Sales2013,Tacconi2013}
PAH  emission is detected in the vicinity (from tens to a few hundreds parsecs) of the harsh environments of some AGN.
Therefore, at least in some galaxies, the PAH molecules are not destroyed (or at least not completely) near the AGN. 
They must be shielded from
the AGN by molecular material with sufficient X-ray absorbing column densities \citep{Voit1992,Miles1994,Watabe2008}. As pointed out by
\citet{Voit1992b}, for PAH features to be absent due to destruction, they have to be fragmented more quickly than they can be rebuilt. In other
words, PAHs will exist if the rate of reaccretion of carbon onto the PAHs is higher than the evaporation rate caused by the AGN. Using
the parameters of the \citet{Voit1991} model, \citet{Miles1994} estimated the column density of hydrogen required to keep the evaporation rate of PAHs below the rate of reaccretion of carbon onto the PAHs. As derived in \citet{Miles1994}, the time scale for X-ray absorption in terms of the hydrogen column density of the intervening material $N_{\rm  H}$(tot), the distance from the AGN fixed in our case by the slit width $D_{\rm agn}$, and the X-ray luminosity of the AGN, can be written as

\begin{equation}
\tau \approx 700\,{\rm yr} \left(\frac{N_{\rm H}{\rm (tot)}}{10^{22}\,{\rm cm}^{-2}}\right)^{1.5} 
\left(\frac{D_{\rm agn}}{\rm
kpc}\right)^2 \left(\frac{10^{44}\,{\rm erg \,s}^{-1}}{L_{\rm X}}\right).
\label{eq:eq_PAHtimescale}
\end{equation}

\citet{Voit1992b} estimated that the time scale needed for reaccretion of a carbon atom on to a fractured  PAH should
be at least 3000 years for the typical conditions of the interstellar medium.

The protecting material, which has to be located between the nuclear sites of SF and the AGN, is likely to be that 
in the dusty torus postulated by the unified  model. 
In a number of works \citep[][]{RamosAlmeida2009,RamosAlmeida2011A,Alonso2011,AlonsoHerrero2012}
we  have demonstrated that the clumpy torus models of \citet{Nenkova2008a, Nenkova2008b}
accurately reproduce the nuclear infrared emission of local Seyfert galaxies. 
These models are defined by six parameters describing the torus geometry and the properties of the dusty clouds.
These are, the viewing angle ($i$) and radial extent ($Y$) of the torus, the angular ($\sigma$) and radial distributions ($q$) of the clouds along with its optical depth ($\tau_{\rm V}$), and the number of clouds along the equatorial direction ($N_{\rm 0}$). The optical extinction of the torus along the line of sight is computed from the model parameters as $A_{V}^{\rm LOS} = 1.086 N_{\rm 0} \tau_{\rm V} e^{(-(i-90)^2/\sigma^2)}$\,mag. According to \citet{Bohlin1978}, the absorbing hydrogen column density is then calculated following $N_{\rm H}^{\rm LOS}/A_{V}^{\rm LOS}$ = $1.9 \times 10^{21} \, {\rm cm}^{-2}\, {\rm mag}^{-1}$.

With the derived torus model parameters, we estimated the
hydrogen column density of the torus material in our line of sight for a sample of Seyferts.
This typically ranges from $N_{\rm H}^{\rm LOS}\simeq 10^{23}$ to a few
times $10^{24}\,{\rm  cm}^{-2}$ \citep[see][for further details]{RamosAlmeida2009,RamosAlmeida2011A,Alonso2011,AlonsoHerrero2012}. 
Using Equation~\ref{eq:eq_PAHtimescale} for the hard X-ray luminosities 
(Table~\ref{table:tab1}) and distances from the AGN probed by our
spectroscopy (see Table~\ref{table:tab3}), and setting $\tau$=3000\,yr as proposed by \citet{Voit1992}, we require
hydrogen column densities of at least a few $10^{23}\,{\rm cm}^{-2}$ to protect the PAHs from the AGN radiation. 
Evidence for such values for the \nh~are found for our sample, as we derived absorbing columns of the order of $10^{23}\,{\rm cm}^{-2}$ or even higher.

In  \citet{RamosAlmeida2011A}, we also demonstrated that Seyfert 2s are more likely to
have higher covering factors than Seyfert 1s.  
Assuming that the nuclear SF occurs inside the torus, the PAH
molecules may be more shielded in the nuclear region of Seyfert 2s. 
However, even the \nh\,values along our line of sight from the torus model fits of Seyfert 1s 
(which would be a lower limit to the total \nh~in the torus) with 
11.3\,\micron~PAH detections are sufficient to protect the PAH carriers.
This is the case for four Seyfert 1s (NGC3227, NGC5506, NGC7469, and NGC7582), as can be seen from the modelling by \citet{Alonso2011}.
Note that the column densities that we refer to are not only those absorbing the X-rays
but also including material much farther away from the accreting BH.

Another interesting aspect to keep in mind from Equation~\ref{eq:eq_PAHtimescale} is that the
column densities needed to protect the PAH molecules from the AGN X-ray emission become higher for
more luminous AGN as well as for distances closer to the AGN. However, we emphasize that for the AGN
luminosities of the RSA Seyferts and distances from the AGN probed by the observations presented here, 
the PAH molecules are likely to be shielded from the AGN by material in the torus residing on smaller scales.
Also, part of the obscuring material even on these
nuclear scales can reside in the host
galaxy as shown by \citet{Gonzalez2013}. Thus, another source of 
opacity that might prevent the PAHs from being destroyed are dust lanes
in galaxies or dust in the nuclear regions of merger systems. 
This might be the case for five sources in our sample, namely NGC\,4945,
NGC\,5128, NGC\,5506, NGC\,7130 and NGC\,7582.

\section{Nuclear Star Formation Rates in Seyfert galaxies}

\subsection{Relation between the $11.3\,\mu$m PAH feature and the [Ne\,{\sc ii}]\,$12.81\,\mu$m emission line on nuclear scales} \label{sec:sec4_1}
In star forming galaxies, the luminosity of the [Ne\,{\sc ii}]~12.81\,\micron~emission line is a good indicator of the SFR
\citep{Roche1991,Ho2007,DiazSantos2010}. We note that these 
measurements are contaminated by the 12.7\,\micron~PAH feature, which
is not easily resolvable. 

In Seyfert galaxies the situation is more complicated because this line can be excited by both SF and AGN
activity. The AGN contribution to the [Ne\,{\sc ii}] varies from galaxy to galaxy in local Seyfert
galaxies and other AGN \citep[see e.g.][]{Melendez2008,PereiraSantaella2010}. 
For the RSA Seyfert sample, \citet{DiamondStanic2010} used IRS
spectroscopy to compare the circumnuclear SFRs  
computed with the [Ne\,{\sc ii}]
line and the 11.3\,\micron~PAH feature as a function of EW of the PAH. They found that the ratio of the two circumnuclear SFRs (on a kpc
scale) is on average unity, with some scatter for galaxies with large PAH EWs $>$ 0.3\,\micron. The most discrepant measurements were for those galaxies with elevated [O\,{\sc iv}]/[Ne\,{\sc ii}] and low EW of the PAH, that is, AGN dominated galaxies.

In Figure~\ref{fig:fig_lineratio_Lagn} (left panel) we show the observed
nuclear PAH/[Ne\,{\sc ii}] ratio as a function of the 
AGN luminosity. 
To correct for the 12.7\,\micron\,PAH contamination, we have used a median ratio of the 11.3 versus the 12.7\,\micron\,PAH features of 1.8, derived in \citet{Smith2007}, and subtracted it from the  \neii\,measurement.
NGC~1365, NGC~1386 and NGC~5347 are not included in the plot
because both lines are undetected and, therefore, the value in the Y axis is completely unconstrained. 
We derived the \neii\,fluxes using the same technique as
explained for the PAH in Section~\ref{sec:sec_measurePAH}, 
integrating the line between 12.6 and 12.9\,\micron.
For seven galaxies, namely IC\,5063, Mrk\,509, NGC\,1068, NGC\,2110, 
NGC\,5135, NGC\,7130 and NGC\,7479, the  \neii\,line falls outside the wavelength range 
covered by our observations. They are not included in the plot except
for NGC\,5135 and NGC\,7130, whose values  
have been extracted from \citet{DiazSantos2010}. Most of the galaxies 
with detections of the nuclear 11.3\,\micron~PAH feature and \neii~show ratios similar to
those of high metallicity star forming galaxies \citep[$\sim 0.7-2$, see e.g.,][]{Roche1991,DiazSantos2010}, even if the [Ne\,{\sc ii}]
fluxes are not corrected for AGN emission. On the other hand, most of the nuclear
spectra with non detections show upper limits to the PAH/[Ne\,{\sc ii}] ratio below 0.5.

To derive the nuclear [Ne\,{\sc ii}] flux solely due to star formation, 
we can use the fractional SF contribution to the [Ne\,{\sc ii}] line within the IRS
aperture estimated by \citet{Melendez2008}. 
For sources with strong AGN contribution (higher than 50\%), we estimated the
[Ne\,{\sc ii}] flux coming from the AGN, which can be subtracted from the 
observed nuclear [Ne\,{\sc ii}]  flux.  This
is shown in Figure~\ref{fig:fig_lineratio_Lagn} (right panel). We note that for NGC~2992, 
NGC~3227, NGC~4151, NGC~5506 and NGC~7172, the estimated AGN 
contribution to the total \neii\, is higher than the nuclear \neii\,value, 
indicating that the AGN [NeII] contributions were overestimated. 
For these galaxies, we did not apply any correction.
Figure~\ref{fig:fig_lineratio_Lagn} shows that for those Seyferts with a nuclear 11.3\,\micron~PAH
detection the PAH/[Ne\,{\sc ii}]$_{\rm SF}$ ratio does not decrease with the
AGN bolometric luminosity. This would be expected if the PAH
emission was to be suppressed. 
Therefore, given that the PAH/[Ne\,{\sc ii}]$_{\rm SF}$ ratio does not
show a dependence on $L_{\rm agn}$, 
we conclude that the 11.3\,\micron~PAH
feature emission can be used to estimate the nuclear SFRs (see next section).

\subsection{Circumnuclear ($\sim 600\,$pc) vs Nuclear ($\sim 65\,$pc) scales}\label{sec:sec4}

In the vicinity of an AGN, we expect the chemistry and/or the
heating as dominated by X-rays from the so-called 
X-ray dominated regions (XDRs). In
principle, XDRs could also contribute to PAH heating through
the photodissociation and photoionization by FUV
photons produced via excitation of H and H$_2$ in collisions
with secondary electrons.  However, as we derived in Section~\ref{sec:sec_PAHshielded}, the torus appears to
provide the appropriate environment to  shield the PAH molecules from
the AGN emission, on typical physical scales of a few
parsecs up to a few tens of parsecs. We thus expect little or no contribution of UV AGN produced photons
to the PAH heating in the nuclear scales of Seyfert galaxies. Hereinafter, we will assume that the aromatic molecules are heated by the
radiation field produced by young massive stars and that the
11.3\,\micron~PAH luminosity can be used to estimate the SFR.

We derived nuclear and circumnuclear SFRs using the PAH
11.3\micron\, feature luminosities and applying the relation 
derived in \citet{DiamondStanic2012} 

\begin{eqnarray}\label{eq:sfr_pah}
{\rm SFR} \left( M_\odot \, {\rm yr}^{-1} \right) = 9.6 \times 10^{-9}
L \left({\rm PAH}_{11.3\,\mu m}, L_{\odot} \right)
\end{eqnarray}

\noindent
using {\tt PAHFIT} measurements 
of galaxies with IR (8--1000\,\micron) luminosities $L_{\rm IR}<
10^{11}\,L_\odot$, using the \citet{Rieke2009} templates and a Kroupa
IMF. This is appropriate for our sample, as the median value of
the IR luminosity of the individual galaxies is $5\times
10^{10}\,L_\odot$\protect\footnote[3]{A few galaxies in our sample have total IR
  luminosities between $10^{11}-5\times 10^{11}\,L_\odot$. However,
  given that we are dealing with nuclear luminosities, we expect those
  being less than the $10^{11}\,L_\odot$  limiting value.}. 
The uncertainties in the derived SFRs using Equation~\ref{eq:sfr_pah}
  are typically 0.28\,dex \citep[see][for full details]{DiamondStanic2012}.

For the 13 galaxies with nuclear 11.3\,\micron~PAH detections, the nuclear
SFRs span two orders of magnitude between
$\sim$0.01--1.2\,$M_\odot\,{\rm yr}^{-1}$ 
for regions of typically $\sim$65\,pc in size. The non-detections
indicate that the nuclear SFR of RSA Seyfert galaxies from the high 
angular resolution spectroscopy can be $\sim0.01-0.2\,M_\odot\,{\rm
  yr}^{-1}$, or even lower. 

The projected nuclear SFR densities are between 2 and
  93\,\msun\,yr$^{-1}\,{\rm kpc}^{-2}$ with a median value
  22\,\msun\,yr$^{-1}\,{\rm kpc}^{-2}$. These are  consistent with the
    simulations of \citet{Hopkins2012b} for similar physical scales. The 
  two galaxies with the largest SFR densities are NGC\,1068 and NGC\,1808 with values of 414 and
329\,\msun\,yr$^{-1}\,{\rm kpc}^{-2}$, respectively.  We
  notice that for those galaxies in common with \citet{Davies2007} (namely Circinus,
    NGC~1068, NGC~3783, NGC~7469, NGC~3227) 
we find quite discrepant values for the SFR density, with ours
lying below those in \citet{Davies2007} except for NGC\,1068.  
It might be due to the use of different SF histories and SFR indicators.
We note that with the 11.3\,\micron\,PAH feature we cannot explore age effects \citep[see][Figure 8]{DiazSantos2008}
as this feature can be excited by both O and B stars, and thus it integrates over ages
of up to a few tens of millions of years \citep{Peeters2004}, unlike the measurements in \citet{Davies2007}  that sample younger populations.
In addition, we detect neither nuclear nor circumnuclear SF in
NGC~3783 based on the PAH measurements. 
On the other hand, PAHs can also be found in the interstellar medium (ISM) as being excited in less-UV rich environments, such as reflection nebulae \citep[e.g.][]{Li2002}. However, the decreased strength of the IR emission features in these objects seems to indicate the low efficiency of softer near-UV or optical photons in exciting PAHs in comparison to SF \citep{Tielens2008}.

The circumnuclear SFRs in our sample  are between
0.2 and 18.4\,\msun yr$^{-1}$ \citep[see also][]{DiamondStanic2012}, and the median
  circumnuclear SFR densities are 1.2\,\msun yr$^{-1}$
  kpc$^{-2}$. These are similar to those of the CfA
  \citep{Huchra1992} and 12\,\micron\,\citep{Rush1993} samples,
  derived using the 3.3\,\micron~PAH feature
  \citep[see][]{Imanishi2003,Imanishi2004}.

The comparison between the nuclear and circumnuclear
SFRs for our sample clearly shows that, in absolute terms, the
  nuclear SFRs  are much lower (see
  Table~\ref{table:tab3}). This is in good
agreement with previous works based on smaller samples of local AGN
\citep[e.g.][]{Siebenmorgen2004,Watabe2008, Hoenig2010,Gonzalez2013}.
The median value of the ratio between the nuclear and circumnuclear SFRs for the detections of the
11.3\,\micron\,feature is $\sim$0.18 (see also
Table~\ref{table:tab3}), with no significant difference for type
  1 and type 2 Seyferts ($\sim 0.18$ and $\sim 0.21$, respectively).

\begin{figure}

\includegraphics[scale=0.37,angle=90]{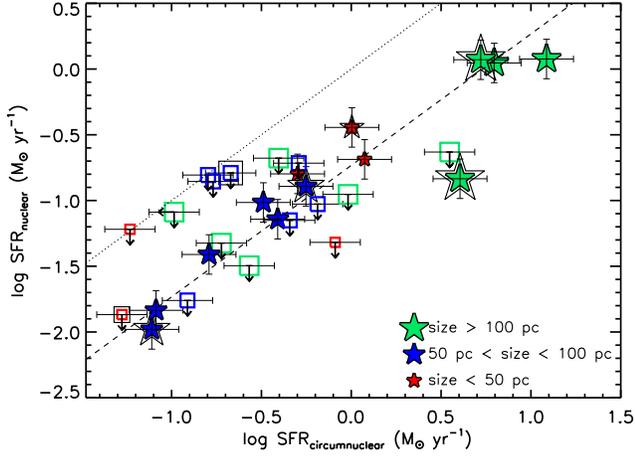} 
\caption{Comparison between the SFR on different scales, where SFR$_{\rm circumnuclear}$ implies typical physical scales of $\sim$0.6\,kpc and SFR$_{\rm nuclear}$ is for $\sim$65\,pc scales. Non-detections are plotted at a $2\sigma$ level. Symbols as in Figure~\ref{fig:fig_Lagn}. The dashed line shows  the median value of the nuclear/circumnuclear SFR ratio for the detections of the nuclear PAH feature (see text and Table~\ref{table:tab3}). The dotted line indicates a nuclear/circumnuclear ratio of one.}

\label{fig:fig_sfr}
\end{figure}

In Figure~\ref{fig:fig_sfr} we plot the nuclear and circumnuclear SFRs probing typical physical
scales of $\sim$65\,pc and $\sim$600\,pc, respectively. Again, non-detections are plotted
as upper limits at the 2$\sigma$ level. Overall, for our detections, the fraction of the
SFR accounted for by the central $\sim 65$\,pc region of our Seyferts ranges between $\sim$5--35\% of that enclosed within the
aperture corresponding to the circumnuclear data.

While the nuclear SFRs are lower than the
  circumnuclear SFRs, the median nuclear projected SFR densities are
  approximately a factor of 20 higher than the circumnuclear ones in our sample (median values
  of 22 and 1\,\msun yr$^{-1}$
  kpc$^{-2}$,
  respectively). This shows that the SF is not
  uniformly distributed. Conversely, it is more highly
  concentrated in the nuclear regions of the RSA Seyferts studied
  here. This is in agreement with 
simulations of \citet{Hopkins2012b}. The molecular gas needed to maintain these nuclear SFR densities 
appears to have higher densities in Seyfert galaxies than those of  quiescent (non Seyferts)
galaxies \citep{Hicks2013}.

\begin{figure}
\center

\includegraphics[scale=0.37,angle=90]{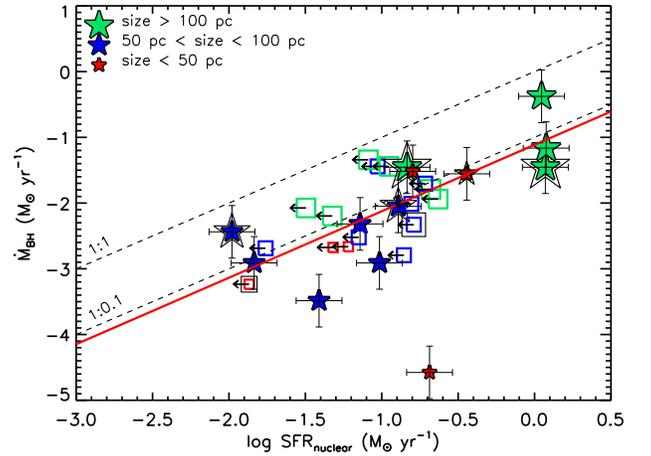} ~

\caption{Observed nuclear SFR vs. $\dot M_{\rm BH}$ relation. Predictions from \citet{Hopkins2010} are shown as dashed lines. We show the $\dot M_{\rm BH}$\,$\approx$\,0.1\,$\times$\,SFR relation, which is expected   for r\,$<$\,100\,pc, and the 1:1 relation which, is expected for the
smallest physical scales ($r<10\,$\,pc). The solid line represents the fit to our detections of the
nuclear  11.3\,\micron\,PAH feature (see text for details).} 
\label{fig:fig_sfr_bhar} 
\end{figure}

\bigskip

\subsection{Nuclear star formation rate vs. black hole accretion rate}\label{sec:sec4}
\citet{Hopkins2010} performed smoothed particle hydrodynamic simulations to study the inflow of gas from galactic scales
($\sim$10\,kpc) down to $\lesssim$0.1\,pc, where key ingredients are gas, stars, black holes (BHs), self-gravity, SF and
stellar feedback. These numerical simulations indicate a relation (with significant scatter) between the SFR and $\dot M_{\rm BH}$ that 
holds for all scales, and that is more tightly coupled for the smaller physical scales.
The model of \citet{Kawakatu2008} predicts that the AGN luminosity should also be tightly correlated with the luminosity of the nuclear
(100\,pc) SF in Seyferts and QSOs, and also that $L_{\rm nuclear,SB}$/$L_{\rm AGN}$ is larger for more luminous AGN.

According to \citet[][and references therein]{Alexander2012}, $\dot M_{\rm BH}$ and AGN luminosities follow the relation

\begin{equation}\label{eq:BHAR}
\dot M_{\rm BH} ({M}_\odot\,{\rm yr}^{-1})=0.15(0.1/\epsilon)(L_{\rm agn}/10^{45}\,{\rm erg\,s}^{-1})
\end{equation}

\noindent
where we used $\epsilon=0.1$ as the typical value for the mass-energy conversion 
efficiency in the local Universe \citep{Marconi2004}. We obtained $\dot M_{\rm BH}$ 
ranging between $5 \times 10^{-6}$ and $0.5\,M_\odot\,{\rm yr}^{-1}$ for our sample.
Uncertainties in the $\dot M_{\rm BH}$ estimations are dominated by those in
$L_{\rm agn}$, i.e. 0.4\,dex, as mentioned in Section~\ref{sec:sample}.

Figure~\ref{fig:fig_sfr_bhar} shows the observed nuclear SFR against $\dot M_{\rm BH}$ 
for the Seyferts in our sample. The different sizes of the symbols indicate 
different physical sizes of the probed regions. The prominent outlier in this figure is 
NGC~1808\protect\footnote[4]{NGC~1808 is a low-luminosity AGN and shows the lowest AGN bolometric
luminosity in our sample. It seems plausible that SF, unlike the AGN as for the 
rest of the sample, is the dominant mechanism contributing to the nuclear mid-IR emission 
\citep[see][]{Mason2012}, as this source is completely off the mid-IR vs hard X-ray 
correlation for AGN \citep[see][]{Gonzalez2013}.}. 
We also show in Figure~\ref{fig:fig_sfr_bhar} as dashed lines predictions from the 
\citet{Hopkins2010} simulations for $r<100\,$pc ($\dot M_{\rm BH}$\,$\,\approx$\,0.1\,$\times$\,SFR)
and $r<10\,$pc ($\dot M_{\rm BH}$\,$\,\approx$\,SFR). These radii encompass 
approximately the physical scales probed by our nuclear SFR. The prediction from 
the \citet{Kawakatu2008} disk model for Seyfert luminosities and BH masses similar to
those of our sample falls between the two dashed lines ($\sim$ 1:0.4 relation).

To derive a possible correlation between the nuclear SFR and $\dot M_{\rm BH}$, 
we applied a simple fit to the nuclear SFR detections (excluding NGC~1808) and obtained
a nearly linear relation (slope of 1.01, and uncertainties of 0.4\,dex in both parameters, Equation~\ref{eq:eq_fit}),
which is close to the 1:0.1 relation  (see
Figure~\ref{fig:fig_sfr_bhar} and below).

\begin{eqnarray}
{\log} \,{\dot M}_{\rm BH} = 1.01 \times {\log \, \rm SFR}_{\rm nuclear} - 1.11
\label{eq:eq_fit}
\end{eqnarray}

Also including the upper limits in the fit we obtained 
a very similar result (slope of 0.95).
In contrast, \citet{DiamondStanic2012} obtained 
a slightly superlinear relation ($\dot M_{\rm BH} \propto {\rm SFR}^{1.6}$), 
when the SFRs are measured in regions of 1\,kpc radius. 
This behaviour (i.e., the relations becoming linear on smaller scales) is 
nevertheless predicted by the \citet{Hopkins2010} simulations.

As can be seen in Figure~\ref{fig:fig_sfr_bhar}, the \citet{Hopkins2010} predictions for $r<100\,$pc reproduce fairly well 
the observed relation for our sample. We do not find a tendency 
for galaxies with SFRs measured in regions closer to the AGN (slit sizes of less than 100\,pc) 
to have larger $\dot M_{\rm BH}$ to SFR ratios (i.e., to lie closer to the 1:1 relation) than
the rest, as predicted by \citet{Hopkins2010}.  It is worth noting that these 
authors caution that their work do not include the appropriate physics for low accretion 
rates ($<< 0.1\,M_\odot \,{\rm yr}^{-1}$). 
The scatter in the theoretical 
estimations and the limited size of our sample of Seyfert galaxies prevent us from further exploring 
this issue. Future planned observations with the mid-IR CanariCam \citep{Telesco2003} instrument on the 10.4\,m Gran 
Telescopio de Canarias (GTC) will allow a similar study for larger samples of Seyfert galaxies.

\section{Summary and conclusions}
We have presented the largest compilation to date of high angular resolution
($0.4-0.8$\arcsec) mid-IR spectroscopy of nearby Seyfert 
galaxies obtained with the T-ReCS, VISIR, and Michelle instruments.
We used the 11.3\,\micron\,PAH feature to study the nuclear SF activity and its relation to
the circumnuclear SF, as well as with $\dot M_{\rm BH}$. The sample includes 29 Seyfert galaxies
(13 Seyfert~1 and 16 Seyfert~2 galaxies)  belonging to the nearby RSA
AGN sample \citep{Maiolino1995,Ho1997}. It covers more than two orders of
magnitude in AGN bolometric luminosity, with the galaxies
located at a median distance of 33\,Mpc. Our data allow us to probe 
typical nuclear physical scales (given by the slit widths) of
$\sim$\,65\,pc. We used the hard X-ray luminosity as a proxy for 
the AGN bolometric luminosity and $\dot M_{\rm BH}$. We used mid-infrared Spitzer/IRS
spectroscopy to study the SF taking place in the circumnuclear regions (a factor of 5--10 larger scales).

The main results can be summarized as follows:

\begin{enumerate}
\item 
The detection rate of the nuclear 11.3\,\micron\, PAH feature in our sample of 
Seyferts is 45\,\% (13 out of 29 sources), at a significance of 2$\sigma$ or higher.  
Additionally, the stacked spectra of six galaxies without a detection of the 11.3\,\micron~PAH feature
and weak silicate features resulted into a positive detection of the 11.3\,\micron\,~PAH 
feature above 2$\sigma$.

\item
There is no evidence of strong suppression of the nuclear 11.3\,\micron~PAH
feature in the vicinity of the AGN, at least for the Seyfert-like 
AGN luminosities and physical nuclear regions (65\,pc median value)
sampled here. In particular, we do not see a tendency for the EW of
the PAH  to decrease for more luminous AGN. 
The hydrogen column densities predicted from clumpy torus model fitting
(a few $10^{23}\,{\rm cm}^{-2}$ up to a few $10^{24}\,{\rm cm}^{-2}$)
would be, in principle, sufficient to shield the PAH molecules from AGN 
X-ray photons in our Seyfert galaxies.

\item
The nuclear SFRs in our sample derived from the 11.3\,\micron~PAH feature luminosities
are between 0.01 and 1.2\,$M_\odot\,{\rm yr}^{-1}$,
where we assumed no
XDR-contribution to the PAH heating.
There is a significant 
reduction of the 11.3\,\micron\,PAH flux from circumnuclear (median size of 600\,pc) 
to nuclear regions (median size of 65\,pc), with a typical ratio of $\sim$5. 
Although this indicates that the SFRs are lower near the AGN in absolute terms, 
the projected SFR rate density in the nuclear regions (median value of 
22\,\msun\,yr$^{-1}\,{\rm kpc}^{-2}$) is approximately 20 higher than in the 
circumnuclear regions. This indicates that the SF activity is highly
concentrated in the nuclear regions in our sample of Seyfert galaxies.

\item
Predictions from numerical simulations for the appropriate physical regions are broadly 
consistent with the observed relation between the  nuclear SFR and $\dot M_{\rm BH}$ in our sample (slope of 1.01$\pm 0.4$).  
Although limited by the relatively small number of sources in our
sample, we do not find decreased nuclear SFR-to-$\dot M_{\rm BH}$ ratios for regions closer to
the AGN, as predicted by the \citet{Hopkins2010} simulations.

\end{enumerate}

\newpage

\small
\acknowledgments
We thank the anonymous referee for comments that helped us to improve the paper, and  E. Piconcelli for useful discussion. PE and AAH acknowledge support from the Spanish Plan Nacional de Astronom\'{\i}a y Astrof\'{\i}sica under grant AYA2009-05705-E. PE is partially funded by Spanish MINECO under grant AYA2012-39362-C02-01. AAH and AHC acknowledge support from the Augusto G. Linares Program through the Universidad de Cantabria. CRA ackowledges financial support from the Instituto de Astrof\'{\i}sica de Canarias and the Spanish Plan Nacional de Astronom\'{\i}a y Astrof\'{\i}sica under grant AYA2010-21887-C04.04 (Estallidos). OGM and JMRE acknowledge support from the Spanish MICINN through the grant AYA2012-39168-C03-01. SFH acknowledges support by Deutsche Forschungsgemeinschaft (DFG) in the framework of a research fellowship (Auslandsstipendium). 

This research has made use of the NASA/IPAC Extragalactic Database (NED) which is operated by the Jet Propulsion Laboratory, California Institute of Technology, under contract with the National Aeronautics and Space Administration. Based on observations obtained at the Gemini Observatory, which is operated by the Association of Universities for Research in Astronomy, Inc., under a cooperative agreement with the NSF on behalf of the Gemini partnership: the National Science Foundation (United States), the National Research Council (Canada), CONICYT (Chile), the Australian Research Council (Australia), Minist\'{e}rio da Ci\^{e}ncia, Tecnologia e Inova\c{c}\~{a}o (Brazil) and Ministerio de Ciencia, Tecnolog\'{i}a e Innovaci\'{o}n Productiva (Argentina). The Cornell Atlas of Spitzer/IRS Sources (CASSIS) is a product of the Infrared Science Center at Cornell University, supported by NASA and JPL.

\bibliography{bibliography,coevalbhNotes.bib}

\begin{thebibliography}{114}
\expandafter\ifx\csname natexlab\endcsname\relax\def\natexlab#1{#1}\fi

\bibitem[{{Akylas} {et~al.}(2001){Akylas}, {Georgantopoulos}, \&
  {Comastri}}]{Akylas2001}
{Akylas}, A., {Georgantopoulos}, I., \& {Comastri}, A. 2001, \mnras, 324, 521

\bibitem[{{Alexander} \& {Hickox}(2012)}]{Alexander2012}
{Alexander}, D.~M., \& {Hickox}, R.~C. 2012, New AR, 56, 93

\bibitem[{{Alonso-Herrero} {et~al.}(2012){Alonso-Herrero}, {Pereira-Santaella},
  {Rieke}, \& {Rigopoulou}}]{AlonsoHerrero2012}
{Alonso-Herrero}, A., {Pereira-Santaella}, M., {Rieke}, G.~H., \& {Rigopoulou},
  D. 2012, \apj, 744, 2

\bibitem[{{Alonso-Herrero} {et~al.}(2011){Alonso-Herrero}, {Ramos Almeida},
  {Mason}, {Asensio R.}, \& et~al.}]{Alonso2011}
{Alonso-Herrero}, A., {Ramos Almeida}, C., {Mason}, R., {Asensio R.}, A., \&
  et~al. 2011, \apj, 736, 82

\bibitem[{{Antonucci}(1993)}]{Antonucci1993}
{Antonucci}, R. 1993, \araa, 31, 473

\bibitem[{{Armus} {et~al.}(2009){Armus}, {Mazzarella}, {Evans}, {Surace},
  {Sanders}, {Iwasawa}, {Frayer}, {Howell}, {Chan}, {Petric}, {Vavilkin},
  {Kim}, {Haan}, {Inami}, {Murphy}, {Appleton}, {Barnes}, {Bothun}, {Bridge},
  {Charmandaris}, {Jensen}, {Kewley}, {Lord}, {Madore}, {Marshall},
  {Melbourne}, {Rich}, {Satyapal}, {Schulz}, {Spoon}, {Sturm}, {U}, {Veilleux},
  \& {Xu}}]{Armus2009}
{Armus}, L., {Mazzarella}, J.~M., {Evans}, A.~S., {et~al.} 2009, \pasp, 121,
  559

\bibitem[{{Asmus} {et~al.}(2011){Asmus}, {Gandhi}, {Smette}, {H{\"o}nig}, \&
  {Duschl}}]{Asmus2011}
{Asmus}, D., {Gandhi}, P., {Smette}, A., {H{\"o}nig}, S.~F., \& {Duschl}, W.~J.
  2011, \aap, 536, A36

\bibitem[{{Ballantyne}(2008)}]{Ballantyne2008}
{Ballantyne}, D.~R. 2008, \apj, 685, 787

\bibitem[{{Bassani} {et~al.}(1999){Bassani}, {Dadina}, {Maiolino}, {Salvati},
  {Risaliti}, {della Ceca}, {Matt}, \& {Zamorani}}]{Bassani1999}
{Bassani}, L., {Dadina}, M., {Maiolino}, R., {et~al.} 1999, \apjs, 121, 473

\bibitem[{{Beckmann} {et~al.}(2006){Beckmann}, {Gehrels}, {Shrader}, \&
  {Soldi}}]{Beckmann2006}
{Beckmann}, V., {Gehrels}, N., {Shrader}, C.~R., \& {Soldi}, S. 2006, \apj,
  638, 642

\bibitem[{{Beifiori} {et~al.}(2009){Beifiori}, {Sarzi}, {Corsini}, {Dalla
  Bont{\`a}}, {Pizzella}, {Coccato}, \& {Bertola}}]{Beifiori2009}
{Beifiori}, A., {Sarzi}, M., {Corsini}, E.~M., {et~al.} 2009, \apj, 692, 856

\bibitem[{{Bianchi} {et~al.}(2005){Bianchi}, {Guainazzi}, {Matt}, {Chiaberge},
  {Iwasawa}, {Fiore}, \& {Maiolino}}]{Bianchi2005}
{Bianchi}, S., {Guainazzi}, M., {Matt}, G., {et~al.} 2005, \aap, 442, 185

\bibitem[{{Bohlin} {et~al.}(1978){Bohlin}, {Savage}, \& {Drake}}]{Bohlin1978}
{Bohlin}, R.~C., {Savage}, B.~D., \& {Drake}, J.~F. 1978, \apj, 224, 132

\bibitem[{{Brightman} \& {Nandra}(2011)}]{Brightman2011}
{Brightman}, M., \& {Nandra}, K. 2011, \mnras, 413, 1206

\bibitem[{{Cid Fernandes} \& {Terlevich}(1995)}]{CidFernandes1995}
{Cid Fernandes}, Jr., R., \& {Terlevich}, R. 1995, \mnras, 272, 423

\bibitem[{{Colina} {et~al.}(2002){Colina}, {Gonzalez Delgado}, {Mas-Hesse}, \&
  {Leitherer}}]{Colina2002}
{Colina}, L., {Gonzalez Delgado}, R., {Mas-Hesse}, J.~M., \& {Leitherer}, C.
  2002, \apj, 579, 545

\bibitem[{{Colling}(2011)}]{Colling2011}
{Colling}, M. 2011, PhD thesis, University of Oxford

\bibitem[{{Dadina}(2007)}]{Dadina2007}
{Dadina}, M. 2007, \aap, 461, 1209

\bibitem[{{Davies} {et~al.}(2007){Davies}, {M{\"u}ller S{\'a}nchez}, {Genzel},
  {Tacconi}, {Hicks}, \& et~al.}]{Davies2007}
{Davies}, R.~I., {M{\"u}ller S{\'a}nchez}, F., {Genzel}, R., {et~al.} 2007,
  \apj, 671, 1388

\bibitem[{{Diamond-Stanic} \& {Rieke}(2010)}]{DiamondStanic2010}
{Diamond-Stanic}, A.~M., \& {Rieke}, G.~H. 2010, \apj, 724, 140

\bibitem[{{Diamond-Stanic} \& {Rieke}(2012)}]{DiamondStanic2012}
{Diamond-Stanic}, A.~M., \& {Rieke}, G.~H. 2012, \apj, 746, 168

\bibitem[{{D{\'{\i}}az-Santos} {et~al.}(2010){D{\'{\i}}az-Santos},
  {Alonso-Herrero}, {Colina}, {Packham}, {Levenson}, {Pereira-Santaella},
  {Roche}, \& {Telesco}}]{DiazSantos2010}
{D{\'{\i}}az-Santos}, T., {Alonso-Herrero}, A., {Colina}, L., {et~al.} 2010,
  \apj, 711, 328

\bibitem[{{D{\'{\i}}az-Santos} {et~al.}(2008){D{\'{\i}}az-Santos},
  {Alonso-Herrero}, {Colina}, {Packham}, {Radomski}, \&
  {Telesco}}]{DiazSantos2008}
{D{\'{\i}}az-Santos}, T., {Alonso-Herrero}, A., {Colina}, L., {et~al.} 2008,
  \apj, 685, 211

\bibitem[{{Ferrarese} \& {Merritt}(2000)}]{Ferrarese2000}
{Ferrarese}, L., \& {Merritt}, D. 2000, \apjl, 539, L9

\bibitem[{{Galliano} {et~al.}(2008){Galliano}, {Madden}, {Tielens}, {Peeters},
  \& {Jones}}]{Galliano2008}
{Galliano}, F., {Madden}, S.~C., {Tielens}, A.~G.~G.~M., {Peeters}, E., \&
  {Jones}, A.~P. 2008, \apj, 679, 310

\bibitem[{Gebhardt(2000)}]{Gebhardt}
Gebhardt, K. 2000, \apj, 539, L13

\bibitem[{{Ghisellini} {et~al.}(1994){Ghisellini}, {Haardt}, \&
  {Matt}}]{Ghisellini1994}
{Ghisellini}, G., {Haardt}, F., \& {Matt}, G. 1994, \mnras, 267, 743

\bibitem[{{Glasse} {et~al.}(1997){Glasse}, {Atad-Ettedgui}, \&
  {Harris}}]{Glasse1997}
{Glasse}, A.~C., {Atad-Ettedgui}, E.~I., \& {Harris}, J.~W. 1997, 2871, 1197

\bibitem[{{Gonz{\'a}lez Delgado} {et~al.}(1998){Gonz{\'a}lez Delgado},
  {Heckman}, {Leitherer}, {Meurer}, {Krolik}, {Wilson}, {Kinney}, \&
  {Koratkar}}]{GonzalezDelgado1998}
{Gonz{\'a}lez Delgado}, R.~M., {Heckman}, T., {Leitherer}, C., {et~al.} 1998,
  \apj, 505, 174

\bibitem[{{Gonz{\'a}lez-Mart{\'{\i}}n}
  {et~al.}(2013){Gonz{\'a}lez-Mart{\'{\i}}n}, {Rodr{\'{\i}}guez-Espinosa},
  {D{\'{\i}}az-Santos}, {Packham}, {Alonso-Herrero}, {Esquej}, {Ramos Almeida},
  {Mason}, \& {Telesco}}]{Gonzalez2013}
{Gonz{\'a}lez-Mart{\'{\i}}n}, O., {Rodr{\'{\i}}guez-Espinosa}, J.~M.,
  {D{\'{\i}}az-Santos}, T., {et~al.} 2013, \aap, 553, A35

\bibitem[{{Goulding} {et~al.}(2012){Goulding}, {Alexander}, {Bauer}, {Forman},
  {Hickox}, {Jones}, {Mullaney}, \& {Trichas}}]{Goulding2012}
{Goulding}, A.~D., {Alexander}, D.~M., {Bauer}, F.~E., {et~al.} 2012, \apj,
  755, 5

\bibitem[{{Guainazzi} {et~al.}(2000){Guainazzi}, {Matt}, {Brandt}, {Antonelli},
  {Barr}, \& {Bassani}}]{Guainazzi2000}
{Guainazzi}, M., {Matt}, G., {Brandt}, W., {et~al.} 2000, \aap, 356, 463

\bibitem[{{Guainazzi} {et~al.}(2004){Guainazzi}, {Rodriguez}, {Fabian},
  {Iwasawa}, \& {Matt}}]{Guainazzi2004}
{Guainazzi}, M., {Rodriguez}, P., {Fabian}, A.~C., {Iwasawa}, K., \& {Matt}, G.
  2004, \mnras, 355, 297

\bibitem[{{Heckman} {et~al.}(1995){Heckman}, {Krolik}, {Meurer}, {Calzetti},
  {Kinney}, {Koratkar}, {Leitherer}, {Robert}, \& {Wilson}}]{Heckman1995}
{Heckman}, T., {Krolik}, J., {Meurer}, G., {et~al.} 1995, \apj, 452, 549

\bibitem[{{Hern{\'a}n-Caballero} \& {Hatziminaoglou}(2011)}]{Hernan2011}
{Hern{\'a}n-Caballero}, A., \& {Hatziminaoglou}, E. 2011, \mnras, 414, 500

\bibitem[{{Hicks} {et~al.}(2013){Hicks}, {Davies}, {Maciejewski}, {Emsellem},
  {Malkan}, {Dumas}, {M{\"u}ller-S{\'a}nchez}, \& {Rivers}}]{Hicks2013}
{Hicks}, E.~K.~S., {Davies}, R.~I., {Maciejewski}, W., {et~al.} 2013, \apj,
  768, 107

\bibitem[{{Ho} {et~al.}(1997){Ho}, {Filippenko}, \& {Sargent}}]{Ho1997}
{Ho}, L.~C., {Filippenko}, A.~V., \& {Sargent}, W.~L.~W. 1997, \apjs, 112, 315

\bibitem[{{Ho} \& {Keto}(2007)}]{Ho2007}
{Ho}, L.~C., \& {Keto}, E. 2007, \apj, 658, 314

\bibitem[{{H{\"o}nig} {et~al.}(2010){H{\"o}nig}, {Kishimoto}, {Gandhi},
  {Smette}, {Asmus}, \& et~al.}]{Hoenig2010}
{H{\"o}nig}, S.~F., {Kishimoto}, M., {Gandhi}, P., {et~al.} 2010, \aap, 515,
  A23

\bibitem[{{H{\"o}nig} {et~al.}(2008){H{\"o}nig}, {Smette}, {Beckert}, {Horst},
  {Duschl}, {Gandhi}, {Kishimoto}, \& {Weigelt}}]{Hoenig2008}
{H{\"o}nig}, S.~F., {Smette}, A., {Beckert}, T., {et~al.} 2008, \aap, 485, L21

\bibitem[{{Hopkins}(2012)}]{Hopkins2012}
{Hopkins}, P.~F. 2012, \mnras, 420, L8

\bibitem[{{Hopkins} {et~al.}(2012){Hopkins}, {Hayward}, {Narayanan}, \&
  {Hernquist}}]{Hopkins2012b}
{Hopkins}, P.~F., {Hayward}, C.~C., {Narayanan}, D., \& {Hernquist}, L. 2012,
  \mnras, 420, 320

\bibitem[{{Hopkins} \& {Quataert}(2010)}]{Hopkins2010}
{Hopkins}, P.~F., \& {Quataert}, E. 2010, \mnras, 407, 1529

\bibitem[{{Houck} {et~al.}(2004){Houck}, {Roellig}, {van Cleve}, {Forrest},
  {Herter}, {Lawrence}, {Matthews}, {Reitsema}, {Soifer}, {Watson}, {Weedman},
  {Huisjen}, {Troeltzsch}, {Barry}, {Bernard-Salas}, {Blacken}, {Brandl},
  {Charmandaris}, {Devost}, {Gull}, {Hall}, {Henderson}, {Higdon}, {Pirger},
  {Schoenwald}, {Sloan}, {Uchida}, {Appleton}, {Armus}, {Burgdorf},
  {Fajardo-Acosta}, {Grillmair}, {Ingalls}, {Morris}, \& {Teplitz}}]{Houck2004}
{Houck}, J.~R., {Roellig}, T.~L., {van Cleve}, J., {et~al.} 2004, \apjs, 154,
  18

\bibitem[{{Huchra} \& {Burg}(1992)}]{Huchra1992}
{Huchra}, J., \& {Burg}, R. 1992, \apj, 393, 90

\bibitem[{{Imanishi}(2003)}]{Imanishi2003}
{Imanishi}, M. 2003, \apj, 599, 918

\bibitem[{{Imanishi} \& {Wada}(2004)}]{Imanishi2004}
{Imanishi}, M., \& {Wada}, K. 2004, \apj, 617, 214

\bibitem[{{Kawakatu} \& {Wada}(2008)}]{Kawakatu2008}
{Kawakatu}, N., \& {Wada}, K. 2008, \apj, 681, 73

\bibitem[{{King}(2010)}]{King2010}
{King}, A.~R. 2010, \mnras, 402, 1516

\bibitem[{{Kormendy} \& {Ho}(2013)}]{Kormendy2013}
{Kormendy}, J., \& {Ho}, L.~C. 2013, ArXiv e-prints

\bibitem[{{Lagage} {et~al.}(2004){Lagage}, {Pel}, {Authier}, {Belorgey},
  {Claret}, {Doucet}, {Dubreuil}, {Durand}, {Elswijk}, {Girardot}, {K{\"a}ufl},
  {Kroes}, {Lortholary}, {Lussignol}, {Marchesi}, {Pantin}, {Peletier},
  {Pirard}, {Pragt}, {Rio}, {Schoenmaker}, {Siebenmorgen}, {Silber}, {Smette},
  {Sterzik}, \& {Veyssiere}}]{Lagage2004}
{Lagage}, P.~O., {Pel}, J.~W., {Authier}, M., {et~al.} 2004, The Messenger,
  117, 12

\bibitem[{{LaMassa} {et~al.}(2012){LaMassa}, {Heckman}, {Ptak}, {Schiminovich},
  {O'Dowd}, \& {Bertincourt}}]{LaMassa2012}
{LaMassa}, S.~M., {Heckman}, T.~M., {Ptak}, A., {et~al.} 2012, \apj, 758, 1

\bibitem[{{Lebouteiller} {et~al.}(2011){Lebouteiller}, {Barry}, {Spoon},
  {Bernard-Salas}, \& et~al.}]{Lebouteiller2011}
{Lebouteiller}, V., {Barry}, D.~J., {Spoon}, H.~W.~W., {Bernard-Salas}, J., \&
  et~al. 2011, \apjs, 196, 8

\bibitem[{{Levenson} {et~al.}(2009){Levenson}, {Radomski}, {Packham}, {Mason},
  {Schaefer}, \& {Telesco}}]{Levenson2009}
{Levenson}, N.~A., {Radomski}, J.~T., {Packham}, C., {et~al.} 2009, \apj, 703,
  390

\bibitem[{{Li} \& {Draine}(2002)}]{Li2002}
{Li}, A., \& {Draine}, B.~T. 2002, \apj, 572, 232

\bibitem[{{Magorrian} {et~al.}(1998){Magorrian}, {Tremaine}, {Richstone},
  {Bender}, {Bower}, {Dressler}, {Faber}, {Gebhardt}, {Green}, {Grillmair},
  {Kormendy}, \& {Lauer}}]{Magorrian1998}
{Magorrian}, J., {Tremaine}, S., {Richstone}, D., {et~al.} 1998, \aj, 115, 2285

\bibitem[{{Maiolino} \& {Rieke}(1995)}]{Maiolino1995}
{Maiolino}, R., \& {Rieke}, G.~H. 1995, \apj, 454, 95

\bibitem[{{Malizia} {et~al.}(2007){Malizia}, {Landi}, {Bassani}, {Bird},
  {Molina}, \& et~al.}]{Malizia2007}
{Malizia}, A., {Landi}, R., {Bassani}, L., {et~al.} 2007, \apj, 668, 81

\bibitem[{{Marco} \& {Brooks}(2003)}]{Marco2003}
{Marco}, O., \& {Brooks}, K.~J. 2003, \aap, 398, 101

\bibitem[{{Marconi} {et~al.}(2004){Marconi}, {Risaliti}, {Gilli}, {Hunt},
  {Maiolino}, \& {Salvati}}]{Marconi2004}
{Marconi}, A., {Risaliti}, G., {Gilli}, R., {et~al.} 2004, \mnras, 351, 169

\bibitem[{{Marinucci} {et~al.}(2012){Marinucci}, {Bianchi}, {Nicastro}, {Matt},
  \& {Goulding}}]{Marinucci2012}
{Marinucci}, A., {Bianchi}, S., {Nicastro}, F., {Matt}, G., \& {Goulding},
  A.~D. 2012, \apj, 748, 130

\bibitem[{{Mason} {et~al.}(2006){Mason}, {Geballe}, {Packham}, {Levenson},
  {Elitzur}, \& et~al.}]{Mason2006}
{Mason}, R.~E., {Geballe}, T., {Packham}, C., {et~al.} 2006, \apj, 640, 612

\bibitem[{{Mason} {et~al.}(2007){Mason}, {Levenson}, {Packham}, {Elitzur},
  {Radomski}, {Petric}, \& {Wright}}]{Mason2007}
{Mason}, R.~E., {Levenson}, N.~A., {Packham}, C., {et~al.} 2007, \apj, 659, 241

\bibitem[{{Mason} {et~al.}(2009){Mason}, {Levenson}, {Shi}, {Packham},
  {Gorjian}, {Cleary}, {Rhee}, \& {Werner}}]{Mason2009}
{Mason}, R.~E., {Levenson}, N.~A., {Shi}, Y., {et~al.} 2009, \apjl, 693, L136

\bibitem[{{Mason} {et~al.}(2012){Mason}, {Lopez-Rodriguez}, {Packham},
  {Alonso-Herrero}, \& et~al.}]{Mason2012}
{Mason}, R.~E., {Lopez-Rodriguez}, E., {Packham}, C., {Alonso-Herrero}, A., \&
  et~al. 2012, \aj, 144, 11

\bibitem[{{Mel{\'e}ndez} {et~al.}(2008){Mel{\'e}ndez}, {Kraemer}, {Schmitt},
  {Crenshaw}, {Deo}, {Mushotzky}, \& {Bruhweiler}}]{Melendez2008}
{Mel{\'e}ndez}, M., {Kraemer}, S.~B., {Schmitt}, H.~R., {et~al.} 2008, \apj,
  689, 95

\bibitem[{{Miles} {et~al.}(1994){Miles}, {Houck}, \& {Hayward}}]{Miles1994}
{Miles}, J.~W., {Houck}, J.~R., \& {Hayward}, T.~L. 1994, \apjl, 425, L37

\bibitem[{{Mueller} {et~al.}(2003){Mueller}, {Madejski}, {Done}, \&
  {Zycki}}]{Mueller2003}
{Mueller}, M., {Madejski}, G.~M., {Done}, C., \& {Zycki}, P.~T. 2003, in
  Bulletin of the American Astronomical Society, Vol.~35, 637

\bibitem[{{Mullaney} {et~al.}(2011){Mullaney}, {Alexander}, {Goulding}, \&
  {Hickox}}]{Mullaney2011}
{Mullaney}, J.~R., {Alexander}, D.~M., {Goulding}, A.~D., \& {Hickox}, R.~C.
  2011, \mnras, 414, 1082

\bibitem[{{Nayakshin} \& {Zubovas}(2012)}]{Nayakshin2012}
{Nayakshin}, S., \& {Zubovas}, K. 2012, \mnras, 427, 372

\bibitem[{{Nenkova} {et~al.}(2008{\natexlab{a}}){Nenkova}, {Sirocky},
  {Ivezi{\'c}}, \& {Elitzur}}]{Nenkova2008a}
{Nenkova}, M., {Sirocky}, M.~M., {Ivezi{\'c}}, {\v Z}., \& {Elitzur}, M.
  2008{\natexlab{a}}, \apj, 685, 147

\bibitem[{{Nenkova} {et~al.}(2008{\natexlab{b}}){Nenkova}, {Sirocky},
  {Nikutta}, {Ivezi{\'c}}, \& {Elitzur}}]{Nenkova2008b}
{Nenkova}, M., {Sirocky}, M.~M., {Nikutta}, R., {Ivezi{\'c}}, {\v Z}., \&
  {Elitzur}, M. 2008{\natexlab{b}}, \apj, 685, 160

\bibitem[{{Panessa} {et~al.}(2006){Panessa}, {Bassani}, {Cappi}, {Dadina},
  {Barcons}, {Carrera}, {Ho}, \& {Iwasawa}}]{Panessa2006}
{Panessa}, F., {Bassani}, L., {Cappi}, M., {et~al.} 2006, \aap, 455, 173

\bibitem[{{Peeters} {et~al.}(2004){Peeters}, {Allamandola}, {Hudgins}, {Hony},
  \& {Tielens}}]{Peeters2004}
{Peeters}, E., {Allamandola}, L.~J., {Hudgins}, D.~M., {Hony}, S., \&
  {Tielens}, A.~G.~G.~M. 2004, in Astronomical Society of the Pacific
  Conference Series, Vol. 309, Astrophysics of Dust, ed. A.~N. {Witt}, G.~C.
  {Clayton}, \& B.~T. {Draine}, 141

\bibitem[{{Peeters} {et~al.}(2002){Peeters}, {Hony}, {Van Kerckhoven},
  {Tielens}, {Allamandola}, {Hudgins}, \& {Bauschlicher}}]{Peeters2002}
{Peeters}, E., {Hony}, S., {Van Kerckhoven}, C., {et~al.} 2002, \aap, 390, 1089

\bibitem[{{Pereira-Santaella} {et~al.}(2010){Pereira-Santaella},
  {Diamond-Stanic}, {Alonso-Herrero}, \& {Rieke}}]{PereiraSantaella2010}
{Pereira-Santaella}, M., {Diamond-Stanic}, A.~M., {Alonso-Herrero}, A., \&
  {Rieke}, G.~H. 2010, \apj, 725, 2270

\bibitem[{{Piconcelli} {et~al.}(2007){Piconcelli}, {Bianchi}, {Guainazzi},
  {Fiore}, \& {Chiaberge}}]{Piconcelli2007}
{Piconcelli}, E., {Bianchi}, S., {Guainazzi}, M., {Fiore}, F., \& {Chiaberge},
  M. 2007, \aap, 466, 855

\bibitem[{{Ramos Almeida} {et~al.}(2013){Ramos Almeida}, {Rodr{\'{\i}}guez
  Espinosa}, {Acosta-Pulido}, {Alonso-Herrero}, {P{\'e}rez Garc{\'{\i}}a}, \&
  {Rodr{\'{\i}}guez-Eugenio}}]{Ramos2013}
{Ramos Almeida}, C., {Rodr{\'{\i}}guez Espinosa}, J.~M., {Acosta-Pulido},
  J.~A., {et~al.} 2013, \mnras, 429, 3449

\bibitem[{{Ramos Almeida} {et~al.}(2009){Ramos Almeida}, {Levenson},
  {Rodr{\'{\i}}guez Espinosa}, {Alonso-Herrero}, {Asensio Ramos}, {Radomski},
  {Packham}, {Fisher}, \& {Telesco}}]{RamosAlmeida2009}
{Ramos Almeida}, C., {Levenson}, N.~A., {Rodr{\'{\i}}guez Espinosa}, J.~M.,
  {et~al.} 2009, \apj, 702, 1127

\bibitem[{{Ramos Almeida} {et~al.}(2011){Ramos Almeida}, {Levenson},
  {Alonso-Herrero}, {Asensio Ramos}, {Rodr{\'{\i}}guez Espinosa}, {P{\'e}rez
  Garc{\'{\i}}a}, {Packham}, {Mason}, {Radomski}, \&
  {D{\'{\i}}az-Santos}}]{RamosAlmeida2011A}
{Ramos Almeida}, C., {Levenson}, N.~A., {Alonso-Herrero}, A., {et~al.} 2011,
  \apj, 731, 92

\bibitem[{{Rieke} {et~al.}(2009){Rieke}, {Alonso-Herrero}, {Weiner},
  {P{\'e}rez-Gonz{\'a}lez}, {Blaylock}, {Donley}, \& {Marcillac}}]{Rieke2009}
{Rieke}, G.~H., {Alonso-Herrero}, A., {Weiner}, B.~J., {et~al.} 2009, \apj,
  692, 556

\bibitem[{{Risaliti} {et~al.}(2005){Risaliti}, {Elvis}, {Fabbiano}, {Baldi}, \&
  {Zezas}}]{Risaliti2005}
{Risaliti}, G., {Elvis}, M., {Fabbiano}, G., {Baldi}, A., \& {Zezas}, A. 2005,
  \apjl, 623, L93

\bibitem[{{Roche} \& {Aitken}(1985)}]{Roche1985}
{Roche}, P.~F., \& {Aitken}, D.~K. 1985, \mnras, 213, 789

\bibitem[{{Roche} {et~al.}(1991){Roche}, {Aitken}, {Smith}, \&
  {Ward}}]{Roche1991}
{Roche}, P.~F., {Aitken}, D.~K., {Smith}, C.~H., \& {Ward}, M.~J. 1991, \mnras,
  248, 606

\bibitem[{{Roche} {et~al.}(2007){Roche}, {Packham}, {Aitken}, \&
  {Mason}}]{Roche2007}
{Roche}, P.~F., {Packham}, C., {Aitken}, D.~K., \& {Mason}, R.~E. 2007, \mnras,
  375, 99

\bibitem[{{Roche} {et~al.}(2006){Roche}, {Packham}, {Telesco}, {Radomski},
  {Alonso-Herrero}, {Aitken}, {Colina}, \& {Perlman}}]{Roche2006}
{Roche}, P.~F., {Packham}, C., {Telesco}, C.~M., {et~al.} 2006, \mnras, 367,
  1689

\bibitem[{{Rush} {et~al.}(1993){Rush}, {Malkan}, \& {Spinoglio}}]{Rush1993}
{Rush}, B., {Malkan}, M.~A., \& {Spinoglio}, L. 1993, \apjs, 89, 1

\bibitem[{{Sales} {et~al.}(2013){Sales}, {Pastoriza}, {Riffel}, \&
  {Winge}}]{Sales2013}
{Sales}, D.~A., {Pastoriza}, M.~G., {Riffel}, R., \& {Winge}, C. 2013, \mnras,
  429, 2634

\bibitem[{{Sales} {et~al.}(2011){Sales}, {Pastoriza}, {Riffel}, {Winge},
  {Rodr{\'{\i}}guez-Ardila}, \& {Carciofi}}]{Sales2011}
{Sales}, D.~A., {Pastoriza}, M.~G., {Riffel}, R., {et~al.} 2011, \apj, 738, 109

\bibitem[{{Sandage} \& {Tammann}(1987)}]{Sandage1987}
{Sandage}, A., \& {Tammann}, G.~A. 1987,

\bibitem[{{Shi} {et~al.}(2007){Shi}, {Ogle}, {Rieke}, {Antonucci}, {Hines},
  {Smith}, {Low}, {Bouwman}, \& {Willmer}}]{Shi2007}
{Shi}, Y., {Ogle}, P., {Rieke}, G.~H., {et~al.} 2007, \apj, 669, 841

\bibitem[{{Siebenmorgen} {et~al.}(2004){Siebenmorgen}, {Kr{\"u}gel}, \&
  {Spoon}}]{Siebenmorgen2004}
{Siebenmorgen}, R., {Kr{\"u}gel}, E., \& {Spoon}, H.~W.~W. 2004, \aap, 414, 123

\bibitem[{{Silk} \& {Rees}(1998)}]{Silk1998}
{Silk}, J., \& {Rees}, M.~J. 1998, \aap, 331, L1

\bibitem[{{Smith} {et~al.}(2007){Smith}, {Draine}, {Dale}, {Moustakas},
  {Kennicutt}, \& et~al.}]{Smith2007}
{Smith}, J., {Draine}, B., {Dale}, D., {et~al.} 2007, \apj, 656, 770

\bibitem[{{Spoon} {et~al.}(2006){Spoon}, {Tielens}, {Armus}, {Sloan},
  {Sargent}, {Cami}, {Charmandaris}, {Houck}, \& {Soifer}}]{Spoon2006}
{Spoon}, H.~W.~W., {Tielens}, A.~G.~G.~M., {Armus}, L., {et~al.} 2006, \apj,
  638, 759

\bibitem[{{Storchi-Bergmann} {et~al.}(2000){Storchi-Bergmann}, {Raimann},
  {Bica}, \& {Fraquelli}}]{StorchiBergmann2000}
{Storchi-Bergmann}, T., {Raimann}, D., {Bica}, E.~L.~D., \& {Fraquelli}, H.~A.
  2000, \apj, 544, 747

\bibitem[{{Tacconi-Garman} \& {Sturm}(2013)}]{Tacconi2013}
{Tacconi-Garman}, L.~E., \& {Sturm}, E. 2013, \aap, 551, A139

\bibitem[{{Telesco} {et~al.}(1998){Telesco}, {Pina}, {Hanna}, {Julian}, {Hon},
  \& {Kisko}}]{Telesco1998}
{Telesco}, C.~M., {Pina}, R.~K., {Hanna}, K.~T., {et~al.} 1998, 3354, 534

\bibitem[{{Telesco} {et~al.}(2003){Telesco}, {Ciardi}, {French}, {Ftaclas},
  {Hanna}, {Hon}, {Hough}, {Julian}, {Julian}, {Kidger}, {Packham}, {Pina},
  {Varosi}, \& {Sellar}}]{Telesco2003}
{Telesco}, C.~M., {Ciardi}, D., {French}, J., {et~al.} 2003, in SPIE Conference
  Series, Vol. 4841, 913--922

\bibitem[{{Thompson} {et~al.}(2005){Thompson}, {Quataert}, \&
  {Murray}}]{Thompson2005}
{Thompson}, T.~A., {Quataert}, E., \& {Murray}, N. 2005, \apj, 630, 167

\bibitem[{{Tielens}(2008)}]{Tielens2008}
{Tielens}, A.~G.~G.~M. 2008, \araa, 46, 289

\bibitem[{{Tielens}(2010)}]{Tielens2010}
{Tielens}, A.~G.~G.~M. 2010,

\bibitem[{{Tueller} {et~al.}(2008){Tueller}, {Mushotzky}, {Barthelmy},
  {Cannizzo}, {Gehrels}, \& et~al.}]{Tueller2008}
{Tueller}, J., {Mushotzky}, R., {Barthelmy}, S., {et~al.} 2008, \apj, 681, 113

\bibitem[{{Uchida} {et~al.}(2000){Uchida}, {Sellgren}, {Werner}, \&
  {Houdashelt}}]{Uchida2000}
{Uchida}, K.~I., {Sellgren}, K., {Werner}, M.~W., \& {Houdashelt}, M.~L. 2000,
  \apj, 530, 817

\bibitem[{{Voit}(1991)}]{Voit1991}
{Voit}, G.~M. 1991, \apj, 379, 122

\bibitem[{{Voit}(1992{\natexlab{a}})}]{Voit1992}
{Voit}, G.~M. 1992{\natexlab{a}}, \mnras, 258, 841

\bibitem[{{Voit}(1992{\natexlab{b}})}]{Voit1992b}
{Voit}, G.~M. 1992{\natexlab{b}}, in Astronomical Society of the Pacific
  Conference Series, Vol.~31, Relationships Between Active Galactic Nuclei and
  Starburst Galaxies, ed. A.~V. {Filippenko}, 87

\bibitem[{{Wada} \& {Norman}(2002)}]{Wada2002}
{Wada}, K., \& {Norman}, C.~A. 2002, \apjl, 566, L21

\bibitem[{{Wang} {et~al.}(2011){Wang}, {Fabbiano}, {Risaliti}, {Elvis},
  {Karovska}, \& et~al.}]{Wang2011}
{Wang}, J., {Fabbiano}, G., {Risaliti}, G., {et~al.} 2011, \apj, 729, 75

\bibitem[{{Watabe} {et~al.}(2008){Watabe}, {Kawakatu}, \&
  {Imanishi}}]{Watabe2008}
{Watabe}, Y., {Kawakatu}, N., \& {Imanishi}, M. 2008, \apj, 677, 895

\bibitem[{{Wild} {et~al.}(2010){Wild}, {Heckman}, \& {Charlot}}]{Wild2010}
{Wild}, V., {Heckman}, T., \& {Charlot}, S. 2010, \mnras, 405, 933

\bibitem[{{Woo} \& {Urry}(2002)}]{Woo2002}
{Woo}, J.-H., \& {Urry}, C.~M. 2002, \apj, 579, 530

\bibitem[{{Young} {et~al.}(2010){Young}, {Elvis}, \& {Risaliti}}]{Young2010}
{Young}, M., {Elvis}, M., \& {Risaliti}, G. 2010, \apj, 708, 1388

\bibitem[{{Young} {et~al.}(2007){Young}, {Packham}, {Mason}, {Radomski}, \&
  {Telesco}}]{Young2007}
{Young}, S., {Packham}, C., {Mason}, R.~E., {Radomski}, J.~T., \& {Telesco},
  C.~M. 2007, \mnras, 378, 888

\end{thebibliography}

\clearpage
\appendix
\renewcommand\thefigure{\thesection A.\arabic{figure}}    
\setcounter{figure}{0}

\begin{figure*}[!b]
\centering
\begin{tabular}{ccc}

{\includegraphics[scale=0.33]{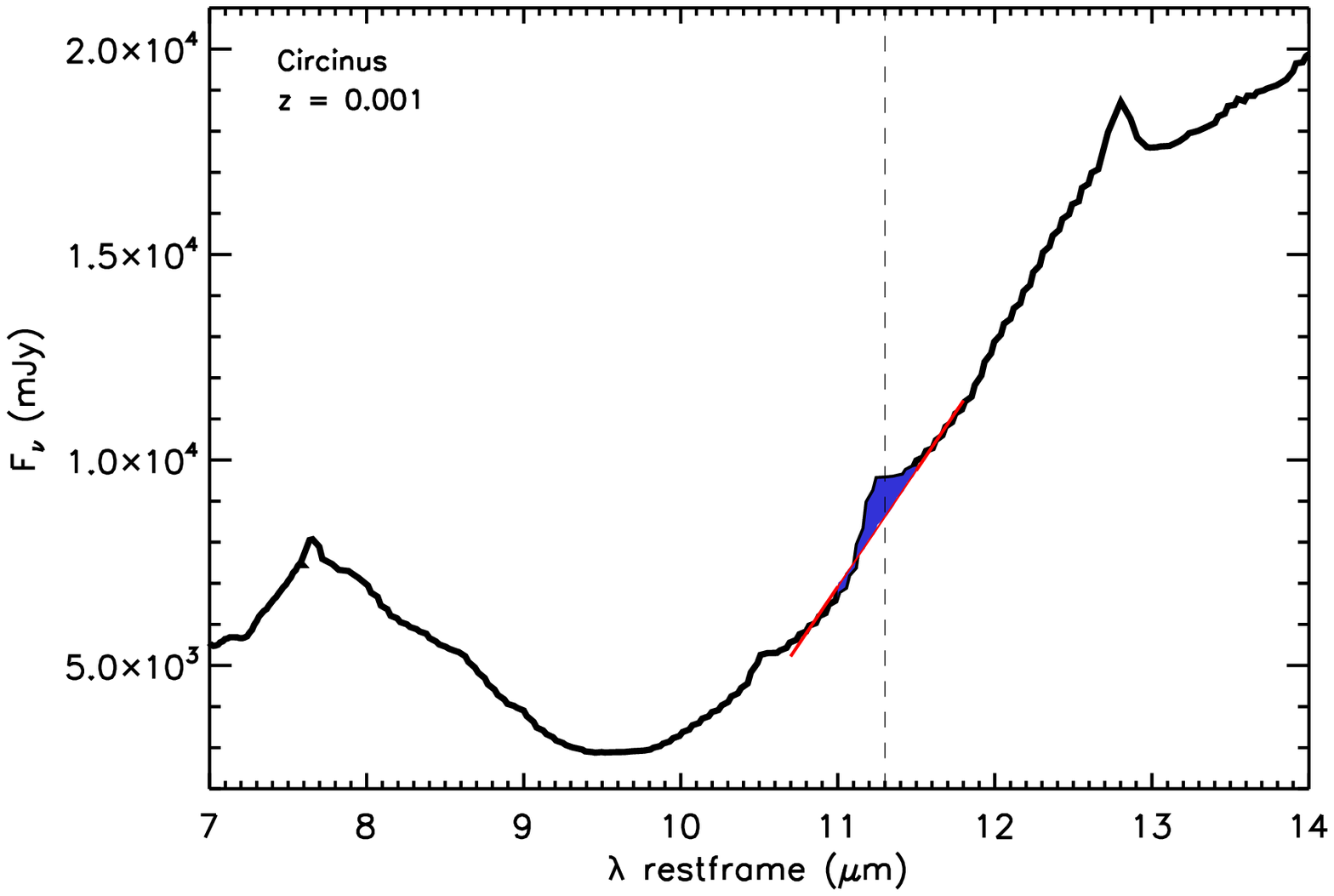}}
{\includegraphics[scale=0.33]{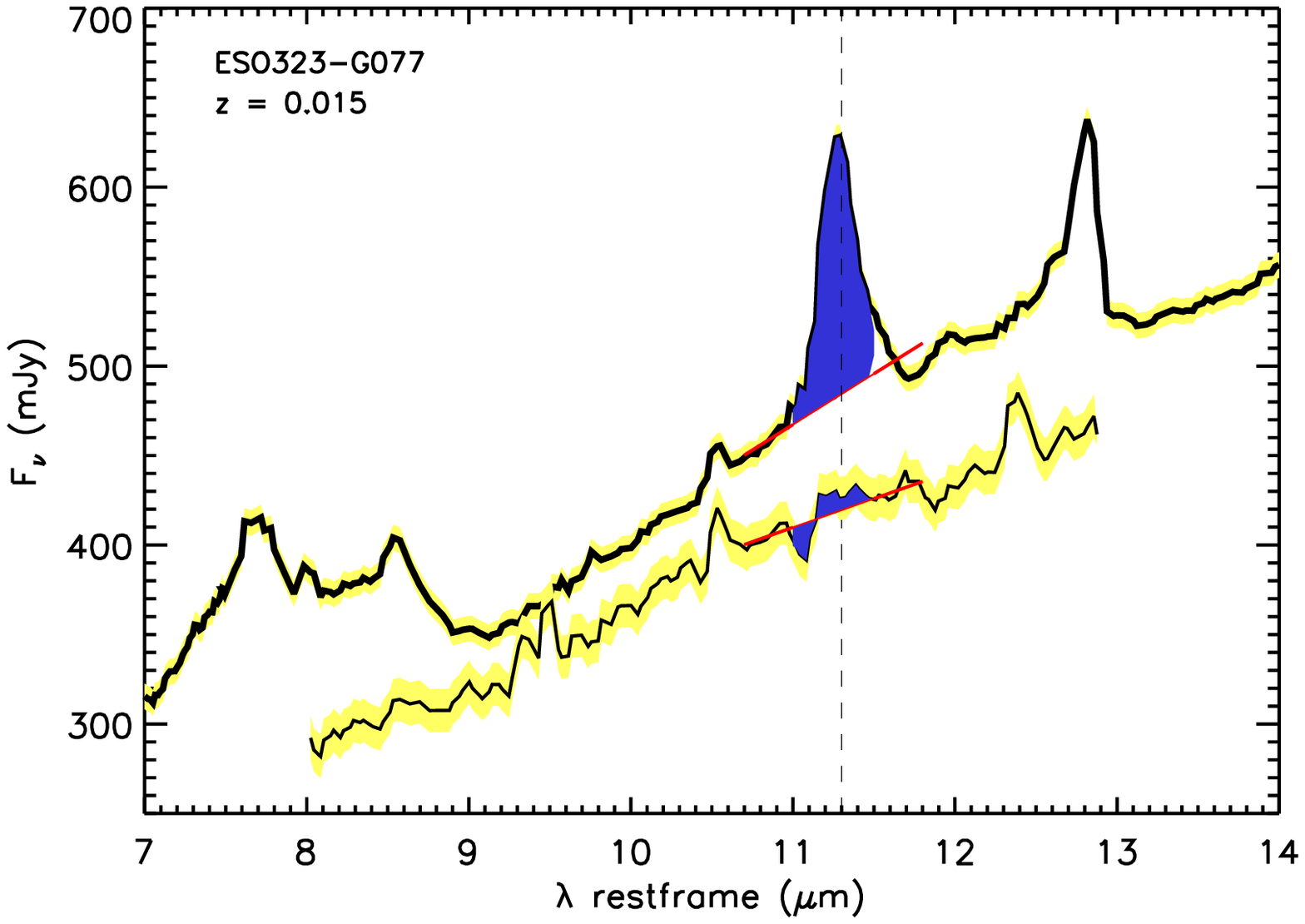}}
{\includegraphics[scale=0.33]{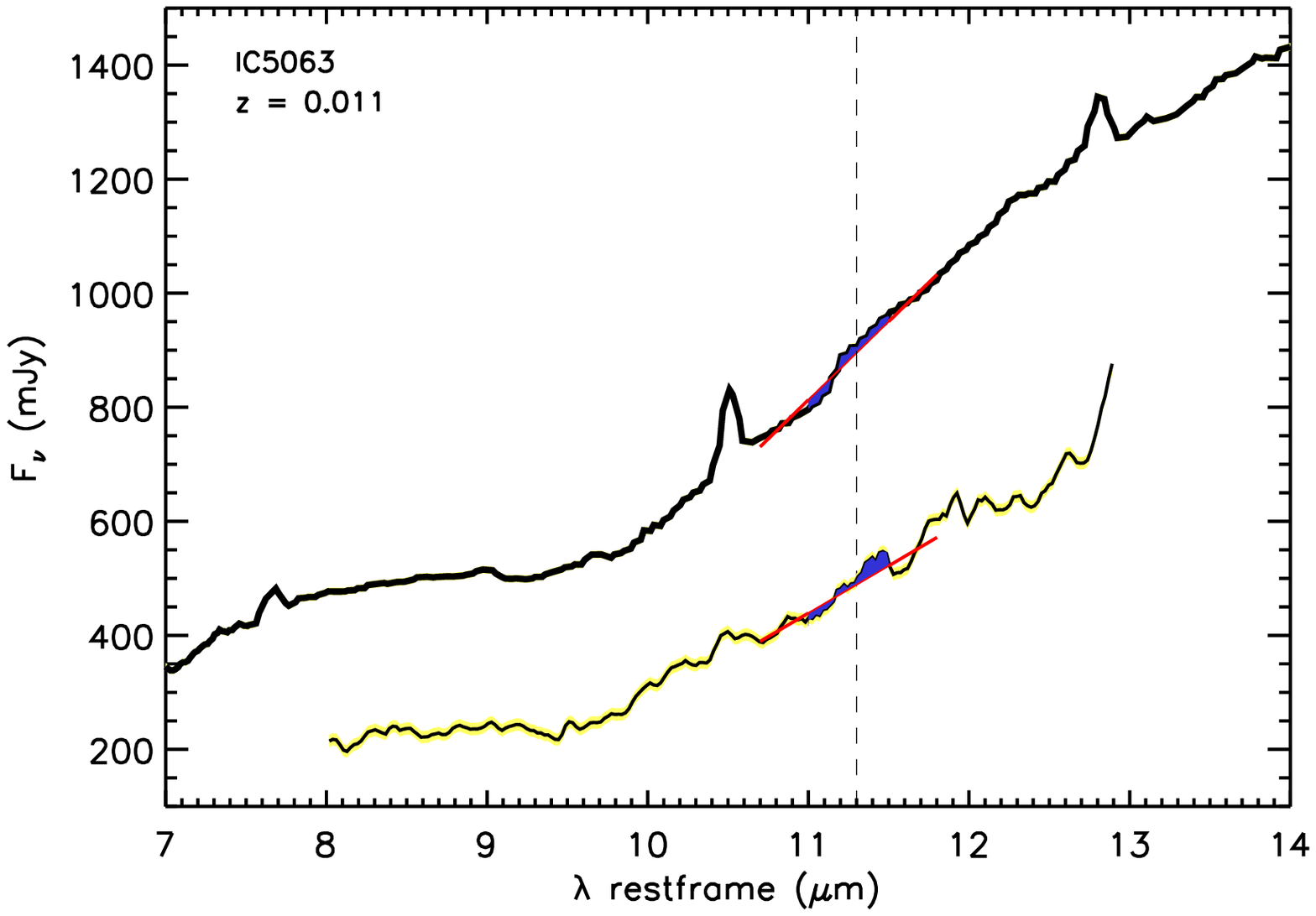}}\\ [-2ex]

{\includegraphics[scale=0.33]{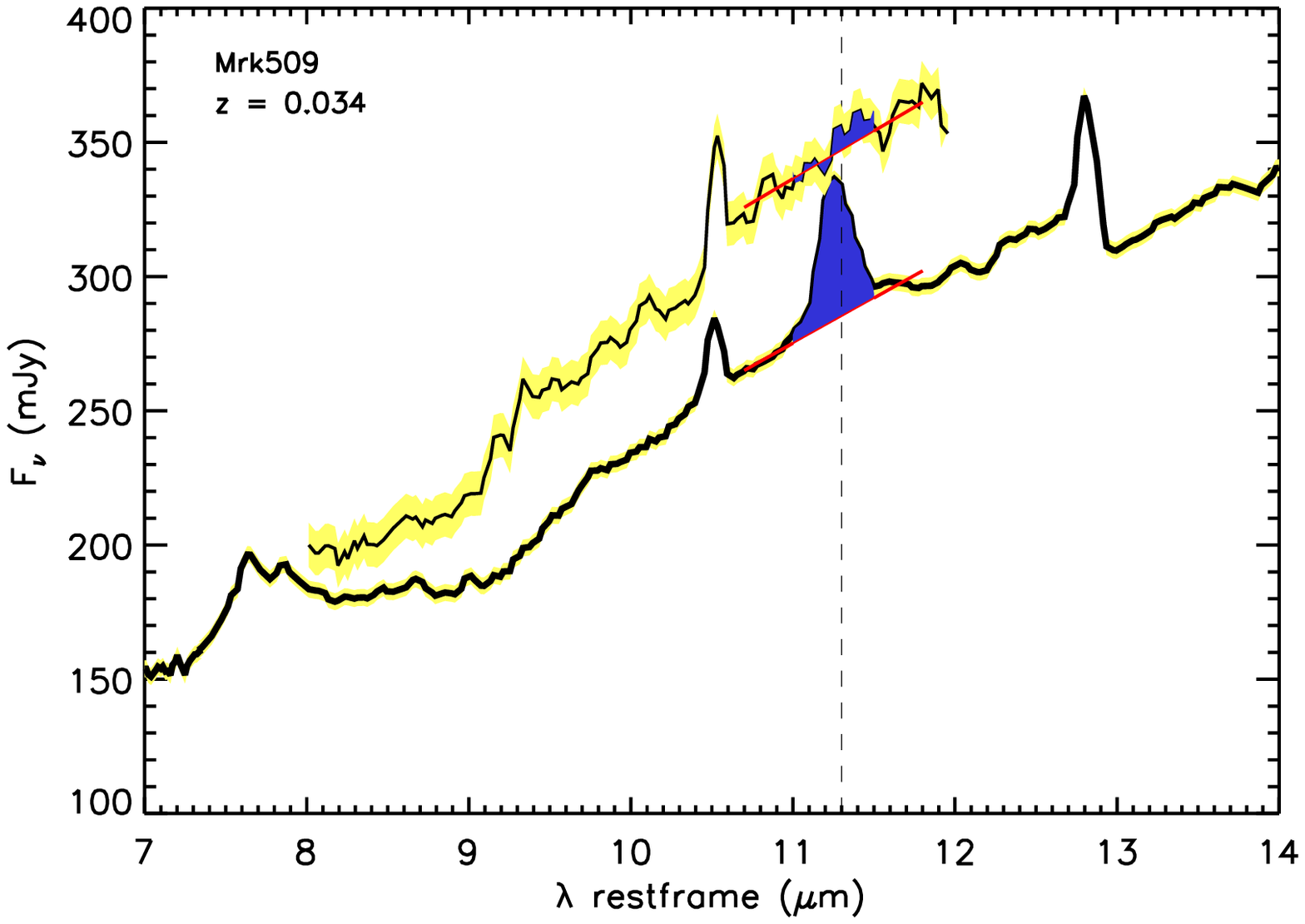}}
{\includegraphics[scale=0.33]{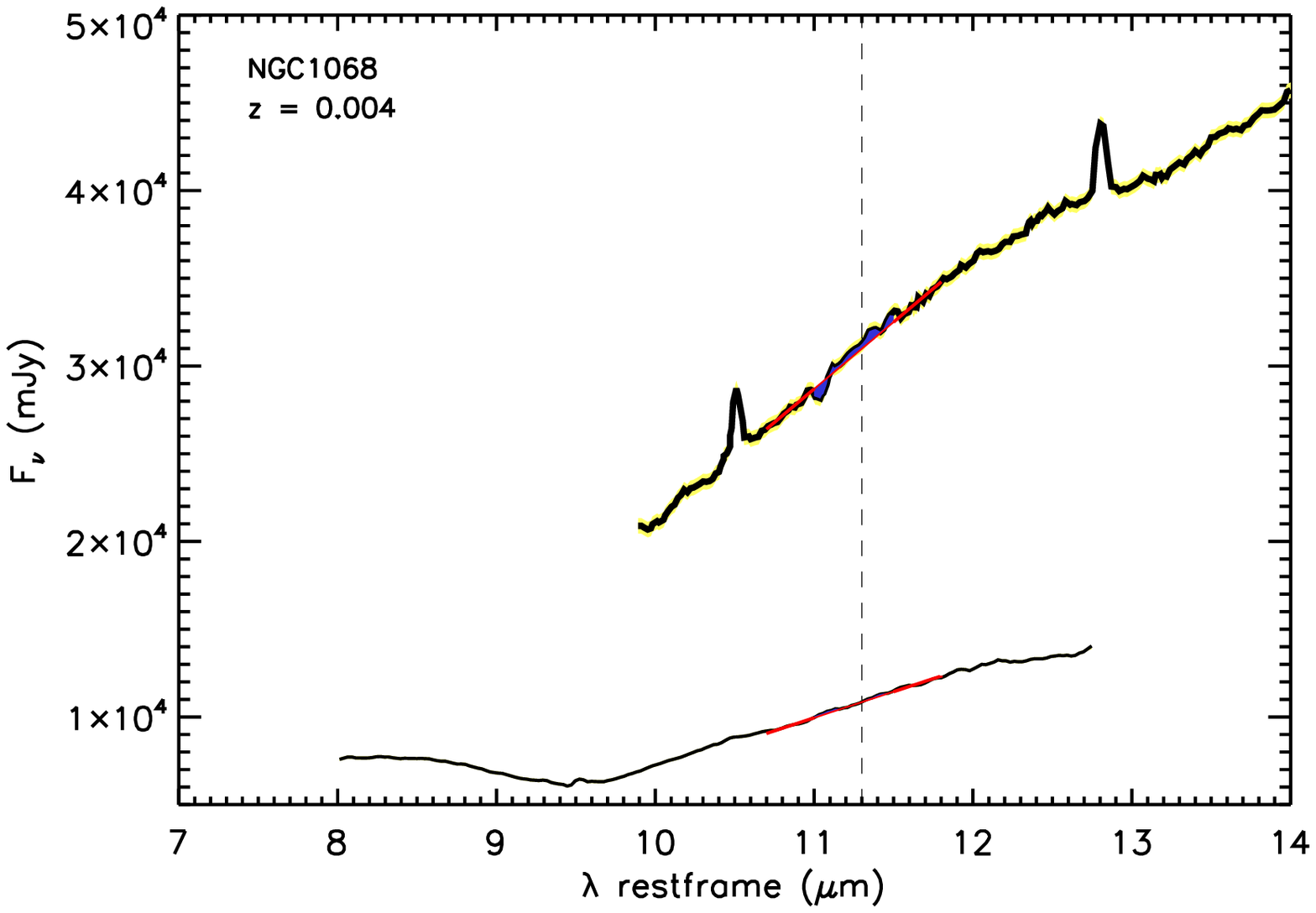}}
{\includegraphics[scale=0.33]{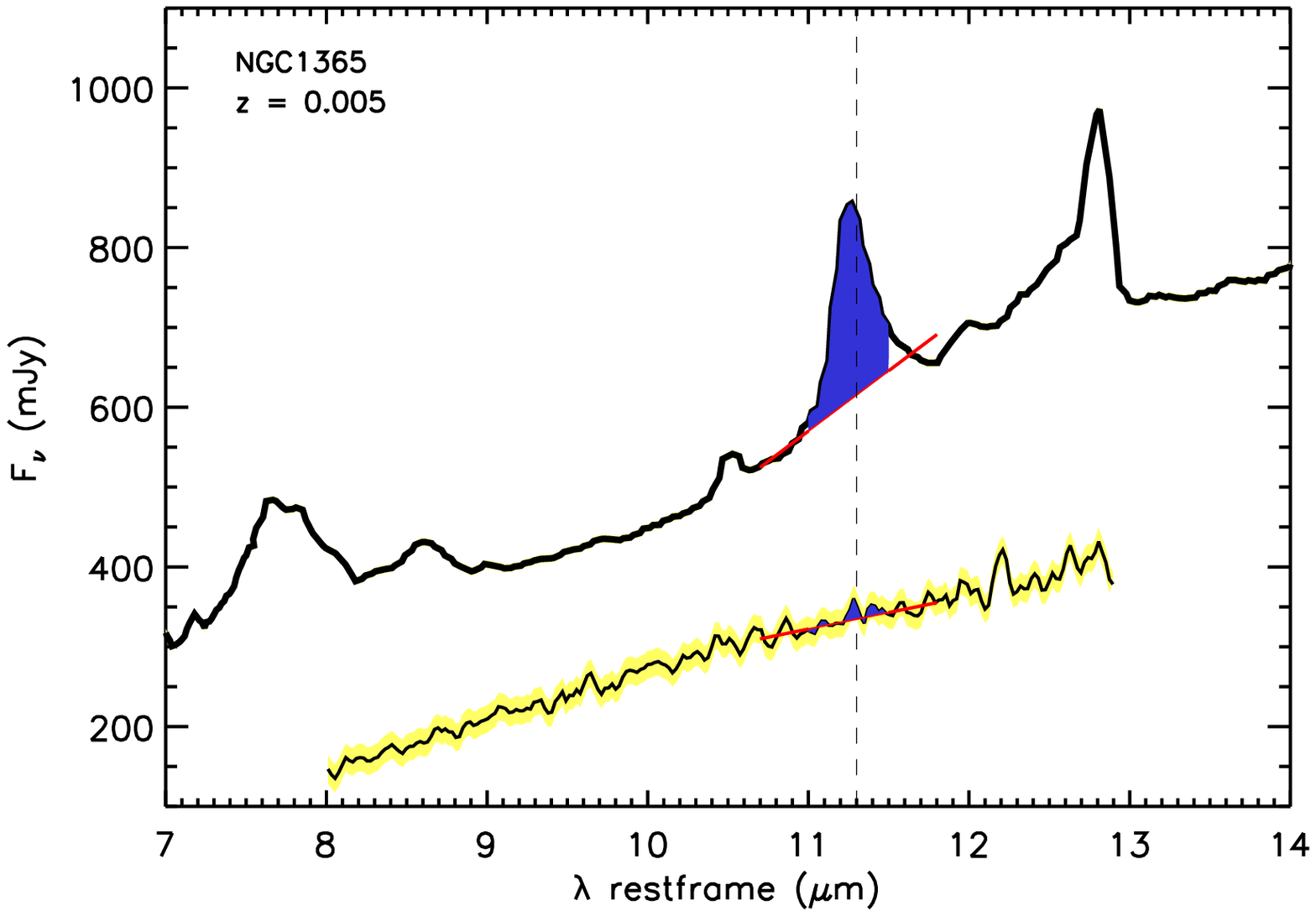}}\\ [-2ex]

{\includegraphics[scale=0.33]{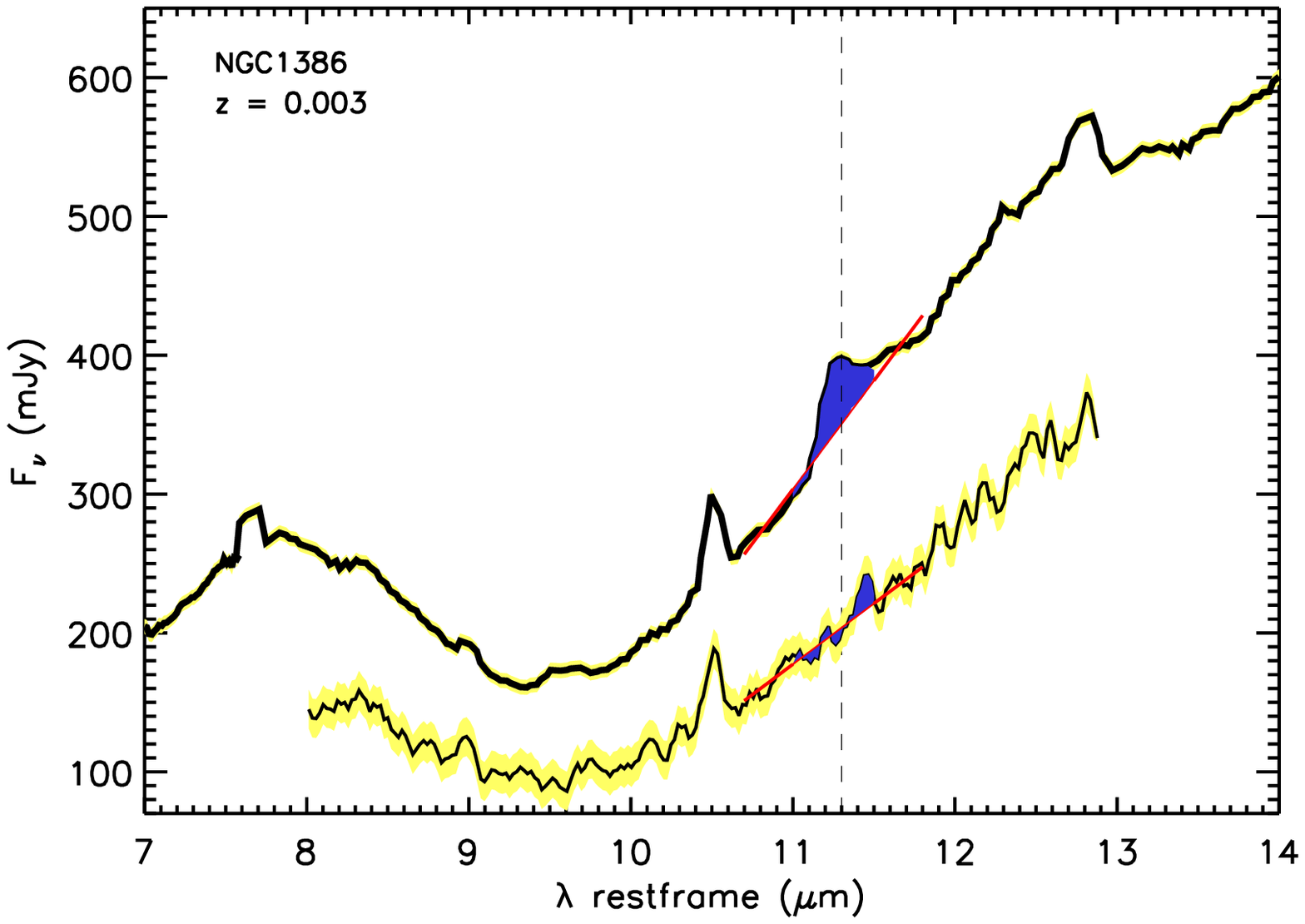}}
{\includegraphics[scale=0.33]{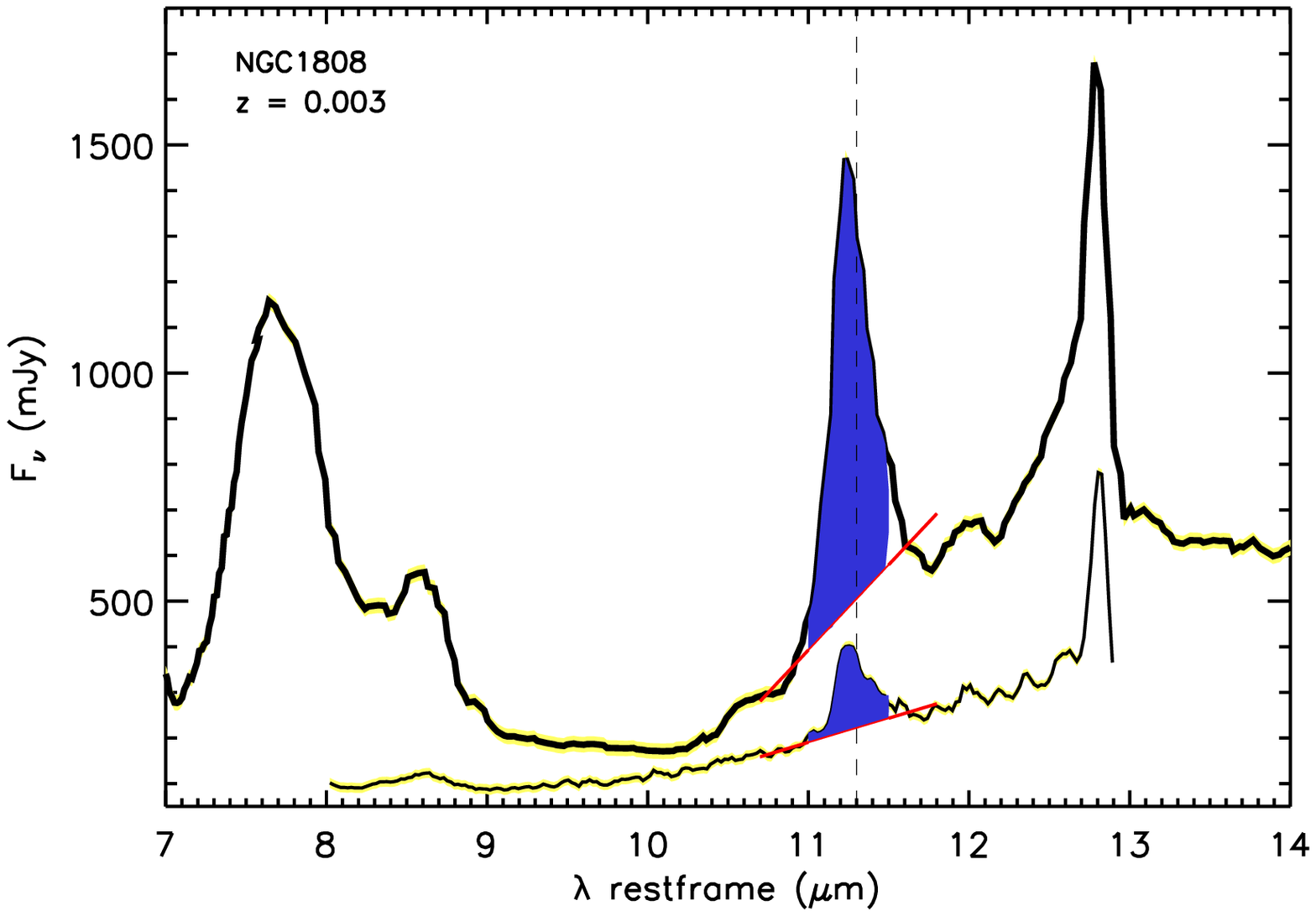}}
{\includegraphics[scale=0.33]{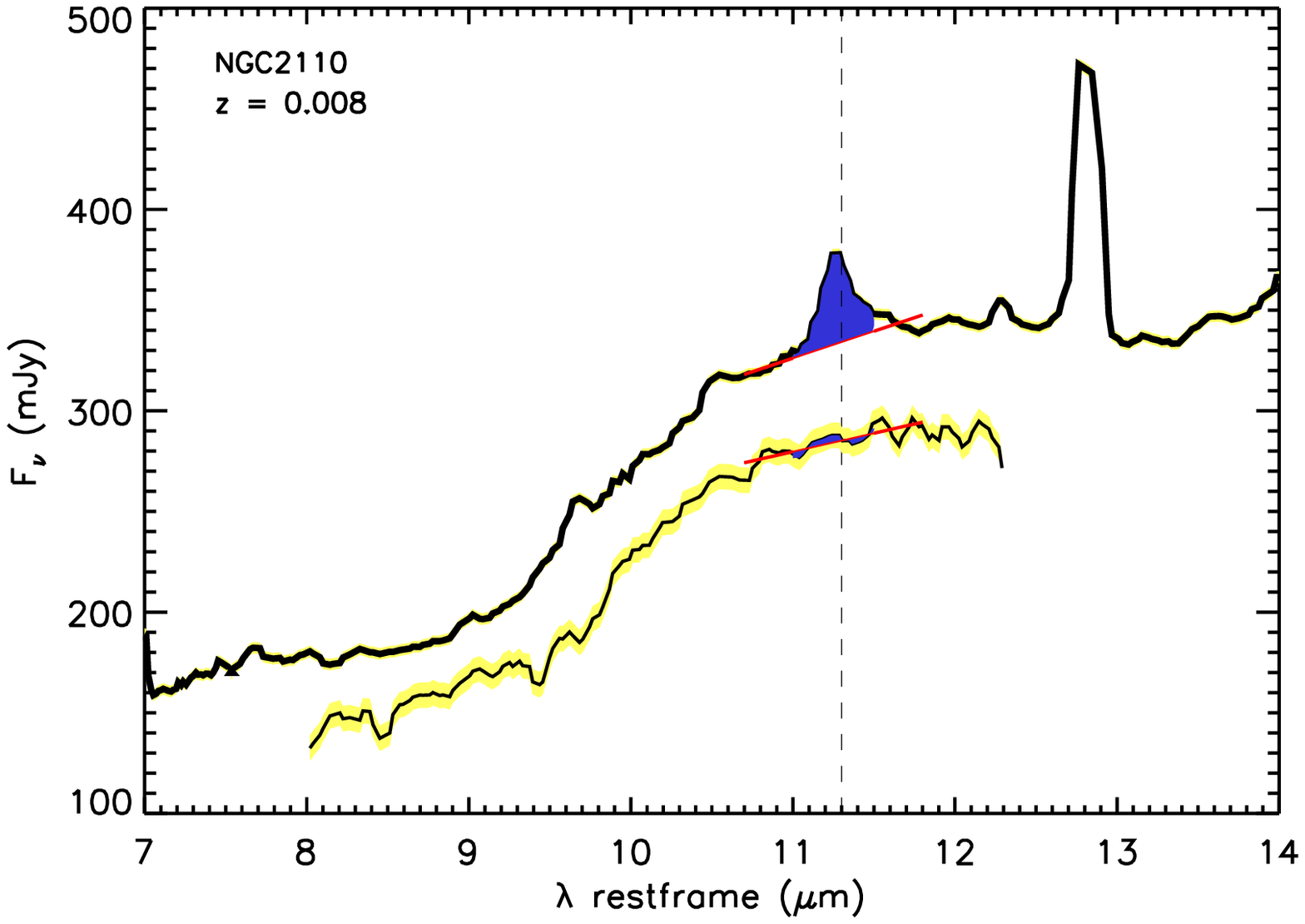}}\\ [-2ex]

{\includegraphics[scale=0.33]{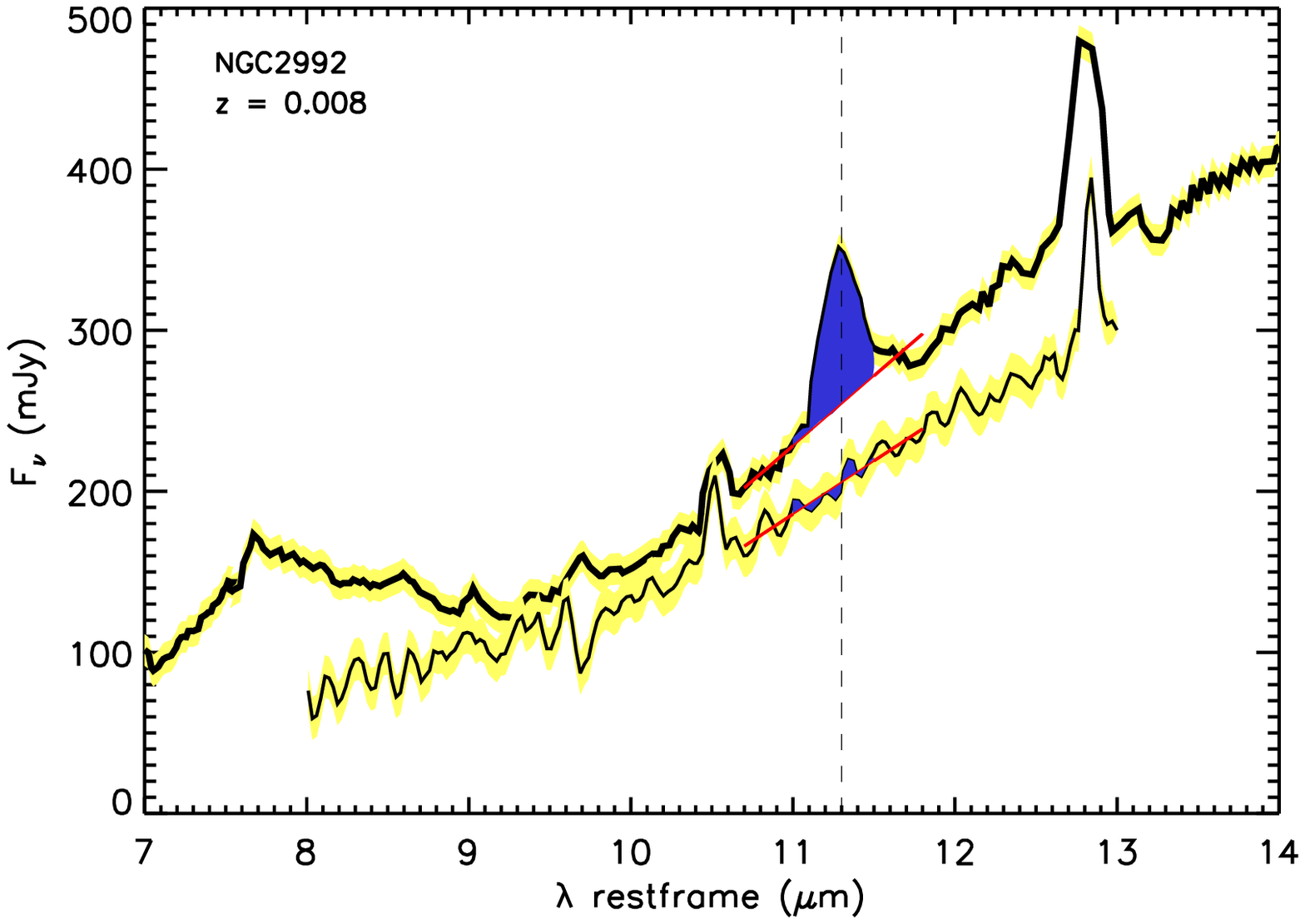}}
{\includegraphics[scale=0.33]{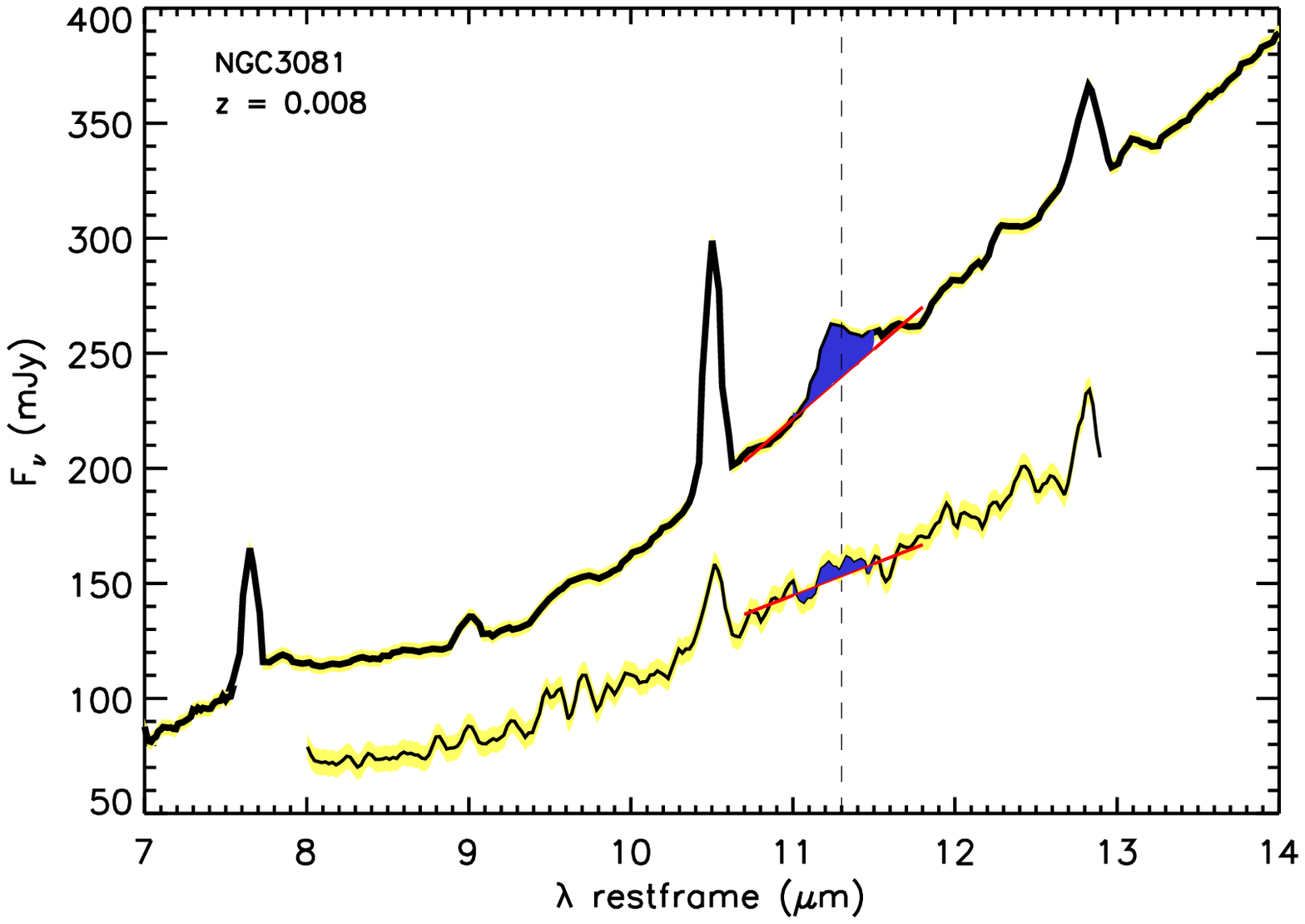}}
{\includegraphics[scale=0.33]{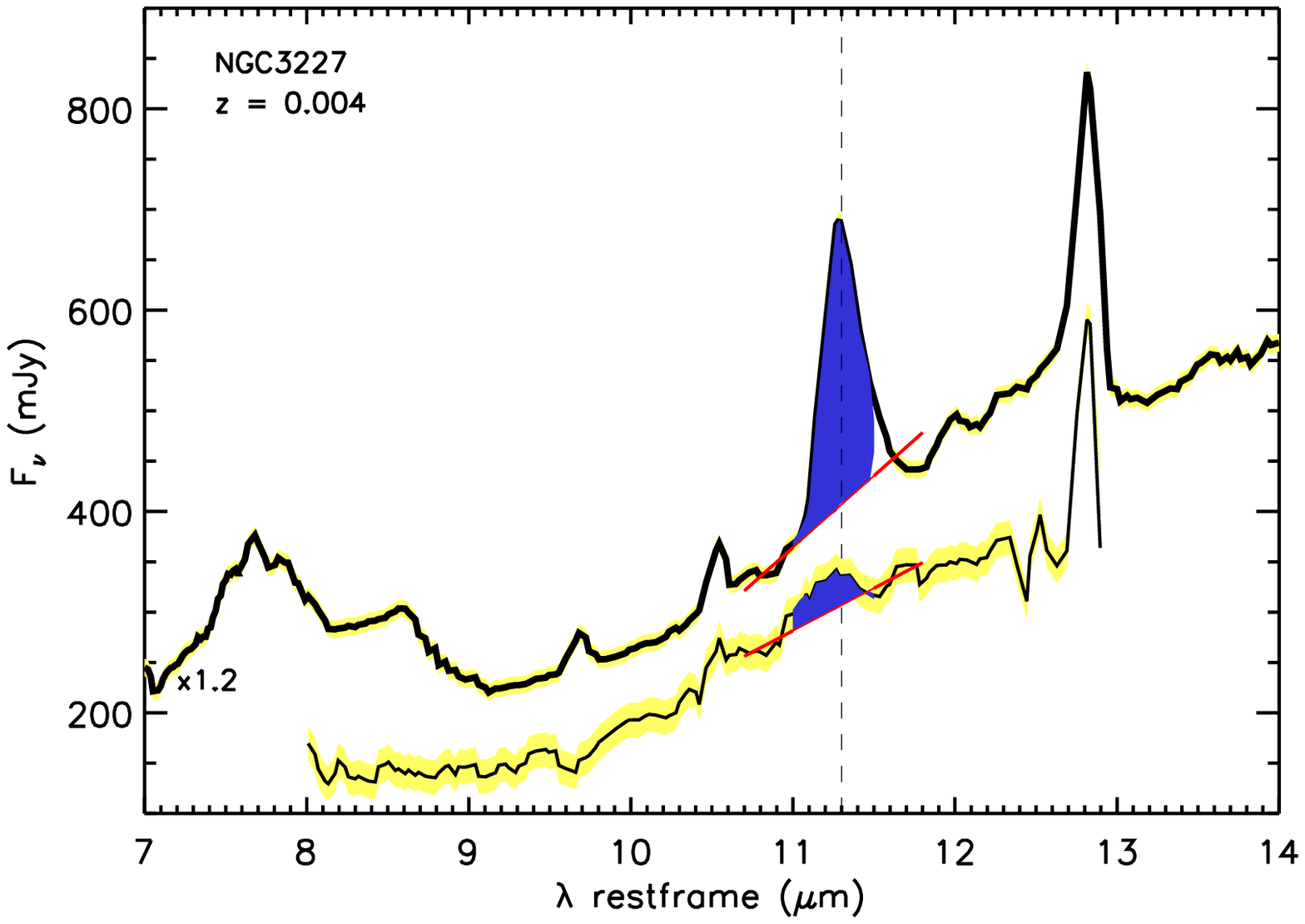}}\\[-1ex]

{\includegraphics[scale=0.33]{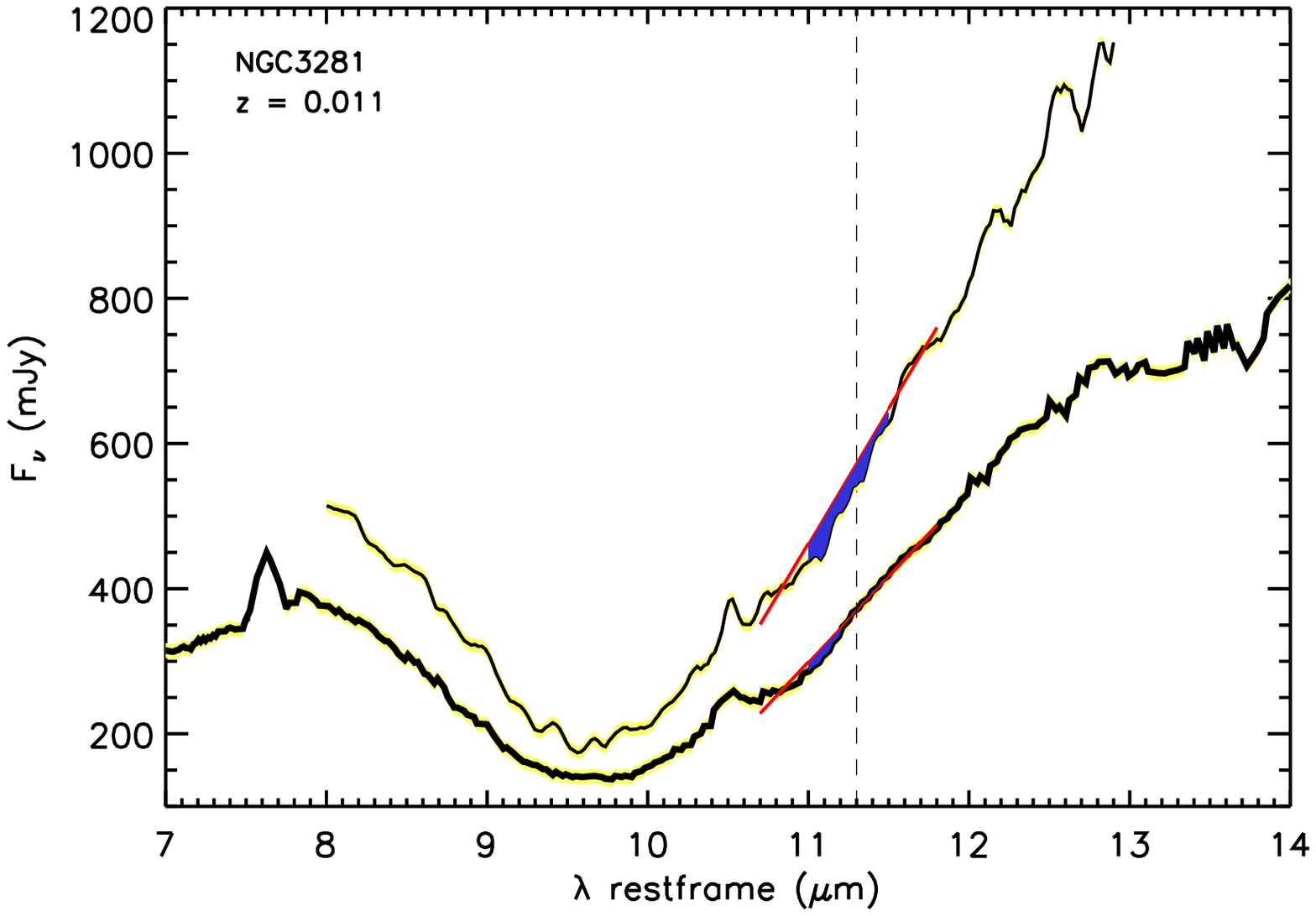}}
{\includegraphics[scale=0.33]{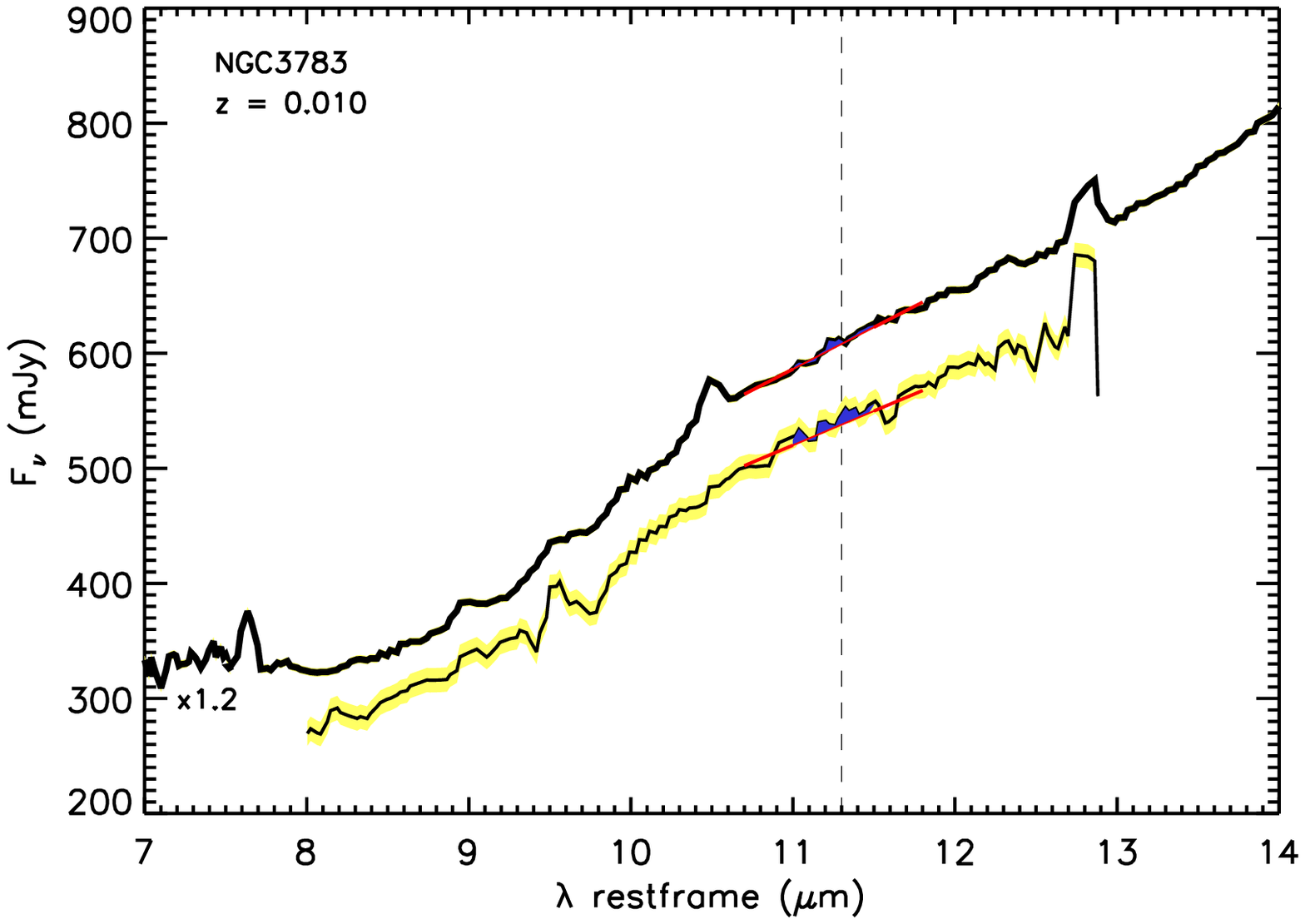}}
{\includegraphics[scale=0.33]{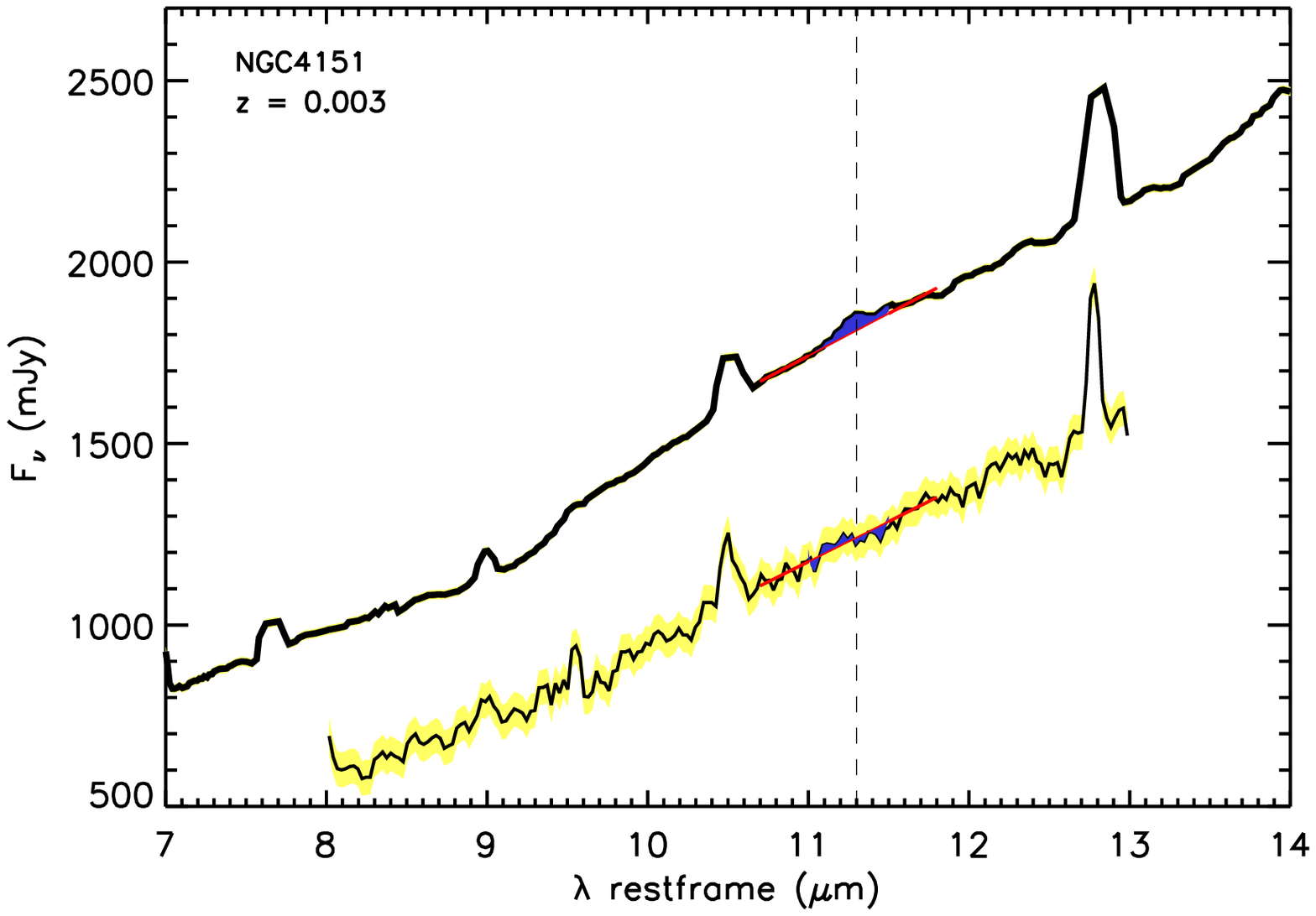}}\\[-1ex]

\end{tabular}
\caption{Spectra of the sample. {\it Spitzer}/IRS SL spectra (thick line) compared with the ground-based
spectra (T-ReCS/VISIR/Michelle: thin line). All observations have been smoothed to the same spectral resolution.
We show the location of the 11.3\,\micron\,PAH feature (dashed line), with
the shaded area indicating the spectral range used for obtaining the integrated flux. 
The red lines are the fitted local continua.
For clarity, we applied a multiplicative factor to the IRS data (shown in the plot) for overlapped spectra.
In a few cases, the nuclear data lie above the circumnuclear data, this can be due to calibration uncertainties.}

\newpage
\label{fig:all_spectra}
\end{figure*}

\begin{figure*}
\centering
\begin{tabular}{cc}

{\includegraphics[scale=0.33]{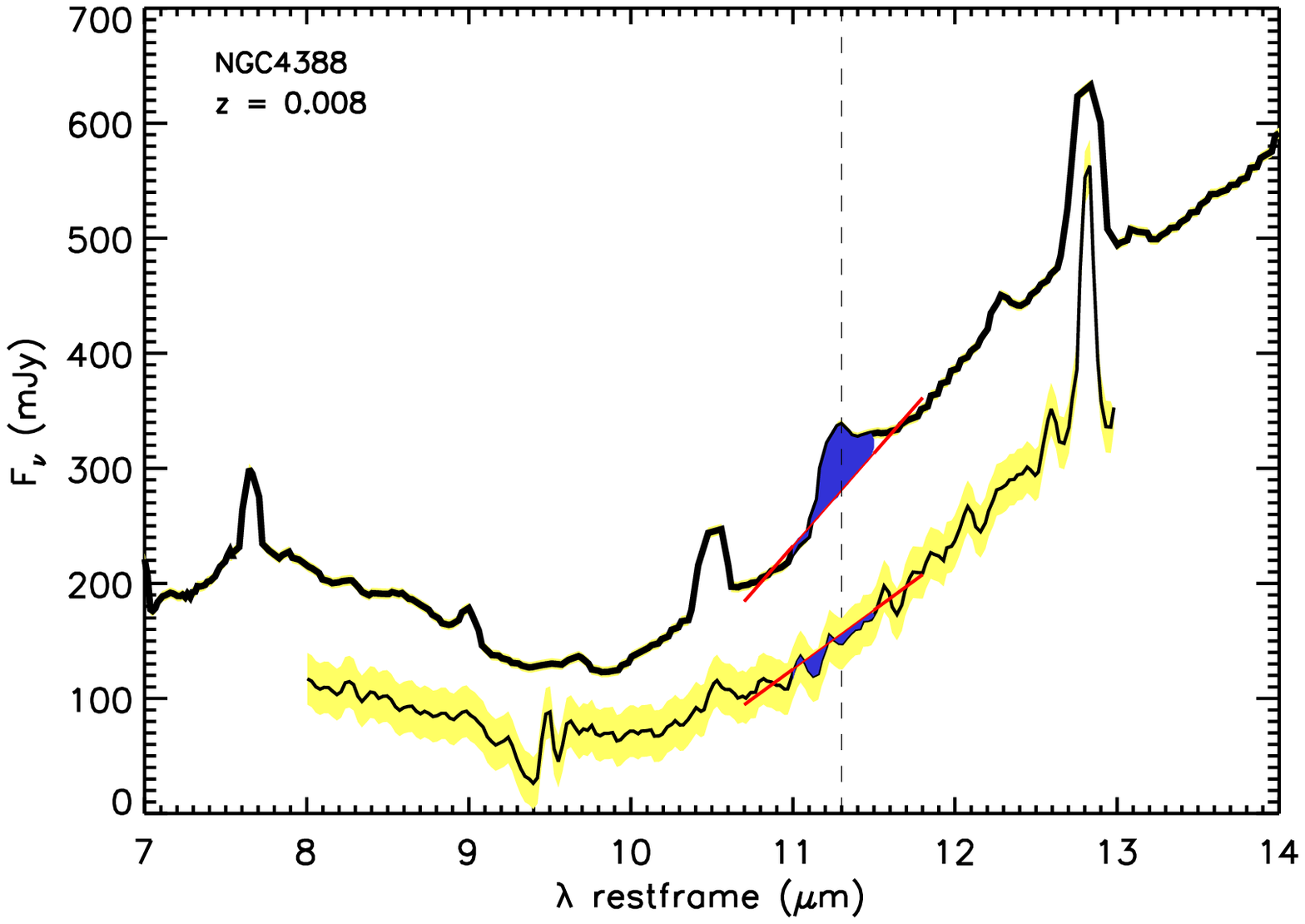}}
{\includegraphics[scale=0.33]{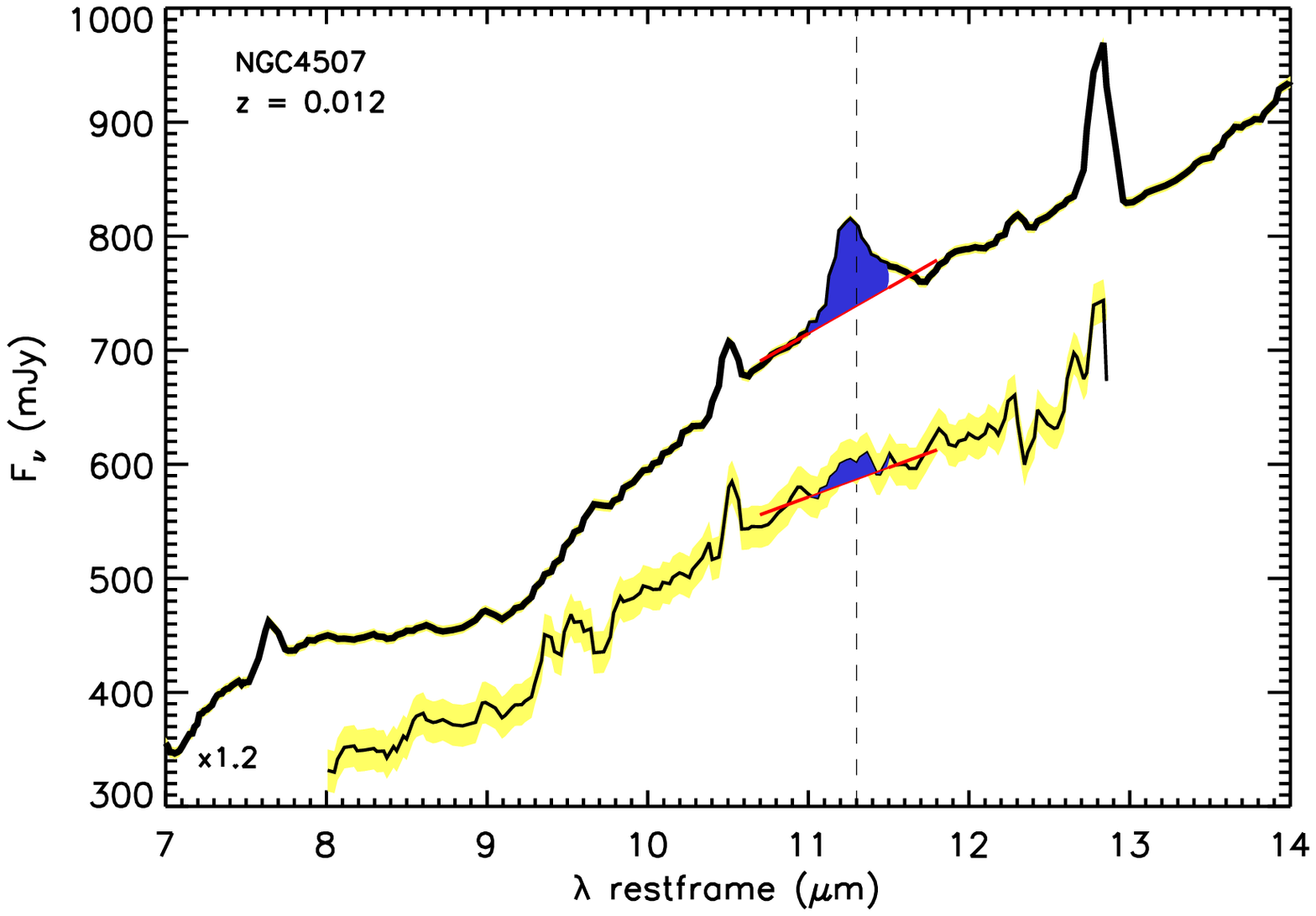}}
{\includegraphics[scale=0.33]{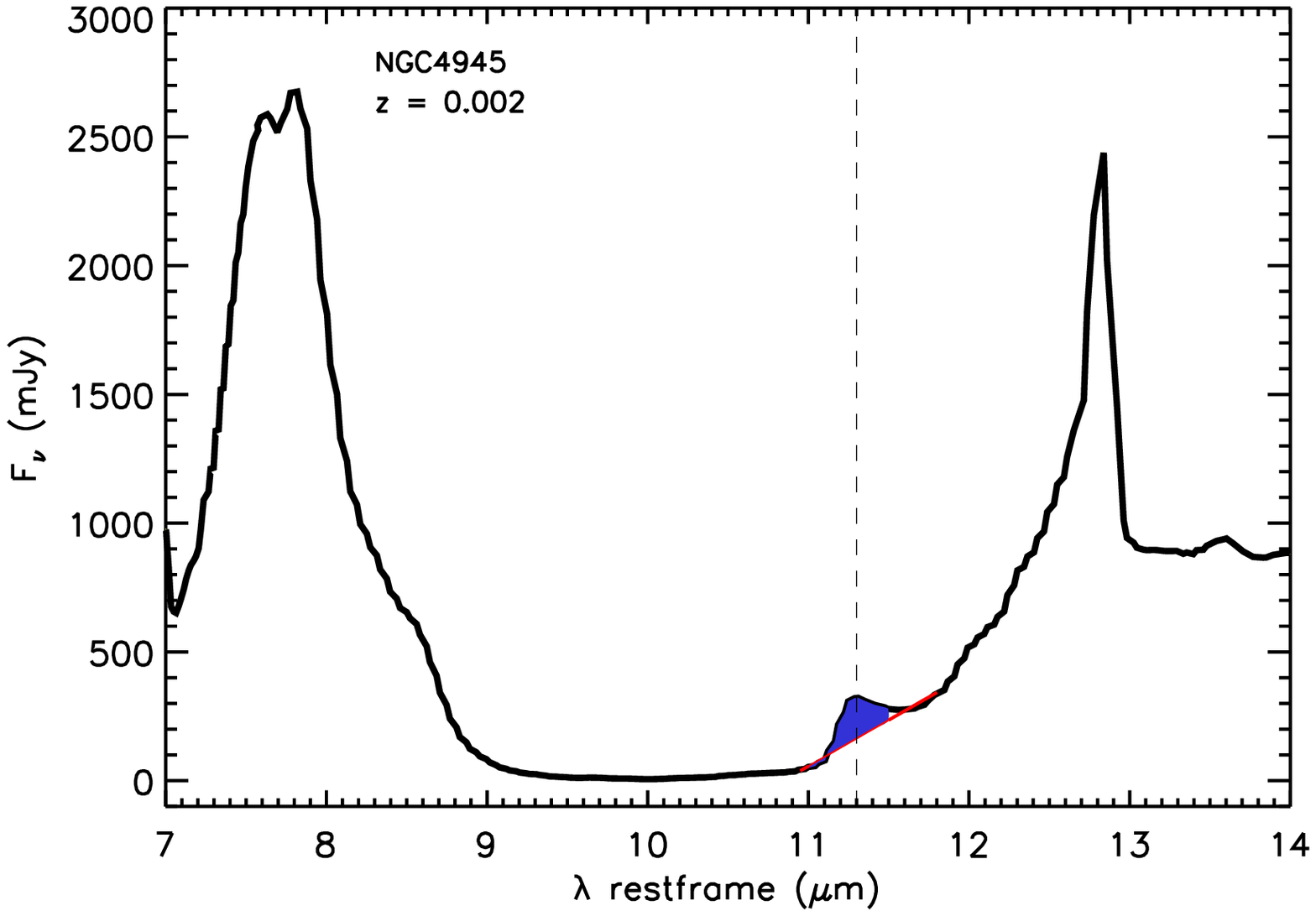}}\\[-1ex]

{\includegraphics[scale=0.33]{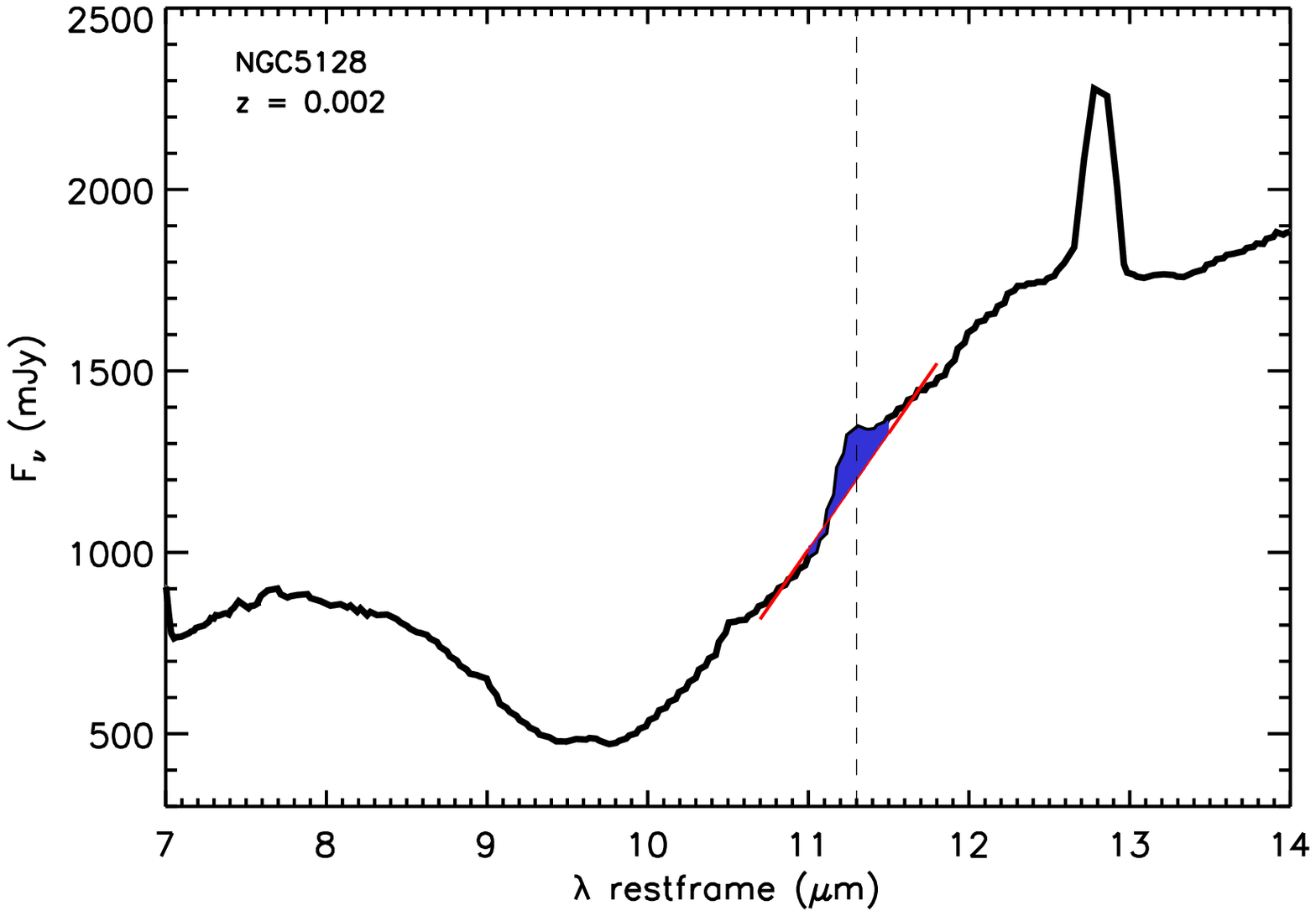}}
{\includegraphics[scale=0.33]{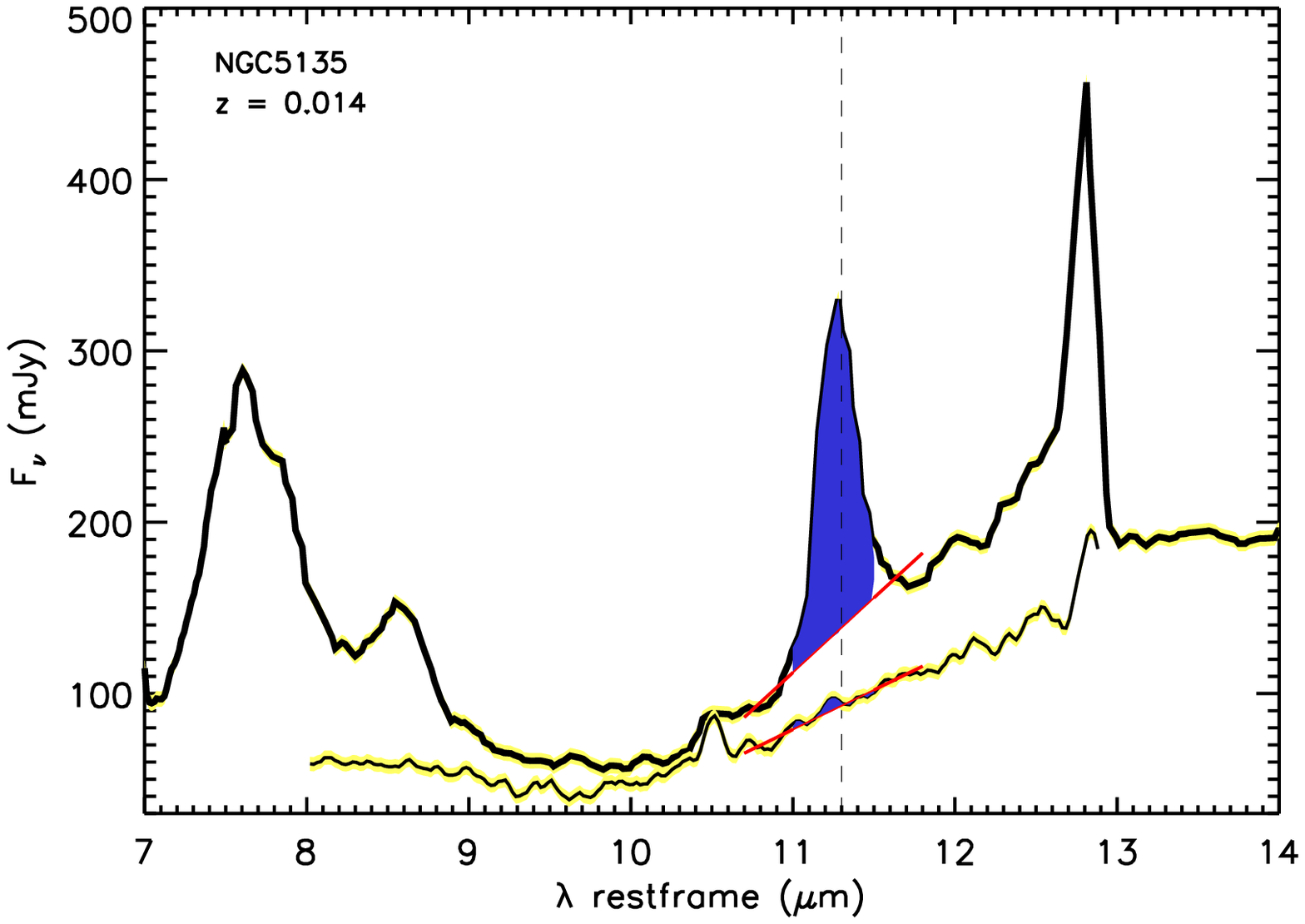}}
{\includegraphics[scale=0.33]{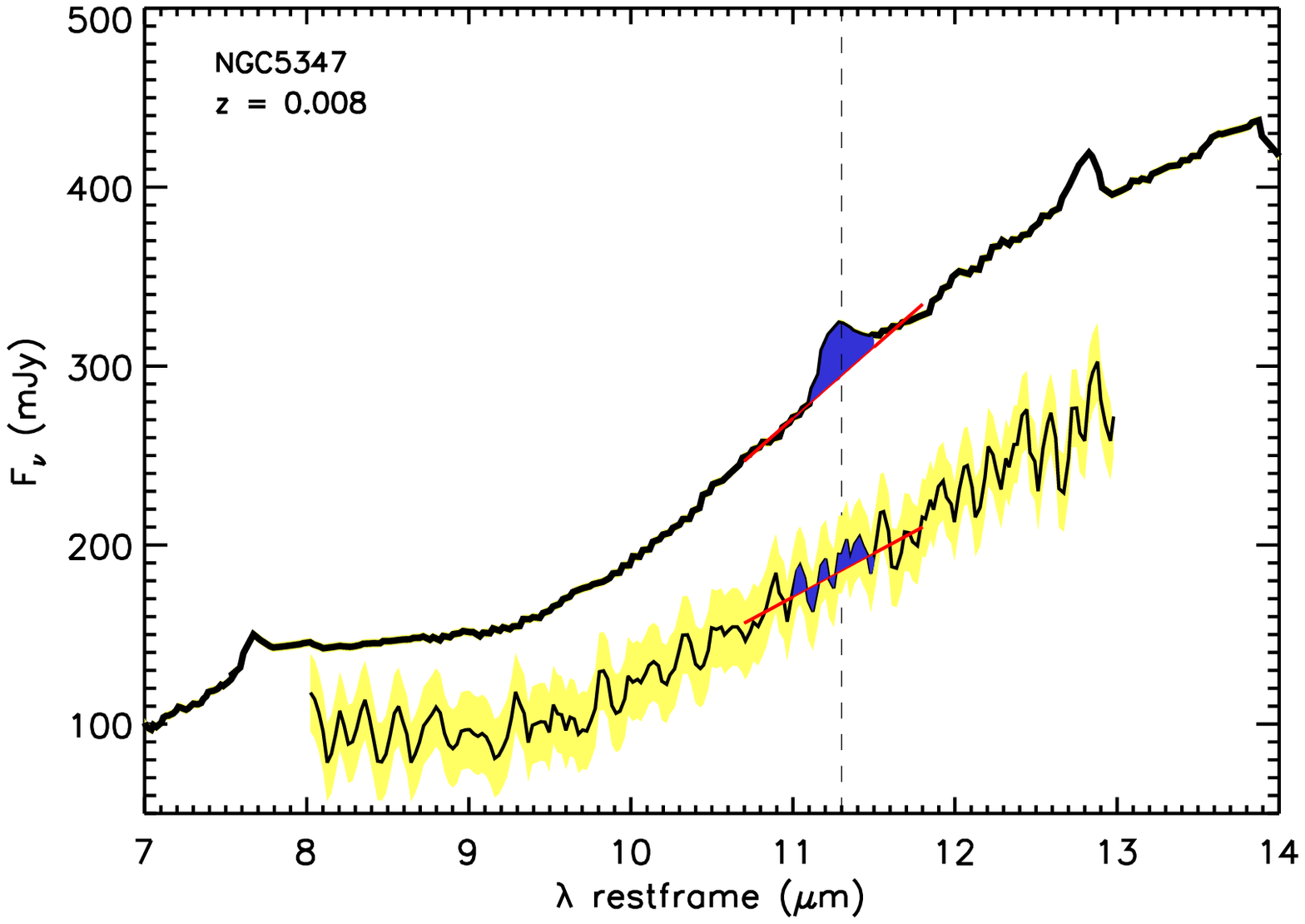}}\\[-1ex]

{\includegraphics[scale=0.33]{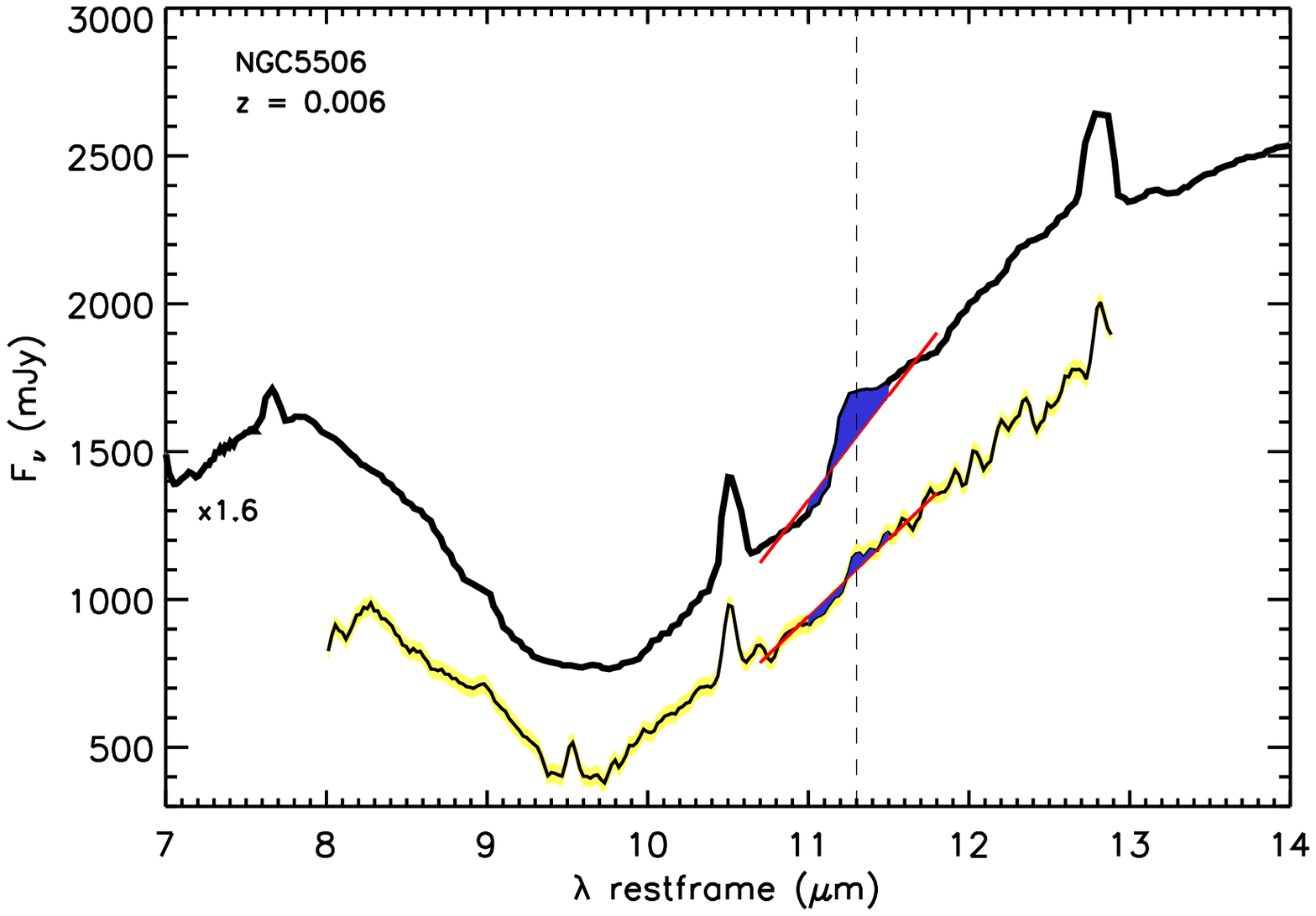}}
{\includegraphics[scale=0.33]{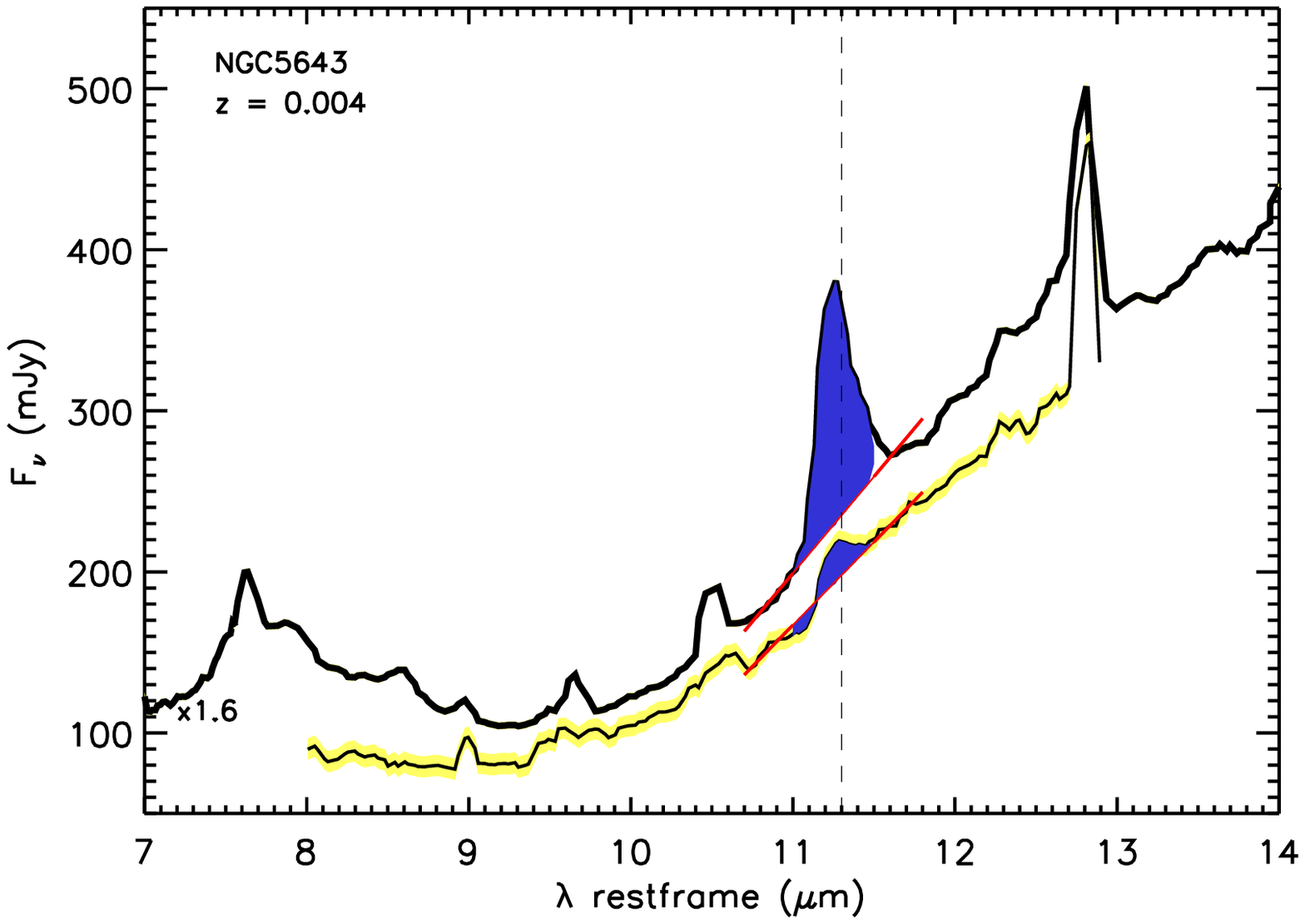}}
{\includegraphics[scale=0.33]{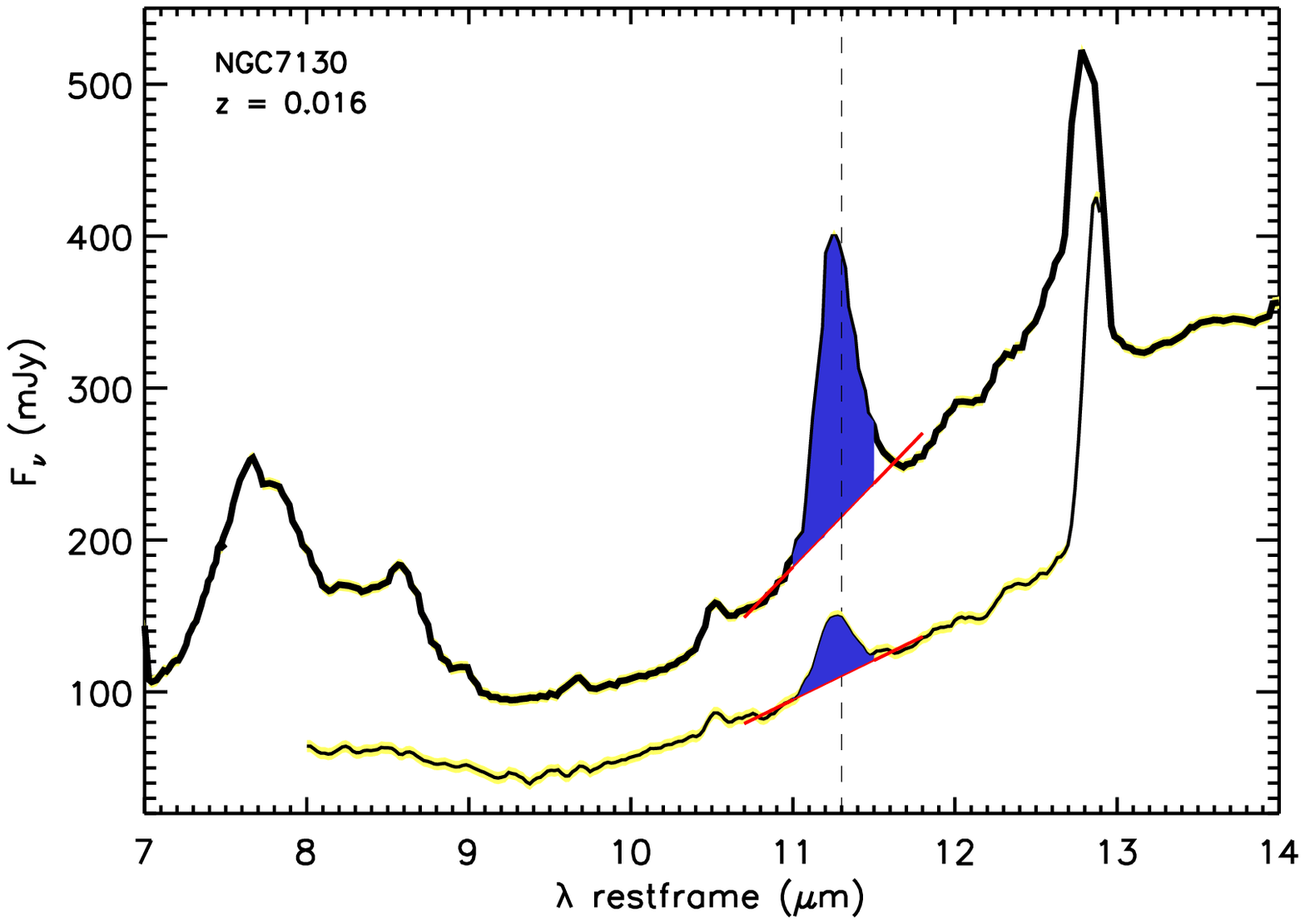}}\\[-1ex]

{\includegraphics[scale=0.33]{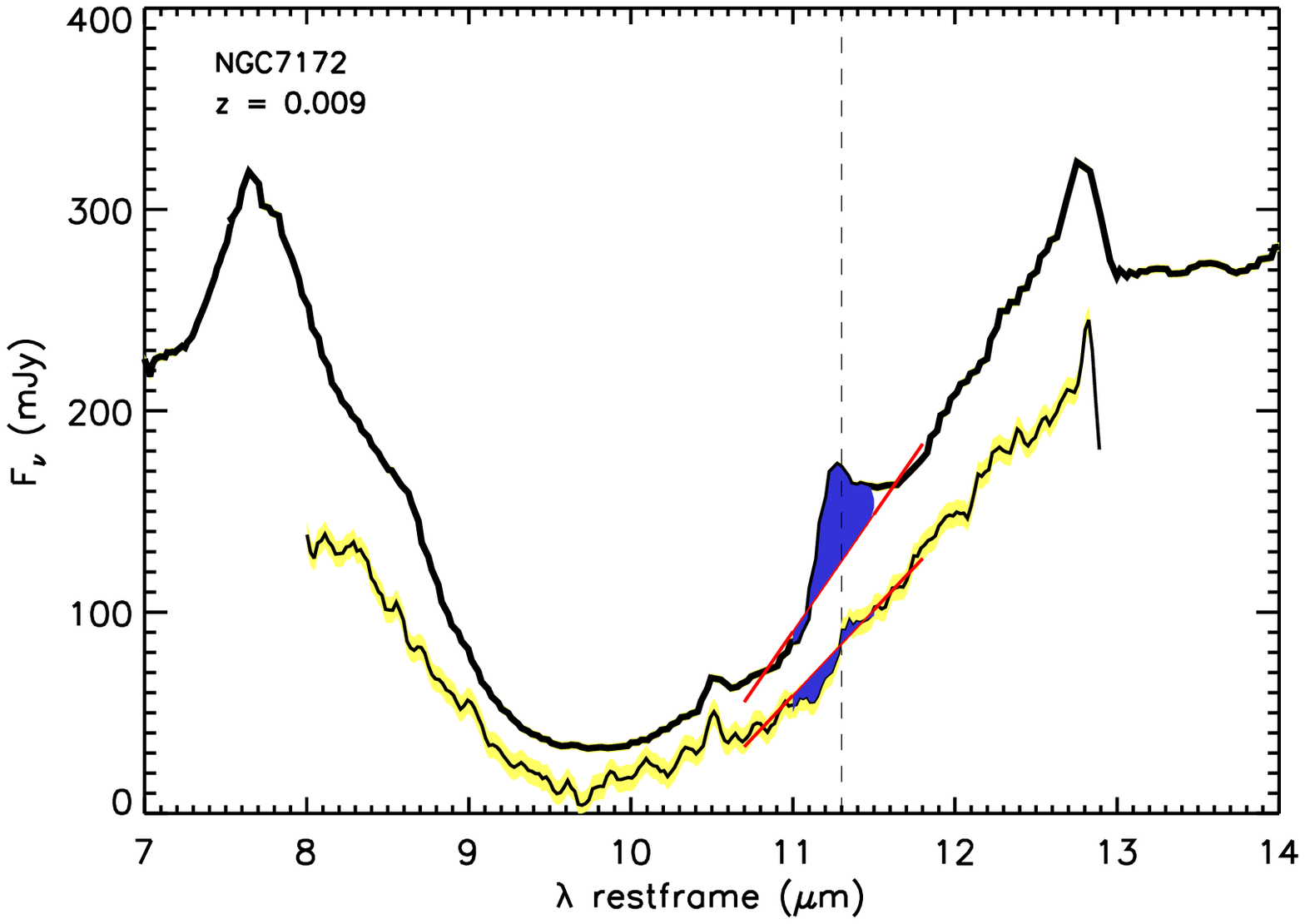}}
{\includegraphics[scale=0.33]{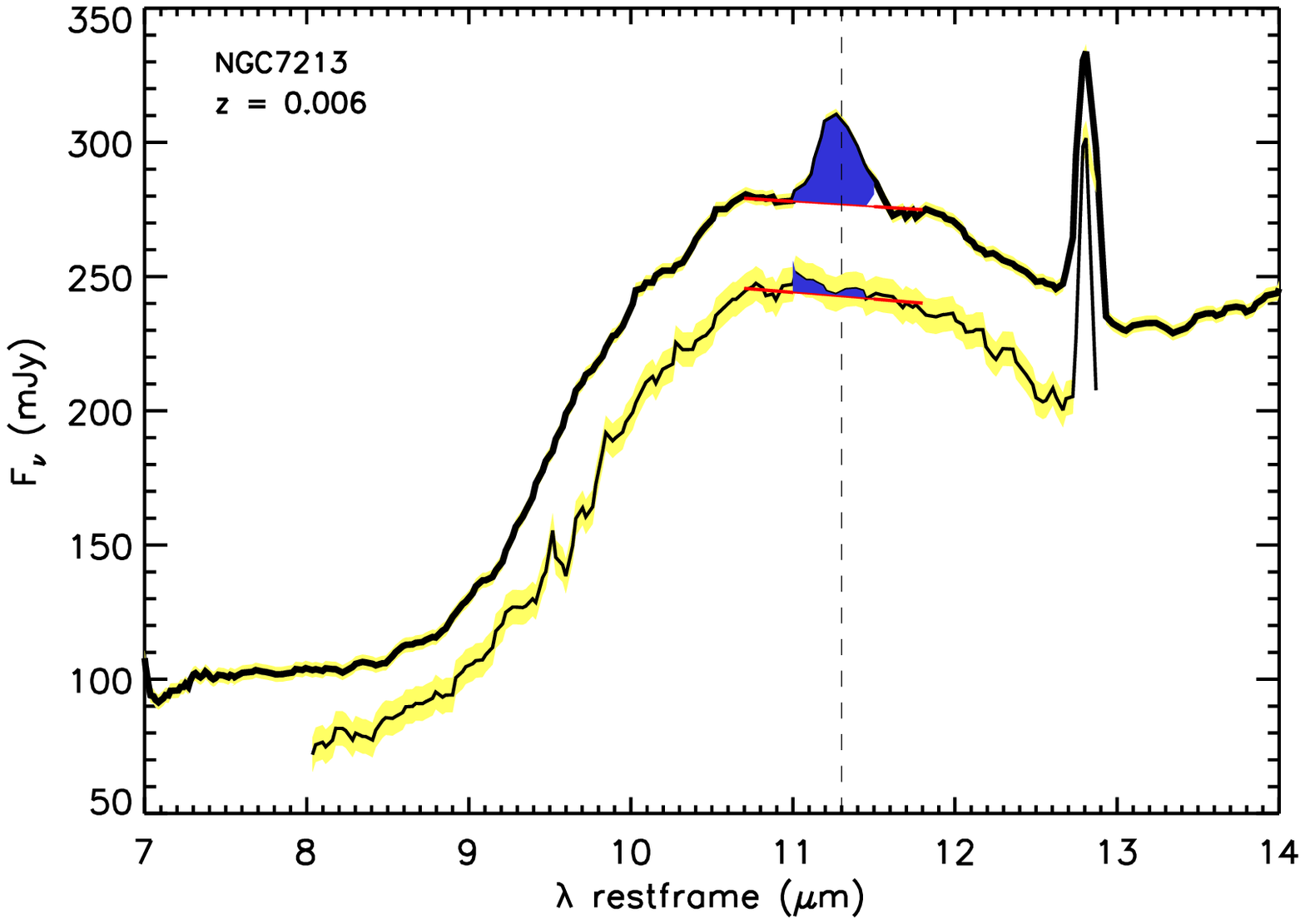}}
{\includegraphics[scale=0.33]{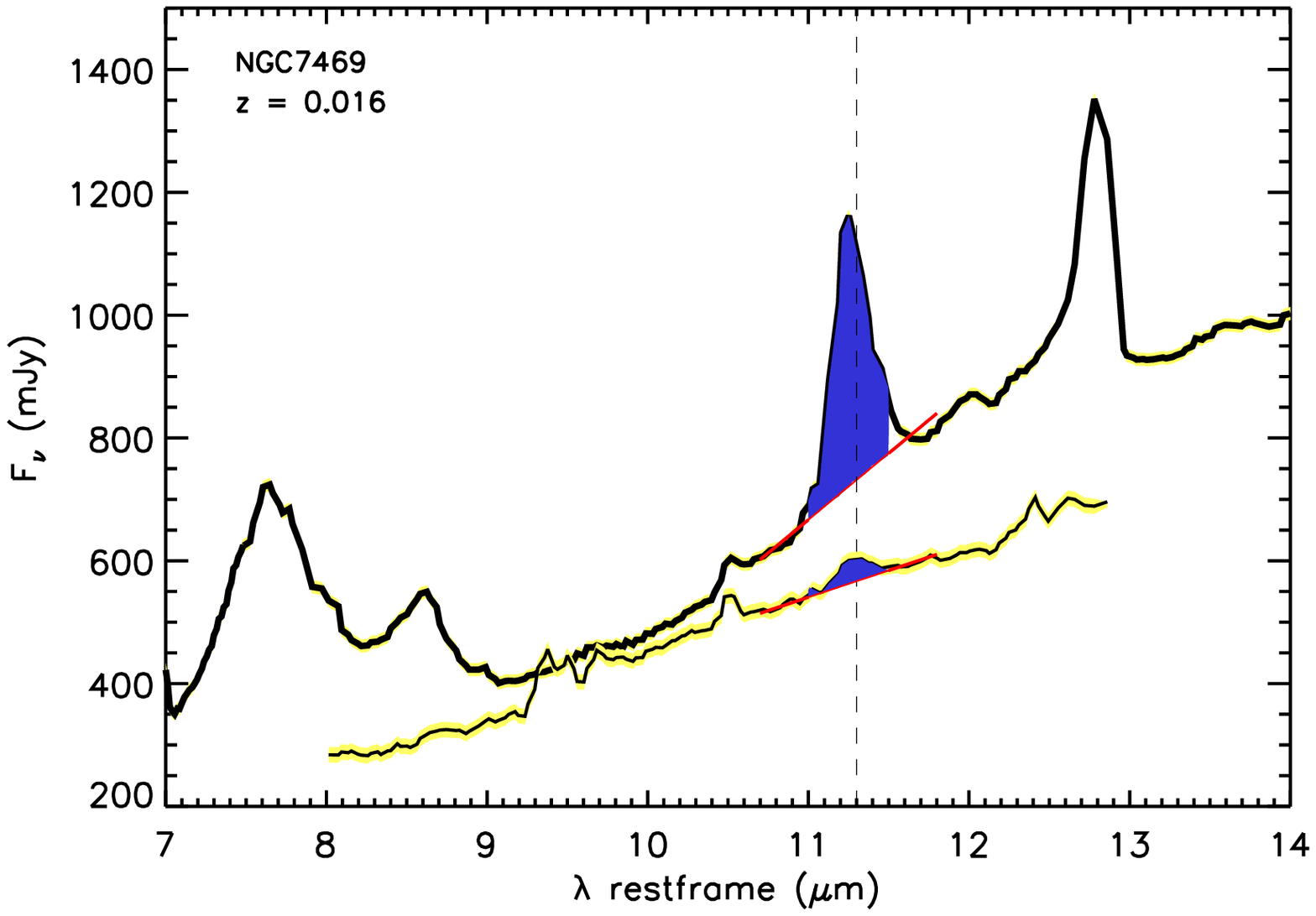}}\\[-1ex]

{\includegraphics[scale=0.33]{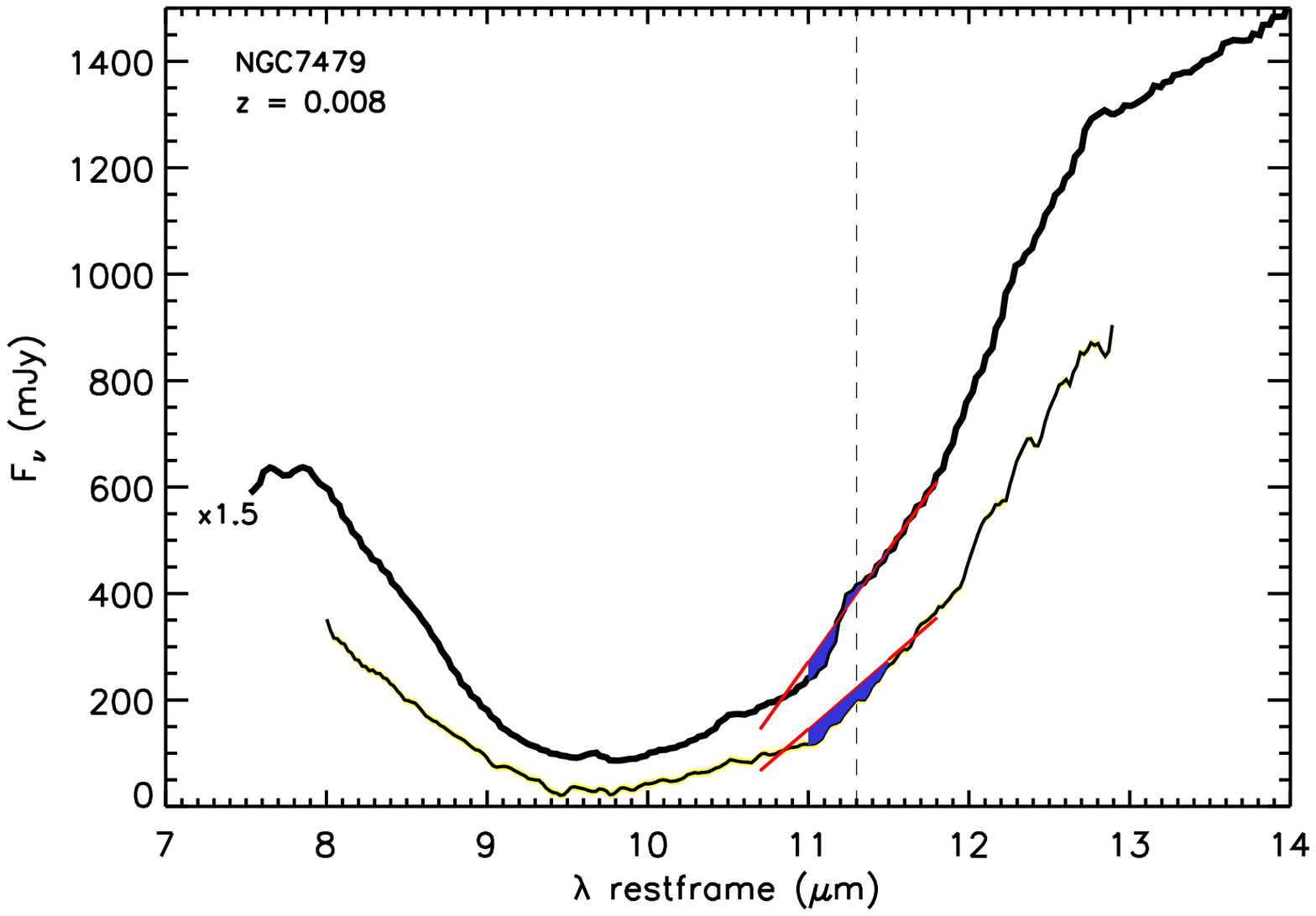}}
{\includegraphics[scale=0.33]{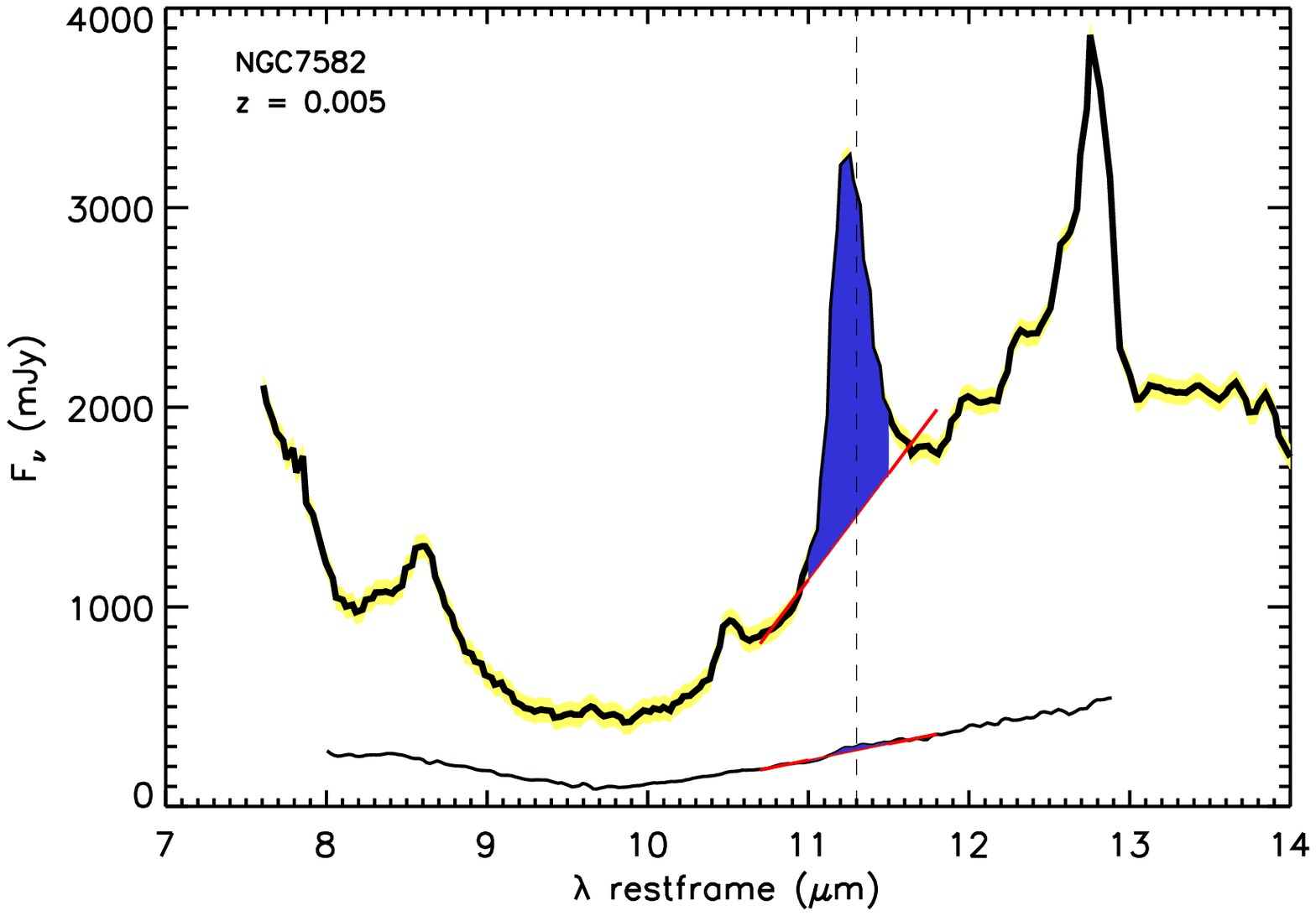}}\\[-1ex]

\setcounter{figure}{0}
\caption{Spectra of the sample. Continued.}
\end{tabular}
\end{figure*}

\end{document}